\documentclass[a4paper,usenames,dvipsnames,11pt]{article}
\pdfoutput=1

\usepackage{jheppub}
\usepackage{soul}

\usepackage{amssymb}
\usepackage{lipsum}
\usepackage{xcolor}
\usepackage{bm}

\usepackage[T2A,T1]{fontenc}
\usepackage[utf8]{inputenc}

\usepackage{upgreek}

\newcommand\ampsq{{\cal M}}
\newcommand\aem{\alpha}
\newcommand\aemotpi{\frac{\aem}{2\pi}}

\def\PDF#1#2{\Gamma_{\!#1/#2}}

\newcommand\sss{\scriptscriptstyle}

\newcommand{\bt}{\bar{t}}

\newcommand{\ord}{{\cal O}}
\newcommand\aNLOs{{\sc\small MG5\_aMC}}
\newcommand\aNLO{{\sc\small MadGraph5\_aMC@NLO}}

\newcommand\hsig{\hat{\sigma}}

\newcommand\MSb{\overline{\rm MS}}

\newcommand{\pt}{p_{\sss T}}

\newcommand{\deltaI}{\delta_{\sss\rm I}}

\newcommand{\LOG}{{\rm LO}_\Gamma}

\newcommand{\NNLOG}{{\rm NNLO}_\Gamma}

\newcommand{\dNNLOG}{\delta{\rm NNLO}_\Gamma}

\newcommand{\figtable}[3]{\begin{tabular}{ll}\multicolumn{2}{l}{#1}\\ #2 & #3\end{tabular}}

\title{Precision phenomenology at multi-TeV muon colliders}

\preprint{TIF-UNIMI-2025-12, IRMP-CP3-25-21, COMETA-2025-21}

\author[a]{Stefano Frixione,}
\author[b,c,d]{Fabio Maltoni,}
\author[d]{Davide Pagani,}
\author[e,f]{Marco Zaro\phantom{,}}

\emailAdd{Stefano.Frixione@cern.ch}
\emailAdd{fabio.maltoni@uclouvain.be}
\emailAdd{davide.pagani@bo.infn.it}
\emailAdd{marco.zaro@unimi.it}

\affiliation[a]{INFN, Sezione di Genova,Via Dodecaneso 33,  Genoa, I-16146, Italy}

\affiliation[b]{
Centre for Cosmology, Particle Physics and Phenomenology (CP3),
Universit\'e Catholique de Louvain, 
Louvain-la-Neuve, 
B-1348, 
Belgium
}

\affiliation[c]{
Dipartimento di Fisica e Astronomia, Università di Bologna, 
Via Irnerio 46, 
Bologna,
 I-40126, 
 Italy
 } 

\affiliation[d]{
INFN, Sezione di Bologna,
Via Irnerio 46, 
 Bologna,
 I-40126, 
 Italy
 }

\affiliation[e]{
 INFN, Sezione di Milano,
Via Celoria 16, 
Milano,
I-20133, 
Italy
 }

\affiliation[f]{
TIFLab, Università degli Studi di Milano,
Via Celoria 16, 
Milano,
I-20133, 
Italy
 }

\abstract{
Future lepton colliders, such as those based on linear $e^+e^-$ or circular
$\mu^+\mu^-$ accelerators, are expected to attain centre-of-mass energies in
the multi-TeV range. In this regime the impact of QED and of weak radiation,
in both the initial and the final state, can become a leading effect.
By employing a general framework presented in a companion paper --
suitable for any flavour of colliding leptons -- we improve next-to-leading
order electroweak predictions by including higher-order contributions, which
encompass, but are not limited to, vector-boson-fusion processes. We apply
this approach to the study of $t\bar{t}$ and $W^+W^-$ production at a
muon collider operating at centre-of-mass energies up to 10~TeV.
We show that such an approach, where both QED and weak contributions are
included at fixed order, in addition to the all-order resummation of
initial-state QED effects, can provide predictions for arbitrary observables
in all of the phase space which are precise at the percent
level.}

\begin{document}
\maketitle

\section{Introduction}
\label{sec:int}

A new generation of lepton colliders based on novel acceleration technologies
-- such as plasma wakefield acceleration~\cite{AWAKE:2022aeo}, and muon
cooling/rapid acceleration~\cite{InternationalMuonCollider:2025sys} -- might
be shown to be viable as early as the next decade. If that will be the case,
the prospect of colliding leptons in the multi-TeV energy range will become
reality before the mid-21$^{st}$ century, providing one with an unprecedented
probe that will help deepen our understanding of fundamental interactions, and
in particular of the electroweak ones, by opening up a new regime of
exploration, which exploits a unique combination of high-precision measurements
and of high-energy observables, with a formidable physics
potential~\cite{Buttazzo:2018qqp,Costantini:2020stv,Buttazzo:2020uzc,
AlAli:2021let,Aime:2022flm,Accettura:2023ked,InternationalMuonCollider:2025sys}.

This exciting possibility has motivated extensive studies of the physics reach
of such machines, both for precision Standard Model (SM) measurements and in
the searches for resonant physics Beyond the SM (BSM) -- see the references in
\cite{InternationalMuonCollider:2025sys} for a recent overview.  A key aspect
which has attracted significant attention is the achievable precision and the
accuracy of SM predictions in this new regime, where approaches employed so
far in $e^+e^-$ simulations will most likely become either unpractical or not
sufficiently precise and/or accurate. One prominent example of this situation
is the proposal to use electroweak (EW) Parton Distribution Functions
(PDFs)~\cite{Chen:2016wkt,Han:2020uid,Han:2021kes,Garosi:2023bvq,
  Capdevilla:2024ydp,Marzocca:2024fqb,Han:2025wdy} to describe vector boson
fusion (VBF) processes, which are expected to play a significant role in
high-energy lepton collisions. This approach aims to account for potentially
large logarithmic corrections of collinear and soft origin in the initial
state, while also simplifying the relevant calculations\footnote{Similar
  methodologies have been proposed to incorporate the effects of final-state
  EW radiation (see e.g.~ref.~\cite{Bauer:2018xag}); we shall not be concerned about them in this work.}.  However,
all techniques of this kind -- while {\em formally} improving certain aspects
of fixed-order results, such as the resummation of EW radiation -- rely on
kinematic approximations rigorously valid only in asymptotic regimes, which
are achieved at logarithmic pace. In addition to that, the current state of
the art for EW PDFs is based on LL resummation and LO matrix-element
computations, and therefore it lacks control of higher orders. At the energy
scales currently envisioned for future muon colliders -- of the order of
10~TeV -- any EW-PDF approach may prove 
to be neither accurate nor precise enough. 
Note that the target precision of a multi-TeV muon collider depends strongly
on the observable under consideration. For instance, Higgs coupling
measurements -- relying on vector-boson-fusion production near threshold --
are expected to reach per-mille-level precision. Conversely, indirect
searches for new physics in the high-energy tails require percent-level
precision\footnote{This corresponds, assuming that new physics effects scale
quadratically with energy, to the target sensitivity of up to $O(10^{-5} -
10^{-6})$ for high-precision observables at electroweak-scale lepton
colliders.}.  It is therefore desirable to set up a theoretical framework
that is systematically improvable and can ultimately achieve the needed level
of precision in all of the phase space.

In view of this, in this work and in its companion paper~\cite{Frixione:2025wsv} 
we adopt a different approach. Firstly, we acknowledge 
the necessity of resumming initial-state collinear QED radiation into suitable 
PDFs, and build upon recent analytical and numerical advancements that 
achieve next-to-leading logarithmic (NLL) accuracy for the 
latter~\cite{Frixione:2019lga,Bertone:2022ktl,Frixione:2023gmf,
Bonvini:2025xxx}. Secondly, rather than attempting to resum the analogous 
weak-radiation effects, we include them order by order by means of 
perturbative calculation of fixed-order predictions, at the 
next-to-leading order (NLO) in the EW coupling.
Thirdly, by means of a general (process- and observable-independent)
framework we supplement such QED-resummed and NLO EW-accurate predictions 
with the contributions stemming from a gauge-invariant and finite
subset of next-to-next-to-leading order (NNLO) corrections, thus 
improving the former predictions in a systematic manner. In particular, 
such a subset leads to the inclusion of the effects due to $\gamma/Z$
neutral VBF processes that appear at the NNLO at the matrix-element level,
which dominate the threshold region.

By analysing two explicit and non-trivial cases -- namely $t \bar{t}$ and
$W^+W^-$ production at a muon collider with a centre-of-mass energy of several
TeVs -- we demonstrate that: {\em i)}~percent-level accurate predictions can
be obtained across the entire accessible phase space in a single step; 
{\em ii)} there is no indication of any perturbative breakdown in fixed-order
results that would necessitate the usage of EW PDFs and/or fragmentation
functions to recover a well-behaved series\footnote{Note, however, that while
  a perturbative breakdown in fixed-order results may occur at multi-TeV
  energies, it is generally not related to EW PDFs or fragmentation
  functions. Instead, it arises from large electroweak Sudakov logarithms due
  to virtual corrections. See the discussion in sect.~\ref{sec:results_sdk}.};
{\em iii)} in the kinematic regions where the approximations underlying the EW
PDFs are valid, the cross section remains relatively small.

Furthermore, we point out that the method we propose is not only process- and
observable-independent, but also systematically improvable. For instance,
charged-current VBF processes can be included currently up to the NLO without
introducing any double counting, while Sudakov suppression effects in the
high-energy regime can also be incorporated, e.g.~by following the approach of
ref.~\cite{Denner:2024yut}.

This paper is organised as follows. In section~\ref{sec:motivation} we outline
our motivations and set our goals; we also give the most straightforward
evidence as to why an EW-PDF approach is unsuitable for high-energy
phenomenology. In section~\ref{sec:app} we present our approach, and in 
section~\ref{sec:res} we discuss the results obtained with it
for $t\bar t$ and $W^+W^-$ production at a muon collider, focusing on the
scenarios where its centre of mass energy is equal to 3 and 10 TeV. 
There, we also discuss two important aspects, i.e.~the factorisation-scheme
dependence and the necessity to include EW Sudakov logarithm resummation effects in some
regions of phase space. We draw our conclusions in section~\ref{sec:con}.

This is a phenomenologically-oriented paper. The reader interested in
the technical aspects of our approach can find them in ref.~\cite{Frixione:2025wsv}.
There is a limited amount of overlap between the two papers as far as
the introductory material is concerned; this is expressed in a language
which reflects the basic nature, technical or otherwise, of the work
to which it belongs.


\section{Motivation and goals\label{sec:motivation}} 
So far, $e^+e^-$ colliders have been operated at 
relatively low energies, where most processes are
initiated by lepton annihilation, and final states are characterised by low
multiplicities. In this regime, theoretical predictions need not account for
complex electroweak partonic dynamics, either in the initial or in the final
states\footnote{Here we have mostly in mind initial-state and final-state
collinear radiation. Soft-radiation effects in the initial state can be very 
relevant for precise measurements in situations where kinematic 
configurations are severely constrained, such as for runs at the $Z$-pole, 
and have been studied in great detail~\cite{Altarelli:1996gh}.}. 
However, proposed future high-energy machines, 
such as multi-TeV muon colliders, pose new challenges. As the collider
centre-of-mass energy $\sqrt{s}$ increases, leptons begin to exhibit partonic
substructure at lower momentum fractions, where the contribution of photons
is increasingly important and becomes dominant, with a rate of growth only
matched, and eventually exceeded, by that of gluons and quarks. Moreover, 
higher energies allow one to access richer phase spaces, potentially 
mimicking hadron-like final states with increased multiplicity.

\begin{figure*}[t!]
	\centering 
    \includegraphics[width=0.8\textwidth, angle=0]{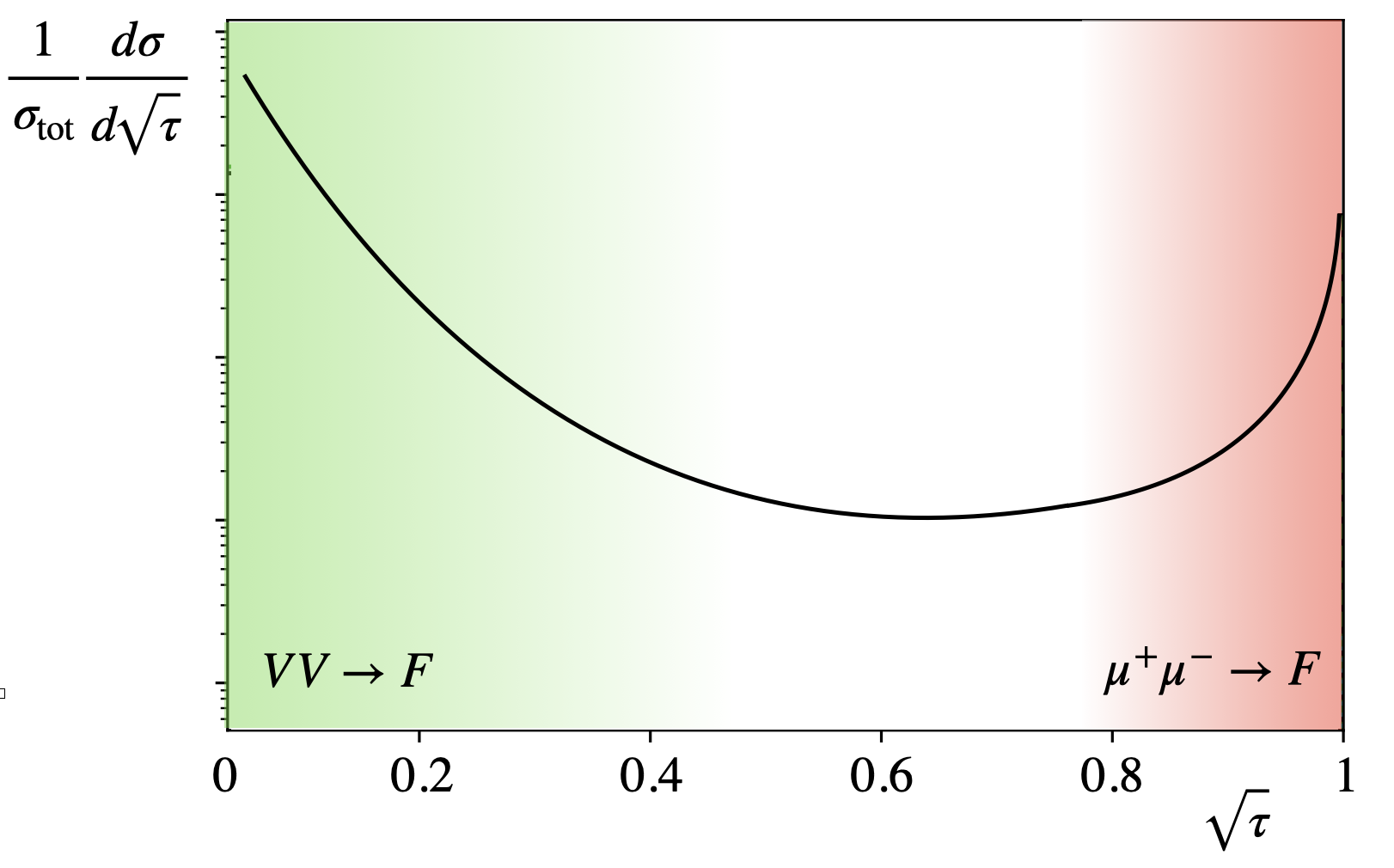}	
\caption{Representative shape of the normalised invariant mass distribution in
$\sqrt{\tau}$ (where \mbox{$\tau=M_F^2 /s$}) of the
final state $F$ at a high-energy muon collider (the actual curve
is taken from $t\bar t$ production at 10 TeV). The threshold region 
(green) is dominated by underlying vector boson fusion $VV\to F$ processes
(with $V=\gamma,Z,W^\pm$); the $\gamma\gamma$ contribution is dominant,
and there are sizeable power-corrections effects when $V=Z$ and
$V=W^\pm$. In the $\sqrt{\tau}\simeq 1$ region (red), the dominant 
process is $\mu^+\mu^-\to F$ annihilation. The region where the 
approximations employed in the EW-PDF LO approach are expected to 
work best (white) is where the cross section is at its minimum. }
	\label{fig:flag}
\end{figure*}

In general, several different mechanisms will contribute to a given final
state. In fig.~\ref{fig:flag} we sketch the typical high-energy behaviour 
of the differential cross section for the production of a final state $F$ 
(with invariant mass $M_F$) with a fraction $\sqrt{\tau}$  
of the total energy of the collision (where $\tau=M_F^2/s$). Two
leading regions, that correspond to local or global maxima of the cross
section, can be clearly identified. The threshold region at small values of
$\sqrt{\tau}$ is typically dominated by processes that feature collinear
emissions of vector bosons from the incoming lepton lines, which bosons 
then fuse into the final state $F$ -- the underlying hard scattering thus 
being $VV\to F$, with $V=\gamma, Z,W^\pm$; there, the $V=\gamma$ contributions 
are increasingly important for decreasing $\tau$ values. Conversely, for 
values of $\sqrt{\tau}$ close to 1 (i.e.~at the kinematic limit), the final 
state carries almost all of the energy of the collider, and therefore it is
$\mu^+\mu^-$ annihilation which gives the dominant contribution.

As one moves from the threshold to the kinematic limit, i.e.~from a
$\gamma\gamma$-dominated to a $\mu^+\mu^-$-dominated situation, one
enters, traverses, and then leaves, a region (depicted in white in 
fig.~\ref{fig:flag}) where  a non-negligible contribution is given by the heavy 
electroweak vector bosons emitted quasi-collinearly by the incoming 
beams, which  by colliding  produce the final state $F$. As is shown in
the figure, this region typically corresponds to a local {\em minimum}
of the cross section; from a diagrammatic viewpoint, the production
mechanism can loosely be associated with VBF topologies.

The cross sections for VBF-like configurations are often approximated 
by using a parton model that treats $W$ and $Z$ bosons as constituents of 
the lepton, based on the so-called Effective Weak-boson Approximation
(EWA)~\cite{Dawson:1984gx,Kane:1984bb,Kunszt:1987tk}. This approach is only
valid in the asymptotic limit of infinite energy, where neglected non-VBF
contributions (which are {\em not} logarithmic in the mass of the 
relevant vector boson) and power corrections vanish\footnote{In this 
paper we focus on initial-state radiation effects. Other analogous effects, 
such as final-state fragmentation, which are also enhanced at high energy, 
can be addressed by means of the same techniques.}. However, for realistic 
collider energies (up to ten(s) of TeV), these contributions generally remain 
significant\footnote{Some of these, subject to the condition that they be
universal, {\em may} be included in the EWA (but not in the EW PDFs) at
the LO. This is expected to improve the accuracy of the approximation,
but to degrade its precision owing to the loss of a clearly-defined
separation between the classes of contributions which are included in
vs excluded from the results. This matter will be further discussed
towards the end of this section.\label{ft:aEWA}}.

Importantly, VBF-like diagrams are just a sub-class of the complete 
gauge-invariant set of graphs 
for a given final state. Thus, if one only considers their contributions
(as one does in methods that rely on the EWA or the EW PDFs) one is liable
to encounter problems since, in order to achieve accurate predictions 
useful for phenomenology, an approach must be valid across the entire phase 
space, with smooth transitions between VBF-dominated and VBF-suppressed regions. 
In any scenario deemed to be realistic in the next century or so, this 
entails exact, gauge-invariant computations that retain full dependence 
on the boson masses without any approximation.

More in detail, while simple in essence the use of EW PDFs is affected
by intrinsic limitations, which need to be carefully considered when 
assessing the reliability of such an approach. In particular, the impact
of the following characteristics must be determined.

\begin{enumerate} 
\item Power-suppressed effects: EW PDFs exactly encode only
logarithmic mass dependence. Effects which are polynomial 
in $m_V^2/Q^2$ (with $V=W,Z$) and/or in its square root are lost despite 
the fact that they can be important, especially near threshold (see footnote~$\ref{ft:aEWA}$).  

\item Theoretical accuracy: EW PDFs are only available at
leading logarithmic (LL) accuracy; in keeping with that, the corresponding 
short-distance coefficients can currently be computed only at the 
leading order. Higher-order corrections to PDFs stemming from heavy bosons,
and the corresponding adjustments at the matrix elements level,
are challenging and not well understood. 

\item Diagrammatic coverage: only the subset of VBF-like 
diagrams is associated with the usage of EW PDFs, and the impact of 
omitted diagrams cannot be quantified in a process- and, crucially,
observable-independent way.

\item Phase space coverage: the kinematical approximation on which 
the EW PDF approach is based is only valid in a very small part of the
phase space, i.e.~far from both the threshold and the kinematical
boundaries; in that region, the rates are small.

\item Simplified kinematics: simulations based on EW PDFs are fully
inclusive over weak boson radiation in the initial state, and in boson decay
products; it is therefore difficult to faithfully mimic realistic 
experimental observables.

\item Error estimation: many uncertainties are not parametric,
and thus cannot be reliably quantified.

\item Weak resummation effects: EW PDFs
differ little from the EWA. Since these differences are typically smaller 
than the combined intrinsic uncertainties, the possible benefits of these 
methods are offset by their limitations.

\end{enumerate}
In order to quantify the impact of the issues mentioned above on
the overall accuracy of an EW-PDF approach, one would need to make
an extensive comparison with an alternative reliable technique. However, since
the goal of this paper is to provide the first phenomenological applications,
and an assessment on its scope and precision, of the alternative method 
introduced in a companion paper (ref.~\cite{Frixione:2025wsv}), we defer such a 
comparison to future work. Here, we limit ourselves
to giving an indication of the consequences stemming from some of the
issues highlighted above. We do so by presenting two simple, yet quite 
compelling, arguments. Firstly, we show a tree-level 
comparison between 
predictions obtained at the matrix-element level and from the EWA, for a
selected class of processes leading to $t\bar t$ and $W^+W^-$ production.
Secondly, by computing parton-parton luminosities, we show that the effects of
the EW resummation as is included in the EW PDFs are mild, and in fact much
smaller than the size of the intrinsic approximation which underpins the LO EW
PDF approach. Taken together, these comparisons support the conclusions that:
{\em a)}~predictions based on either the EWA or the EW PDFs may not be
accurate enough to be used in phenomenology, and one would need to check 
their reliability process by process and observable by observable; 
{\em b)}~the improvement resulting from trading the EWA for the EW PDFs 
is much smaller than the effects which are neglected by considering only 
VBF topologies. In that, one finds an additional motivation for the 
definition and the implementation of an approach alternative to them.

\begin{figure*}[t!]
	\centering 
    \includegraphics[width=\textwidth, angle=0]{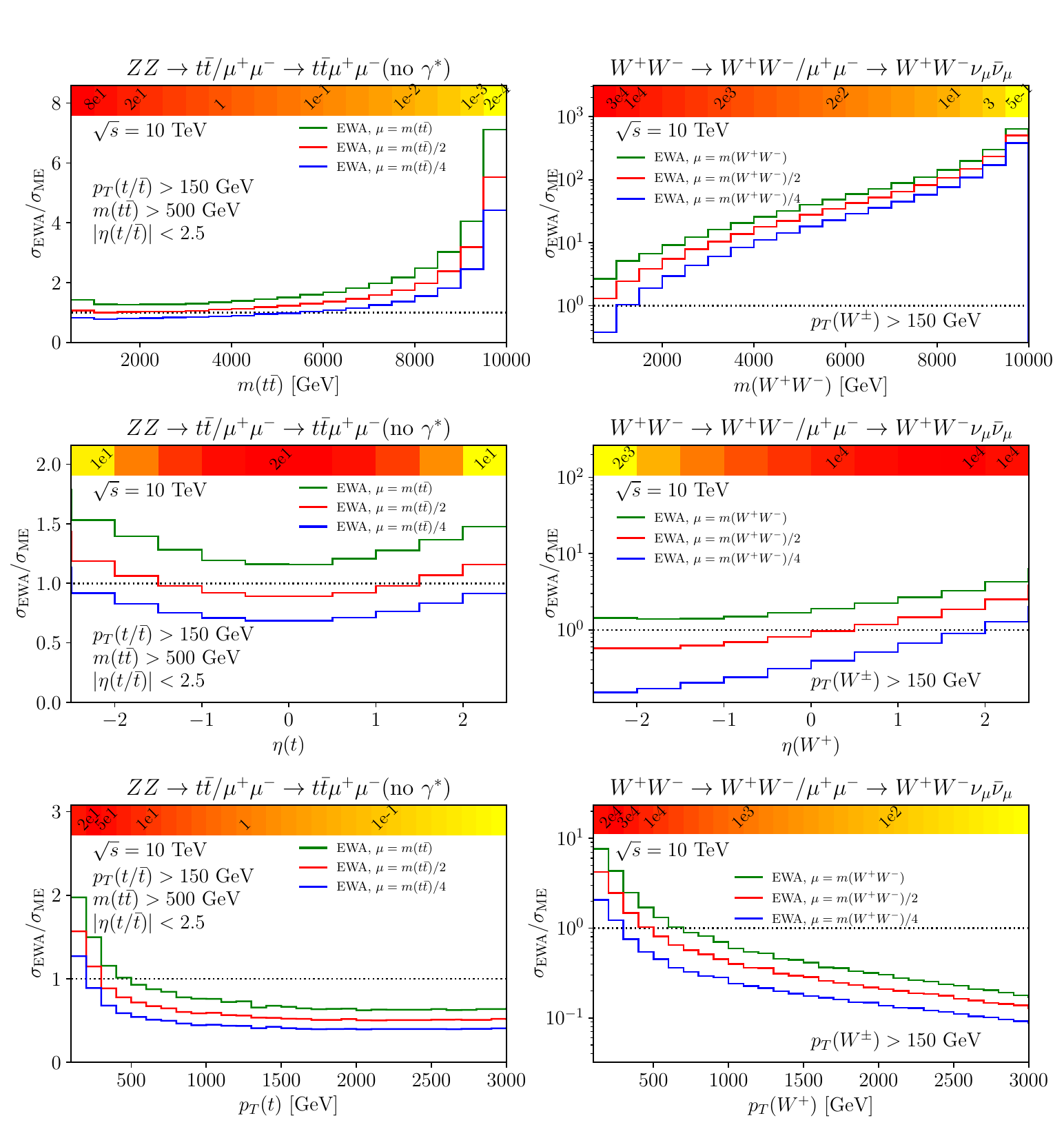}	
\caption{Ratios of the  EWA predictions $(2\to 2)$ over those of the
 matrix elements $(2\to 4)$ for $m(t\bar t)$, $\eta(t)$, and $\pt(t)$ 
(left panels), and their analogues  in $W^+W^-$ production
(right panels) -- see the text for details.}
	\label{fig:mevspdf}
\end{figure*}

Let us then consider the production\footnote{For a phenomenological application at hadron colliders,
    see also ref.~\cite{Dittmaier:2023nac}.} of an $(X,\bar X)$ pair with
either $(X,\bar X)=(t,\bar t)$ or $(X,\bar X) = (W^+, W^-)$, and analyse 
the differences between differential distributions obtained by means of:
\begin{itemize}
\item The $V_1V_2\to X\bar X$ matrix elements, where $V_i$ are either
$W$'s or $Z$'s (kept on their physical mass shell), 
convoluted with the relevant EWA functions, henceforth referred to
simply as ``EWA''.
\item The corresponding $\mu^+\mu^-\to X\bar{X}\mu^+\mu^-$ 
matrix elements, henceforth possibly referred to as ``ME''.
\end{itemize}
In fig.~\ref{fig:mevspdf} we present the ratios of the EWA predictions, 
obtained with three typical choices for the factorisation scale that enters 
the EWA functions, over those stemming from the $2\to 4$ matrix elements,
with the c.m.~energy set equal to 10~TeV.
The results for $t\bt$ and $W^+W^-$ production are on the left and on the
right panels, respectively, with the top, middle, and bottom panel relevant
to the pair invariant mass $m(X\bar{X})$, the pseudorapidity of the
positively-charged heavy particle $\eta(X)$, and its transverse 
momentum $p_T(X)$. The coloured band on top of each plot is indicative 
of the differential rate of the matrix-element-based distribution, 
ranging from red (large cross section) to yellow tones (small cross
section). Actual values for the binned cross section (in pb per bin) are
written in a few bins. All of the results have been produced with 
\aNLO\ (\aNLOs\ henceforth)~\cite{Alwall:2014hca,Frederix:2018nkq,
Frixione:2021zdp,Bertone:2022ktl} 
whose default EWA implementation is that of ref.~\cite{Ruiz:2021tdt}.

More precisely, the $t\bt$ results in the left column of
fig.~\ref{fig:mevspdf} have been obtained as follows. The $2\to 2$
process which underpins the EWA result is $ZZ \to t\bar{t}$; for the
EWA functions we adopt the parametrisation of sect.~3.4 of 
ref.~\cite{Garosi:2023bvq}, and this in order to be as self-consistent
as possible, since later we shall compare luminosities obtained from 
the EWA with those from the EW PDFs of that same ref.~\cite{Garosi:2023bvq}.
Likewise, for a comparison which is as close as possible between the
$2\to 2$ and $2\to 4$ results, in the $\mu^+\mu^- \to t\bar{t}\mu^+\mu^-$ 
matrix elements we exclude the annihilation channels\footnote{This is a 
gauge-invariant class. Technically, it can be obtained by requiring the 
matrix element generator to treat $\mu^+$ and $\mu^-$ as if they had 
different flavours, e.g.~by replacing $\mu^-$ with $e^-$ both in the 
initial and final state. In this way, topologies such as 
\mbox{$\mu^+\mu^-\rightarrow t \bar t (\gamma/Z \rightarrow)\mu^+\mu^-$}
are safely excluded. We shall use the same method also for $W^+W^-$ 
production, to be discussed below. \label{foot:flav}}. In addition to that,
in order to remove the contributions of intermediate photons (and thus their 
interference with the $Z$ bosons, which is not accounted for in the EWA),
we veto all QED interactions\footnote{While in general this operation would
violate gauge-invariance, as e.g.~in the case of 
\mbox{$ \mu^+ \mu^- \to W^+ W^-\mu^+ \mu^- $}, this does not happen in
the present case, thanks to the fact that
\mbox{$ \mu^+ \mu^- \to t\bar{t} \mu^+ \mu^- $} diagrams do not feature
$W$ bosons. Having said that, we note that at muon colliders with energies 
up to tens of TeV, the $Z$-pair induced contribution to $t\bar t$ production 
has a very small rate with respect to the photon-induced 
channel.\label{foot:zztt}}. Simple acceptance cuts are imposed (see
the labels in fig.~\ref{fig:mevspdf} or eq.~(\ref{eq:cuts})), which
limit the contribution from regions of the phase space close to its
borders---regions where the discrepancies between the EWA and ME approaches
would otherwise be even more pronounced. In particular, in this setup
additional diagrams contributing to the full $ 2 \to 4 $ process (such as
the radiation of a $Z/\gamma^*$ from one muon line, which then branches 
into a $t\bar t$ pair) are kinematically suppressed, and the dominant 
contribution stems from an effective $ZZ \to t\bar t$ topology.
This is a deliberately academic scenario: both the process definition 
(diagram selection) and the cuts have been chosen so that differences 
are minimised; in other words, this is a best-case situation.

One starts by observing that the EWA exhibits a significant
scale dependence, which in itself precludes a precise comparison with the full
result. This dependence is due to the transverse polarisations of the
initial-state $Z$ bosons, since the longitudinal modes do not depend on the
scale. Focusing on the top panel on the left column, one sees that for 
smaller scale choices and at moderate invariant masses (approximately 
smaller than 6~TeV), the EWA and the ME agree at the 20--30\% level. 
However, the shapes of the pseudorapidity distributions are considerably different, 
leading to discrepancies of a similar magnitude. The situation is even
worse in the case of $\pt(t)$, where the discrepancies between the
EWA and the ME predictions can be as large as 100\%, even under the 
most favourable scale choices.

The bottom line is this: while the EWA and ME descriptions can be made
to agree under selective choices of kinematical configurations and scale 
parameters (and still within significant uncertainties), this is at best 
true in some one-dimensional projections of the phase space, i.e.~for one
observable at a time. When one considers the phase space in its true
multi-dimensional nature, even this rough agreement breaks down.

We now turn to discussing the case of $W^+W^-$ production, whose results
are shown in the right-hand panels of fig.~\ref{fig:mevspdf}.
In this case, our goal is to compare the EWA and ME descriptions emerging
from charged-current (sub)processes. Thus, in the case of the EWA 
we select $W^+W^-$ fusion, whereas for the full process 
\mbox{$\mu^+ \mu^- \to W^+ W^- \nu_\mu \bar \nu_\mu$} neutral currents
contributions involving $\mu^+\mu^-$ or $\nu_\mu\bar \nu_\mu$ annihilation
and/or creation are excluded (see footnote~\ref{foot:flav}). 
At variance with the case of $t\bt$ production,
we aim to explore a larger fraction of the phase space in order to
assess the full spectrum of possibilities. 
So while we relax the acceptance cuts, these cannot be completely
eliminated, in view of the presence of a pole in the $W^+W^- \to W^+W^-$
scattering amplitude, due to the exchange of a photon in the $t$ channel;
this, we avoid by imposing \mbox{$\pt(W^\pm)>150$~GeV}. Indeed, in general
for a given partonic c.m.~energy $\sqrt{s}$, by requiring
$\pt(W^\pm)>\pt^{\rm cut}$, in the limit $\pt^{\rm cut} \to 0$ 
one obtains a cross section proportional 
to \mbox{$(\sqrt{s}/\pt^{\rm cut})^2$}.
It is important to note that such a pole is actually an artefact of the 
EWA (and EW PDF) approach, because it is not present in the full 
$2\to 4$ matrix element. There, the $W$'s stemming from the muon branchings
(which enter the sub-diagrams where they play the same role as the
initial-state $W$'s in the $2\to 2$ matrix elements employed by the EWA)
are space-like, and therefore off-shell, and thus automatically regulate
the potential $t$-channel divergence. In conclusion, if no cut
is imposed, the EWA simply leads to a divergent cross section, and
therefore cannot be used sensibly. Even by imposing a $\pt(W^\pm)$ 
cut, as we have done,  results are still IR-sensitive, owing to
the $\pt^{\rm cut}$ dependence mentioned above. Indeed, the inspection
of fig.~\ref{fig:mevspdf} shows extremely large discrepancies between
the EWA and the ME approaches in both shapes and rates -- the EWA 
approximation is not under control. 

\begin{figure}
	\centering 
\includegraphics[width=0.9\textwidth, angle=0]{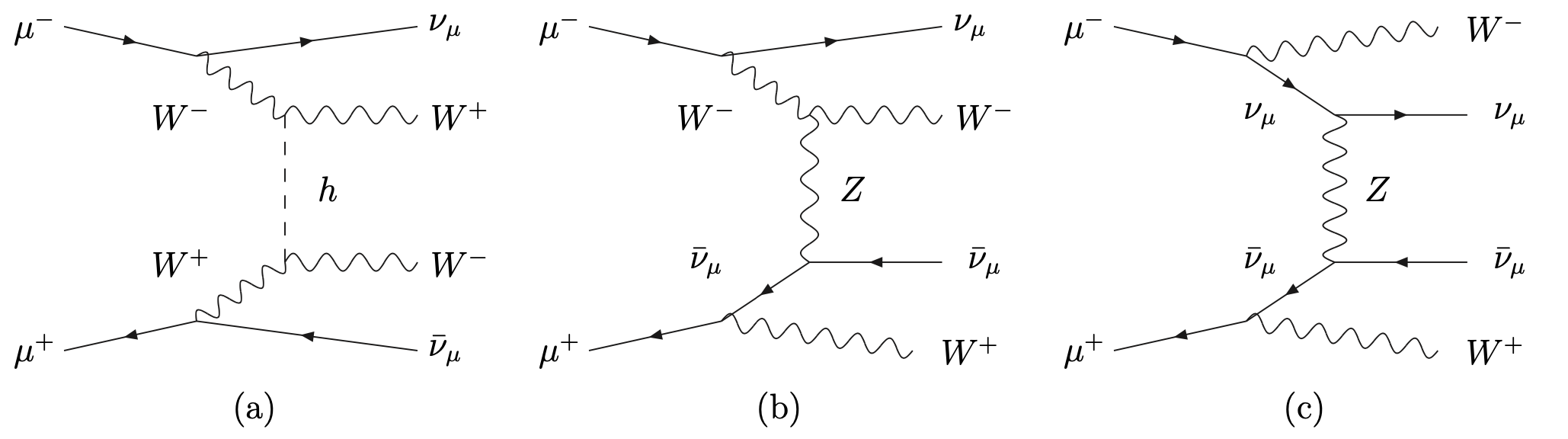}	
\caption{Representative Feynman diagrams of the three classes of contributions
to \mbox{$\mu^+\mu^-\to W^+W^- \bar \nu_\mu \nu_\mu$} which are kinematically 
enhanced at high energy, namely: (a) VBF scattering (reduced process $W^+ W^- \to W^+ W^-$); 
(b) single-$W$ radiation, either from initial or final state 
(reduced process $f W\to fW$); (c) double-$W$ radiation, either from initial or final state
(reduced process$ f \bar f \to f \bar f$). Here, $f=\mu^\pm,\nu_\mu,\bar \nu_\mu$;
typically, the impact of these contributions is largest for (a), and smallest
for (c).}
\label{fig:vbf_no_vbf}
\end{figure}

We note that even in the simplified setup we have adopted, where the 
incoming muons are treated as if they had different flavours (see
footnote~\ref{foot:flav}), \mbox{$\mu^+\mu^-\to W^+W^- \nu_\mu \bar\nu_\mu $}
features three main classes of graphs which result in possible
kinematic enhancements; these classes can in turn be loosely identified with 
underlying $2\to 2$ reduced processes, namely (see fig.~\ref{fig:vbf_no_vbf}):
\begin{enumerate} 
\item[(a)] vector boson scattering $W^+ W^- \to W^+ W^-$; this leads to the
  largest enhancement, owing to the two spin-1 particles exchanged in the $t$
  channels of the $2\to 4$ diagrams (a.k.a.~VBF).
\item[(b)] $f W \to f W$ scattering, stemming from one (in either the 
final or initial state) muon branching into a $W$ boson and one branching
into one fermion (a.k.a.~single-$W$ radiation).
\item[(c)] $f\bar f\to f\bar f$, stemming from two (in either the 
final or initial state) muon branchings into a $W$ boson
(a.k.a.~double-$W$ radiation).
\end{enumerate}
While the $\pt(W^\pm)$ cut we have imposed is expected to suppress
such contributions, some leftover will unavoidably remain, hampering
the agreement between the EWA and the full matrix elements results.
Again, this is a best-case scenario: indeed, if we considered the
physical case where $\mu^+$ and $\mu^-$ do have the same flavour,
we would get {\em also} annihilation channels, such as  
\mbox{$\mu^+\mu^-\to W^+W^- (Z \to )\bar \nu_\mu \nu_\mu$}. 
None of these is VBF-like, and therefore the agreement between the EWA 
and the full matrix elements predictions cannot but degrade further. 

\begin{figure*}[t!]
	\centering 
    \includegraphics[width=0.98\textwidth, angle=0]{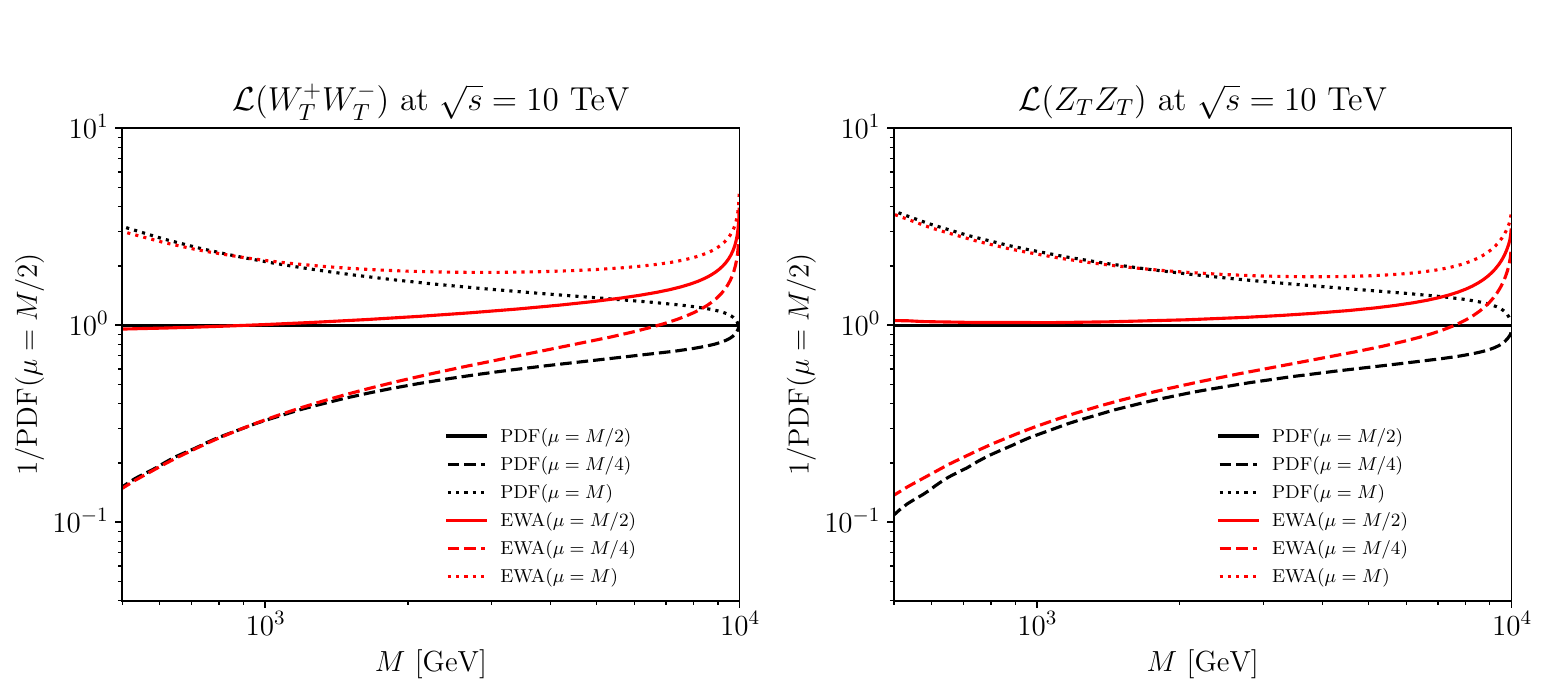}
\caption{Comparison of typical parton-parton luminosities computed with
the EWA (red curves) and the EW PDFs (black curves).}
	\label{fig:EVAPDF}
\end{figure*}
In summary, it is clear that in situations such as the one discussed for
$W^+W^-$ production (i.e.~those which feature other topologies in addition to
the VBF one), any approach capable of retaining the full information
associated with the exact $2\to 4$ matrix element is superior to the EWA based
on the $2\to 2$ matrix elements; this conclusion extends to the EW-PDF
approach, which employs the same matrix-element framework as the EWA.

Conversely, for what concerns cases such as $t\bt$ production, and thus its
underlying $ZZ$-initiated neutral-current contribution, the situation is
arguably more favourable, which can easily be understood as follows. 
$W^+W^-$ production features particles in the final state (the $W$'s)
which also directly enter the EW-PDF resummation procedure, which is in
turn related to the fact that they can be produced (in particular, 
collinearly) through different parton-parton scatterings as well --
see fig.~\ref{fig:vbf_no_vbf}(c). This implies that a merging of different 
final-state multiplicities would be mandatory in order for the $W$'s to be 
fully accounted for in all of the phase space. On the other hand, the $t\bt$
pair can only be generated in high-$Q^2$ interactions.  One therefore
wonders whether by ``upgrading'' the EWA description of $t\bt$ production
by means of an EW-PDF approach, an improvement in the precision of the 
cross section would follow, whose impact would be larger than the loss 
of accuracy due to the approximations inherent to the 2 $\to$ 2 scattering. 
In other words we pose the question: is the difference between the EWA and the
EW-PDF results larger  than that between the EWA and the $2\to 4$
matrix elements results?  If the answer is negative, the upgrade is
phenomenologically irrelevant even in this best-case scenario, and it shows
that the resummation included in the EW PDFs is subleading and does not
provide one with a tangible improvement.

In order to answer this question (indirectly, given that at present \aNLOs\ 
does not support the publicly-available computer codes for EW PDFs), 
we estimate the impact of the resummation included by construction in 
the EW PDFs, but not present in the EWA, on the luminosities relevant to 
the processes under consideration -- this exercise is significant in
that the EWA functions can be seen as the initial conditions for
EW-PDF evolution, and therefore differ from the latter only owing
to higher-order terms. Thus, we compare the luminosities computed
by using the EWA functions with those obtained from the EW PDFs.
In fig.~\ref{fig:EVAPDF} we present the results for two such comparisons,
relevant to the transversely-polarised $W^+W^-$ pair (left panel) and 
$ZZ$ pair (right panel); we remind the reader that, for the self-consistency 
reason mentioned before, both the EWA functions and the EW PDFs 
are those of ref.~\cite{Garosi:2023bvq}. The plots display the 
luminosities (again, we have considered a 10~TeV muon collider)
as a function of the invariant mass $M$ of a putative final-state
system (e.g.~the $t\bt$ pair), which are all normalised (i.e.~divided
by) the luminosity obtained with the EW PDFs evaluated at the scale 
$\mu=M/2$, taken to be our central choice. The black and
red curves give the EW-PDF and EWA results, respectively. The predictions
obtained by setting $\mu=M$ and $\mu=M/4$, are shown as well (with the
same normalisation as before).

It is manifest that the impact of the resummation of the EW logarithms (the
difference between PDFs and EWA curves at the same scale) is minimal compared
to the dependence of the luminosities on the factorisation scale, which is
consistent with the one shown in fig.~\ref{fig:mevspdf} for the invariant mass
distributions. Their hierarchy is reverted only at $m(X)\gtrsim 5~{\rm TeV}$, where on the other hand weak VBF processes are by
far subdominant. Most importantly, in the range considered, the impact of the resummation of the EW
logarithms is much smaller than the difference between $2\to 4$ matrix
elements and the EWA approximation.
 
Even in the favourable case of $t\bt$ production, the situation of the EW PDFs
can be compared to an example relevant to LHC physics, where one would use a
pure LL resummed prediction for the $Z$ transverse momentum in comparison to
experimental data.  Such a comparison would utterly fail, for two reasons:
insufficient logarithmic accuracy, and lack of any non-trivial matrix element
information; the former prevents the agreement with data in the peak region
and below ($\pt(Z)\lesssim 4$~GeV), and the latter in the complementary region
(where hard recoils are important). The only positive feature of such a
prediction would be qualitative, in that it would confirm the existence of a
small-$\pt(Z)$ peak, and the finiteness of the cross section at $\pt(Z)\to
0$. The methods needed to improve this prediction and its comparison to data
are well known, and all include an increase of the logarithmic accuracy, and a
matching with matrix elements.

Exactly as for the LL-accurate $\pt(Z)$ unmatched resummation just
mentioned, the current EW PDFs results only retain the LLs, do not 
include any matching with non-trivial matrix elements, and work for a 
single observable at a time. Therefore our results, summarised by 
fig.~\ref{fig:mevspdf}, are not surprising\footnote{While 
fig.~\ref{fig:mevspdf} features the EWA, and not the EW-PDF, results,
it is legitimate to use it to draw conclusions on the latter. Specifically,
in the context of the present example, while the $\pt(Z)$ resummation is 
mandatory ($\pt(Z)$ tends to zero), that performed by the EW PDFs is not 
(it takes a lot of energy for $m_Z$ to be negligible w.r.t.~the hard scale 
of the process).}: the best one can hope for is a qualitative description. 
Precisely as for the $\pt(Z)$ example, the situation can be improved by 
increasing the logarithmic accuracy {\em and} by matching to matrix 
elements\footnote{We point out that, even if not strictly necessary,
higher-logarithmic accuracy would be needed for a better matching with
higher-order matrix elements.}, neither of which options
appears to be around the corner.

For what only concerns the EWA, if one stipulates that its functions are
not to be used as EW-PDF initial conditions (in which case they would be
subject to more stringent constraints), then one has a certain freedom
to include in it non-logarithmic terms -- in the context of the $\pt(Z)$
example given above, this would amount to comparing the EWA with 
fixed-order $Z$-production matrix elements. While it is clear that
the inclusion of non-logarithmic terms would improve this comparison,
the non-universality of such terms, and/or the presence of analogous terms
that do not factorise (and therefore {\em can not} be employed in the 
EWA) imply an immediate loss of control over the accuracy of the
approximation. This can only be established {\em a posteriori} through
a comparison with the very matrix elements whose computations one
seeks to avoid by introducing the EWA in the first place -- thus,
one ends up with a circular argument\footnote{A recent implementation 
of the EWA which includes some subleading universal terms in the vector-boson 
mass~\cite{Bigaran:2025rvb} does not appear to improve the quality of the 
EWA vs matrix-element agreement in a substantial manner, in particular for 
those processes where additional (non-VBF) topologies are present.}.

Given the limitations of the EWA and EW-PDF results outlined above, 
we are compelled to seek an alternative approach. Let us specify the 
requirements that such an approach should have.

Firstly, it needs to employ exact matrix elements, incorporating without any
approximation mass effects for the weak bosons and other particles (possibly
bar the truly light ones, i.e.~electrons, muons, and quarks of the first and
second families) at the electroweak scale, and thereby including the complete
set of contributing topologies and diagrams. Secondly, it must allow, at least
in principle, the inclusion of higher-order corrections -- specifically, of
next-to-leading order (NLO) and beyond.  Thirdly, it should feature the
resummation of small-lepton-mass logarithms, using available PDFs for leptons,
photons, and, optionally, quarks and gluons, up to next-to-leading logarithmic
(NLL) accuracy. Finally, it must provide one with the full and smooth 
coverage of the phase space, ensuring no loss of information across the 
entire energy range -- from the threshold to asymptotic regimes.

With these criteria in mind, we now present a strategy (introduced in
ref.~\cite{Frixione:2025wsv}) designed to overcome the limitations of the 
conventional EWA- and EW-PDF-based approaches. This strategy isolates
VBF-like configurations that emerge at the NNLO in a gauge-invariant and
theoretically controlled manner, framing them as an extension of NLO
predictions. Such configurations correspond to double-real diagrams,
and are matched to collinear resummation for light particles.


\section{VBF-improved NLO electroweak cross sections}
\label{sec:app}
In this section we summarise a few key ingredients of the method introduced
in ref.~\cite{Frixione:2025wsv}; the reader is encouraged to check that paper for
more details on the procedure and its numerical validation.

We begin by stating the fact that the automatic computation of the complete
NLO EW corrections for an arbitrary process at lepton colliders is 
available in \aNLOs~\cite{Frederix:2018nkq,Frixione:2021zdp}. 
Here, we show how the corresponding predictions can be augmented to 
consistently account for all NNLO double-real processes that include
VBF topologies in an exact manner, while avoiding double 
counting and resumming collinear lepton-mass logarithms of QED origin. 
We shall briefly go through these features in the remainder of this
section; the reader must bear in mind that the following discussion is 
kept intuitive intentionally. 

\begin{figure}
	\centering 
\includegraphics[width=0.8\textwidth, angle=0]{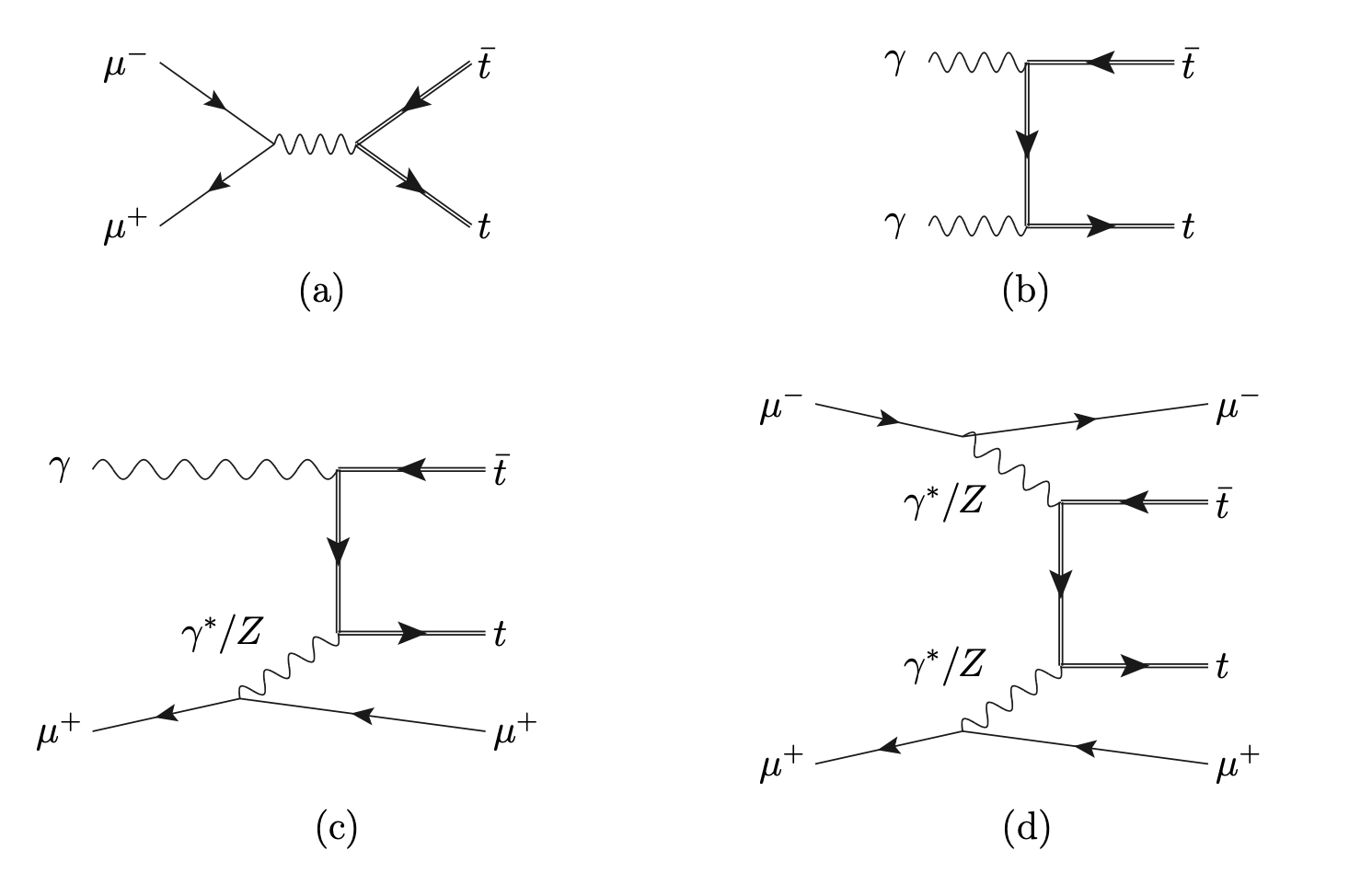}	
	\caption{Representative Feynman diagrams for the various classes of
	contributions to $t \bar t$ production. (a) and (b) correspond to the two production channels
	at LO. Diagram (c) appears as NLO real correction, while diagram (d)
	is formally an NNLO double-real correction.  }
	\label{fig:feynman}
\end{figure}
In order to be definite, let us consider $t\bar t$ production at a muon
collider. Lest confusion should arise, we use a different notation
to tell the physical muon that constitutes the beam (${\upmu}$) from the 
muon parton ($\mu$) which is inside the former.  At a mass scale equal 
to the muon mass, the ``probability density of finding a parton muon in
the physical muon'' is just equal to a Dirac delta, its argument a function 
of the parton momentum fraction $\zeta$, thus:
$\Gamma_{\mu^\pm/\upmu^\pm}(\zeta,m_\upmu^2)=\delta(1-\zeta)$. At higher scales
$\Gamma_{\mu^\pm/\upmu^\pm}$ evolves into a different distribution which has
a $\zeta<1$ tail. In addition to that, a non-zero probability of finding a 
photon in the physical muon $\upmu$, $\Gamma_{\gamma/\upmu^\pm}(\zeta)$,
exists, which instead peaks at small values of $\zeta$. Initial conditions 
and explicit expressions for these parton distribution functions are now 
available at the NLO+NLL accuracy~\cite{Frixione:2019lga,Bertone:2022ktl,
Frixione:2023gmf,Bonvini:2025xxx} for both electrons and muons.

In this framework, at the LO $t\bar t$ production proceeds through two partonic 
channels, \mbox{$\mu^+ \mu^-\to t\bar t$} (annihilation) and 
\mbox{$\gamma\gamma\to t\bar t$} (fusion) --
see fig.~\ref{fig:feynman}(a) and (b), respectively. At the NLO in the
EW theory, real and virtual diagrams enter. Among the real corrections, a 
new class of diagrams arises, e.g., $\gamma \mu^-\to  t\bar t \mu^-$ with a $Z$
or a $\gamma$ exchanged in a $t$-channel, fig.~\ref{fig:feynman}(c).  
In the kinematic configurations enhanced by collinear emissions, this 
contribution can be seen as ``half" of a VBF-type contribution underpinned
by a \mbox{$(Z/\gamma^*) \, (Z/\gamma^*) \to t \bar t$} process, as in
fig.~\ref{fig:feynman}(d). At the NNLO two-loop-virtual, one-loop-real, 
and double-real diagrams enter; among the latter, one finds the VBF-type 
contributions mentioned above and depicted in fig.~\ref{fig:feynman}(d). 
When squaring the amplitudes, these contributions mix diagrams featuring 
$Z$- and $\gamma$-exchanges in the $t$-channels which cannot be disentangled 
-- in other words, $Z$- and $\gamma$-exchange diagrams interfere with each
other. However, in the collinear limits, where the outgoing muons are either 
forward or backward w.r.t.~their parent particles, the photon poles can be 
uniquely identified as coming from two different fermion lines\footnote{This
is so owing to the fact that collinear configurations are semi-classical.}, 
and the ensuing collinear divergences can then be reabsorbed into the 
corresponding muon PDFs. While at the NLO one would need to consider 
only one such divergence at a time, at the NNLO double divergences
appear where there are two incoherent NLO-type singular structures, one
for each of the two incoming fermion lines. This implies that the full
massless-muon matrix element \mbox{$\mu^+ \mu^-\to  t\bar t \mu^+ \mu^-$}
can be rendered finite by means of a suitable set of insertions of collinear 
counterterms, and subsequently integrated over the phase space. These
insertions are compensated by analogous terms that are
included in the PDFs, that thus effectively resum collinear logarithms 
of QED origin. Conversely, VBF-type contributions mediated by weak charged 
currents that contribute to 
\mbox{$\mu^+\mu^- \to t\bar t \nu_\mu \bar \nu_\mu$} 
do not feature any collinear divergences, and can simply be added (at the
squared-amplitude level) to all of the other contributions to the
cross section, without needing any counterterms. 

With the help of the basic elements listed just above, the essence of
our approach is the following: we start with an NLO EW result for the
process of interest, and consider the full set of amplitudes that first
appear at the NNLO. If these do not feature QED-type initial state
divergences, such as those mediated by charged weak currents, one can just 
square and add them to the NLO result. Otherwise, one includes the complete
amplitude squared (thus including $Z/\gamma$ interference effects)
plus a suitable set of collinear counterterms for the photon currents that 
render the former amplitude squared infrared-finite, while at the same time 
avoiding double counting with terms of NLO origin. Working out the details 
of this procedure is not straightforward, and one needs to take care of 
several technical aspects. As was already mentioned, these are described 
fully in ref.~\cite{Frixione:2025wsv}; here we limit ourselves to presenting the 
final result, restricting the notation to the case of a muon collider.

Following ref.~\cite{Frixione:2025wsv}, the differential cross section for 
producing a given final state in $\ \upmu^+\upmu^- $ collisions is
written thus:
\begin{eqnarray}
d\sigma(P_1,P_2)&=&d\zeta_1\,d\zeta_2\,
\sum_{ij}\PDF{i}{\upmu^+}(\zeta_1)\,\PDF{j}{\upmu^-}(\zeta_1)\, 
d\hat \sigma_{ij}(\zeta_1 P_1,\zeta_2 P_2) \,,
\label{convolution}
\end{eqnarray}
where $P_i$ are the momenta of the physical muons ($\upmu^\pm$), for the 
summation indices we use $i=\mu^+,\gamma$ and
$j=\mu^-,\gamma$\footnote{This is for simplicity; predictions aiming at
reproducing as closely as possible an experimental setup should use all
parton types (i.e.~all light leptons, all quarks, the photon, and the gluon);
the formalism does not require any modifications to account for this
extension.}, and the short-distance cross sections can be conventionally
written in terms of the electroweak coupling $\alpha$ as follows:
\begin{eqnarray}
d\hat \sigma_{ij} &=& 
d\hat \sigma_{ij}^{[0]}+
\aemotpi\,d\hat \sigma_{ij}^{[1]}+
\left(\aemotpi\right)^2\,d\hat \sigma_{ij}^{[2]} + O(\alpha^3)\,,
\label{eq:exp}
\end{eqnarray}
where all of the $d\sigma_{ij}^{[k]}$ coefficients are understood to be 
of the same perturbative order as the Born amplitude squared, i.e.~$\alpha^b$
with $b$ a process-dependent integer (e.g., $b=2$ for $t\bar{t}$ and $W^+ W^-$ production).

The keeping of only the two leftmost terms in eq.~(\ref{eq:exp}) corresponds
to an NLO computation, which as we said we perform automatically with
\aNLOs. Conversely, the rightmost term is the NNLO
contribution. Its full computation in a process- and observable-independent
way within the EW theory is presently unfeasible. However, as was 
sketched in the discussion above, some of its terms (in particular,
those stemming from VBF topologies which, as we shall show, are non-negligible
for the phenomenology of high-energy muon collisions) can be singled out
and rendered finite~\cite{Frixione:2025wsv}. More in detail, a well-defined
gauge-invariant\footnote{Gauge invariance is ensured by not discarding any diagram for the double real-emission
process.} subset of the NNLO corrections in the partonic $\mu^+\mu^-$
channel can be identified together with suitable subtractions; this contribution,
which essentially consists in the double real-emission from the initial state,
we call
$d\hat{\Sigma}_{\dNNLOG}$:
\begin{eqnarray} 
\left(\aemotpi\right)^2 d\hsig_{\mu^+\mu^-}^{[2]} & \ni &
d\hat{\Sigma}_{\dNNLOG}(p_1,p_2)\,,
\end{eqnarray}
and we write it thus (see eq.~(7.3) of ref.~\cite{Frixione:2025wsv})
\begin{eqnarray} 
d\hat{\Sigma}_{\dNNLOG}(p_1,p_2) & \equiv & 
\left(\frac{1}{1-y_1}\right)_{\deltaI}\!
\left(\frac{1}{1-y_2}\right)_{\deltaI}\!
\left((1-y_1)(1-y_2)\ampsq_{\mu^+\mu^-}^{(m+2)}\right)
d\phi_{m+2}\big(p_1,p_2\big) \nonumber\\*&+& \aemotpi\,{\cal
Q}_{\gamma\mu}^{(\deltaI)^\prime}(z_2)
\left(\frac{1}{1-y_1}\right)_{\deltaI}\!
\left((1-y_1)\ampsq_{\mu^+\gamma}^{(m+1)}\right)
d\phi_{m+1}\big(p_1,z_2p_2\big)\,dz_2 \nonumber\\*&+& \aemotpi\,{\cal
Q}_{\gamma\mu}^{(\deltaI)^\prime}(z_1)
\left(\frac{1}{1-y_2}\right)_{\deltaI}\!
\left((1-y_2)\ampsq_{\gamma\mu^-}^{(m+1)}\right)
d\phi_{m+1}\big(z_1p_1,p_2\big)\,dz_1 \nonumber\\*&+&
\left(\aemotpi\right)^2\, {\cal Q}_{\gamma\mu}^{(\deltaI)^\prime}(z_1)\, {\cal
Q}_{\gamma\mu}^{(\deltaI)^\prime}(z_2) \ampsq_{\gamma\gamma}^{(m)}\,
d\phi_{m}\big(z_1p_1,z_2p_2\big)\,dz_1\,dz_2\,,
\label{final00NNLO2rep}
\end{eqnarray}
where $p_i$ are the momenta of the partonic muons.  

Equation~(\ref{final00NNLO2rep}) includes all neutral VBF-like 
contributions, and is structured so that it can be added to an NLO result 
without double counting. Here $m$ is the number of final state particles 
of the Born process (e.g.~$m=2$ for $t\bar t$ or $W^+W^-$ production). 
\mbox{${\cal Q}^{(\deltaI)'}_{\gamma \mu}(z)$} is a function emerging
from the phase-space integration of a splitting function, plus a term
that removes any double counting (see eq.~(5.20) of ref.~\cite{Frixione:2025wsv}),
while \mbox{${\cal M}^{(m+0,1,2)}_{ij}$} are the matrix elements for
the production of the final-state system of interest, in the collisions
of partons $i$ and $j$ with $0$, $1$, or $2$ extra light particles in
the final state; by $d\phi_n$ we denote an $n$-body phase space.
Finally, the quantities in the large round brackets are plus distributions
(which implement the subtractions stemming from the collinear counterterms),
modified in such a way that the subtractions are only performed in a restricted
region, defined by the arbitrary parameter $\deltaI$. We point out that 
this is the same parameter that enters the 
\mbox{${\cal Q}^{(\deltaI)'}_{\gamma \mu}$} functions; by design,
eq.~(\ref{final00NNLO2rep}) is independent of this parameter, which
constitutes a useful tool to check the correctness of any numerical
implementation of this formula. 
The \mbox{${\cal Q}^{(\deltaI)'}_{\gamma \mu}$} functions and the
subtractions are directly connected with compensating terms included
in the PDFs with which the short-distance cross sections are convoluted.
This connection, technically called a factorisation scheme, is controlled
by certain other functions, which the physical cross section must be 
independent of -- we shall return to this point in 
sect.~\ref{sec:results_scheme}.

In summary, eq.~(\ref{final00NNLO2rep}) incorporates the $2\to 2+m$ matrix 
elements where QED initial state collinear divergences have been subtracted,
and included in the PDFs with which one convolutes the cross section according
to eq.~(\ref{convolution}). Equation~(\ref{final00NNLO2rep}) enables the
exact retention of both logarithmic and power-suppressed terms in the $Z$
boson mass at relative $\ord(\aem^2)$, as well as that of all the effects 
due to $Z/\gamma$ interference. At the same time, this framework resums 
all light-fermion mass effects via lepton PDFs, at the logarithmic accuracy 
at which such PDFs are defined. Overall, eq.~(\ref{final00NNLO2rep}) allows
one to significantly improve any NLO EW predictions in those kinematic 
regions characterised by VBF-like topologies or, more broadly, by partonic
interactions underpinned by  $\gamma\gamma$, $\gamma Z$, and $ZZ$
fusion. 

We conclude by pointing out that the subtractions that appear in
eq.~(\ref{final00NNLO2rep}) only correspond to a subset of the complete
NNLO collinear counterterms; therefore, we understand the presence of
collinear cutoffs, without which one would still find singularities.
This point is discussed at length in ref.~\cite{Frixione:2025wsv}: the bottom
line is that such cutoffs are harmless physics-wise, and their presence
can therefore be ignored in the context of phenomenological simulations.


\section{Results}
\label{sec:res}

In order to illustrate the practical consequences of the method presented
in sect.~\ref{sec:app}, we now consider two processes in detail, namely
the inclusive production of a $t\bar t$ pair, and that of a $W^+W^-$ pair. 
The parameters employed in our simulations are set as follows:
\begin{equation}
m_t = 173.3~\textrm{GeV}\,,\quad
m_W = 80.419~\textrm{GeV}\,,\quad
m_Z = 91.188~\textrm{GeV}\,,\quad
m_H = 125~\textrm{GeV}\,.
\end{equation}
All decay widths are set equal to zero. We work in the $G_\mu$ renormalisation
scheme~\cite{Sirlin:1980nh}, where the EW parameters are determined after
one chooses the value of the Fermi constant, which we set thus
\begin{equation}
G_\mu=1.16639 \cdot 10^{-5} ~\textrm{GeV}^{-2}\,.
\end{equation}
The muon and photon 
PDFs inside the physical muon are computed at the NLO+NLL accuracy, following
refs.~\cite{Frixione:2019lga,Bertone:2019hks,Bertone:2022ktl,Bonvini:2025xxx},
in the so-called $\Delta$ factorisation scheme~\cite{Frixione:2012wtz}; 
technically, they are based on \textsc{eMela}~\cite{Bertone:2022ktl,
Frixione:2023gmf}. The factorisation scale is set equal to the invariant
mass of the pair:
\begin{equation}
\mu_F=m(X\bar X)\,,
\label{factscale}
\end{equation}
with $(X,\bar X) = (t,\bar t)$ or $(X,\bar X) = (W^+, W^-)$.
Two different centre-of-mass energies are considered, namely 
\mbox{$\sqrt{s}=3$~TeV} and \mbox{$\sqrt{s} = 10$~TeV}; in the latter
case, in addition to the predictions obtained by integrating over the
whole phase space, we also present those that stem from imposing some
fiducial cuts, whereby we require:
\begin{equation}
    p_T(X),\,\,p_T(\bar X) > 150~\textrm{GeV}\,,\quad
    m(X\bar X) > 500~\textrm{GeV}\,,\quad
    |y(X)|,\,\,|y(\bar X)| < 2.5 \,.
    \label{eq:cuts}
\end{equation}
As is discussed at length in ref.~\cite{Frixione:2025wsv}, some technical cuts are 
necessary in order to screen uncancelled singularities arising in the 
$\delta$NNLO$_\Gamma$ corrections; we employ an invariant-mass cut on 
the outgoing muons:
\begin{equation}
    m(\mu^+\mu^-) > 200~\textrm{GeV}\,.
\end{equation}
As is demonstrated in ref.~\cite{Frixione:2025wsv}, the effect of such a cut (and of 
a more general class of technical cuts that possibly include (pseudo)rapidity 
cuts on the outgoing muons) on physical distributions is phenomenologically 
negligible.

While presenting our results, we focus on different aspects in turn.
We begin with a general discussion is sect.~\ref{sec:results_gen}; 
process-specific results are then introduced in sects.~\ref{sec:results_tt}
and~\ref{sec:results_ww} for $t\bt$ and $W^+W^-$ production, 
respectively\footnote{Some results at NLO accuracy for these (and other) 
processes, computed by neglecting any photon-initiated contribution,
have been discussed in refs.~\cite{Bredt:2022dmm,Ma:2024ayr}. In the former, 
the software {\sc Whizard}~\cite{Kilian:2007gr} has been employed;
in the latter, the kinematic region of interest has been restricted 
to the one dominated by muon-muon annihilation.}.
In sect.~\ref{sec:results_sdk} we show how the predictions for some observables 
can be further improved by including higher-order, yet sizeable, effects 
due to EW Sudakov logarithms. Finally, we discuss the impact of the choice 
of factorisation scheme in sect.~\ref{sec:results_scheme}.  

\subsection{Differential predictions: general considerations}
\label{sec:results_gen}

In this section, we discuss the general features which are common to
differential observables in $t\bar t$- and $W^+W^-$-pair production, 
while process-specific characteristics are presented in 
sects.~\ref{sec:results_tt} and~\ref{sec:results_ww}.

In keeping with eqs.~(\ref{convolution}) and~(\ref{eq:exp}), and with the
notation introduced in ref.~\cite{Frixione:2025wsv}, we remind the reader that by
NLO and NNLO we understand the predictions accurate {\em up to} those 
respective orders, i.e.~NLO results include both the 
$\ord(\aem^b)$ and the $\ord(\aem^{b+1})$ contributions, whereas the
NNLO results are obtained by adding the $\ord(\aem^{b+2})$ contributions 
to the NLO predictions. Conversely, we denote by $\delta$NLO and
$\delta$NNLO the sole $\ord(\aem^{b+1})$ and $\ord(\aem^{b+2})$ terms,
respectively.  Furthermore, the LO, $(\delta)$NLO, and $(\delta)$NNLO 
contributions which emerge from the procedure of ref.~\cite{Frixione:2025wsv} 
are denoted by $\LOG$, $(\delta)$NLO$_\Gamma$, and $(\delta)$NNLO$_\Gamma$,
respectively\footnote{See sect.~6 of ref.~\cite{Frixione:2025wsv} for precise
definitions and further details. Notably, the LO$_\Gamma$ term coincides
with the pure $\gamma\gamma$-initiated contribution and can therefore also
be denoted as LO$_{\gamma\gamma}$.}.

All of the figures of sects.~\ref{sec:results_tt} and~\ref{sec:results_ww}
have the same layout, thus: there are three panels with the results
obtained at 3~TeV (top left), 10~TeV (bottom left), and 10~TeV with
the acceptance cuts of eq.~(\ref{eq:cuts}) (bottom right). A fourth panel 
(top right) reports the labels relevant to the various histograms.

The layouts of the three panels that constitute a figure are also 
identical to one another, and feature a main frame and three insets.
In the main frame, we show the LO, LO$_{\gamma \gamma}$, NLO, and 
NNLO$_\Gamma$ differential distributions, as pink-dotted, purple-dotted,
green-dashed and black-solid histograms, respectively. In addition
to these, we also show (by using full symbols) the LO-accurate results 
for processes which lead to the same $X\bar X$ final state, but 
are \emph{not} included in the NNLO$_\Gamma$ predictions\footnote{We opt for 
a LO simulation for simplicity, although there is no technical limitation
which prevents one from simulating such processes at the NLO with 
\aNLOs.}.
Such processes are the associated production of the $X\bar X$ pair of
relevance with either a neutrino pair ($X\bar X \nu\bar\nu$), or an on-shell
$Z$ boson ($X\bar X Z$), and are of interest because they account for 
effects due to EW radiation and to charged-current contributions (i.e., 
$W^+W^-$-fusion). For the 
neutrino-pair associated production, we consider both the case where no
cuts are imposed on the neutrinos (cyan circles), and the one where we
require the neutrinos to be muonic and  
 that they have an invariant mass larger than $200$~GeV (fuchsia 
diamonds). The $Z$-associated production results are displayed as yellow
stars. The rationale for considering neutrino-pair associated production
both without and with invariant mass cuts is because in the latter case
the contributions stemming from a resonant $Z$ boson (and which are
already included in the $X\bar X Z$ results) are effectively discarded,
and what is left is expected to be dominated by $WW$-fusion.

The top inset shows the relative impact of the LO, LO$_{\gamma \gamma}$, 
$X\bar X \nu\bar\nu$, and $X\bar X Z$ contributions, defined as the
ratio between these predictions and the NNLO$_\Gamma$ one -- the same
plotting patterns as in the main frame are employed here. Similarly,
the middle inset presents the relative impact (defined as before) of 
the $\delta$NLO and $\delta$NNLO$_\Gamma$ corrections (green-dashed and
black-solid histograms, respectively, as is done in the main frame);
the latter is multiplied by a factor of $2$ in order to enhance its
visibility. In addition to those, we also display the $\delta$NLO$_\Gamma$ 
(blue-dotted) and the $\delta$NLO corrections (red-dashed), the latter
of which computed by setting the finite part (according to the FKS
conventions) of the virtual amplitudes equal to zero (and hence denoted 
by $\delta$NLO-no$V$) -- the latter two predictions help understand the 
``origin'' of EW corrections, as we shall see in the following. Finally, 
in the bottom inset, we plot the uncertainty bands obtained by varying 
the factorisation scale up and down by a factor of two around its central 
value (see eq.~(\ref{factscale})) for the LO (dotted pink), NLO (shaded 
green) and NNLO$_\Gamma$ (shaded grey) predictions.


\subsubsection{$t\bar t$ inclusive production}
\label{sec:results_tt}

\begin{figure*}
    \centering 
    \figtable
        {\includegraphics[height=0.5\textwidth, angle=0]{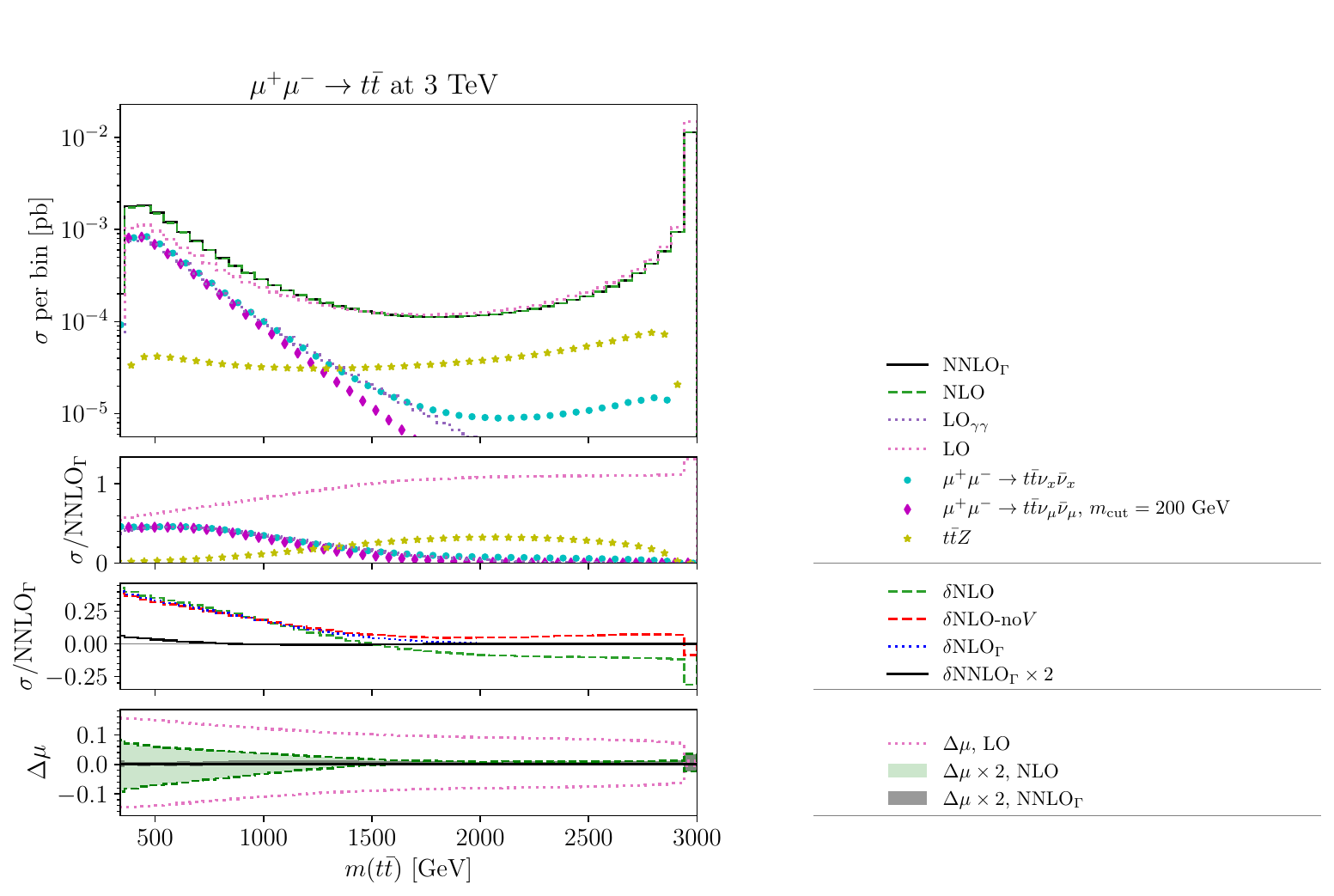}}
        {\includegraphics[height=0.5\textwidth, angle=0]{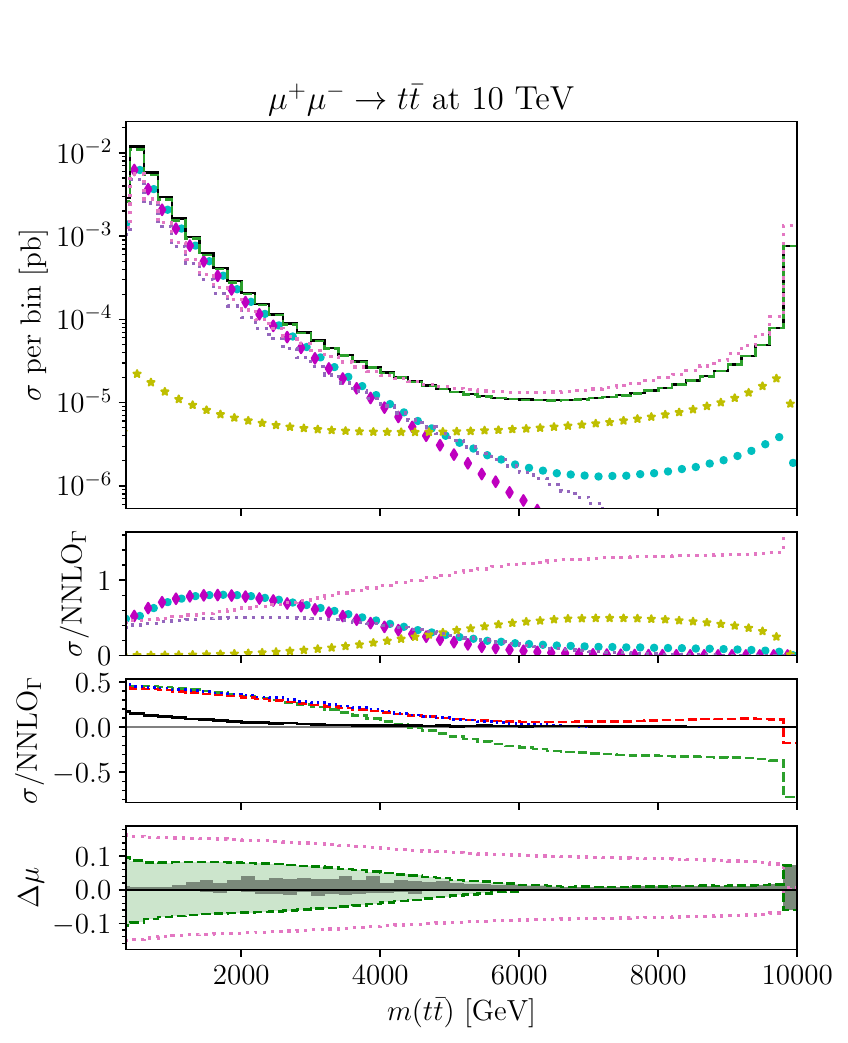}}
        {\includegraphics[height=0.5\textwidth, angle=0]{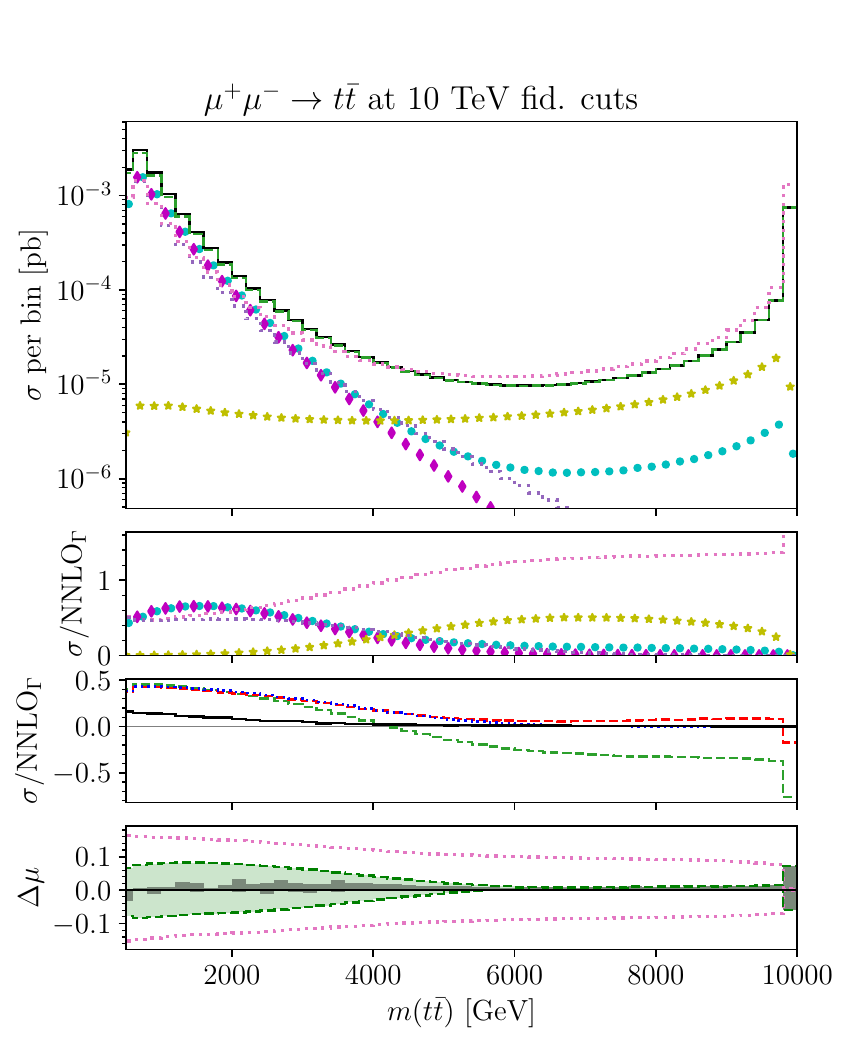}}	
	\caption{Invariant mass distribution of the top-antitop pair, $m(t \bar t)$, in $t\bar t$ production at a muon collider with $\sqrt s=$3, 10 TeV. At 10 TeV, both the cases without and with acceptance cuts defined in eq.~\eqref{eq:cuts} are displayed. See the text for details. } 
	\label{fig:mtt}%
\end{figure*}

\begin{figure*}
    \centering 
    \figtable
        {\includegraphics[height=0.5\textwidth, angle=0]{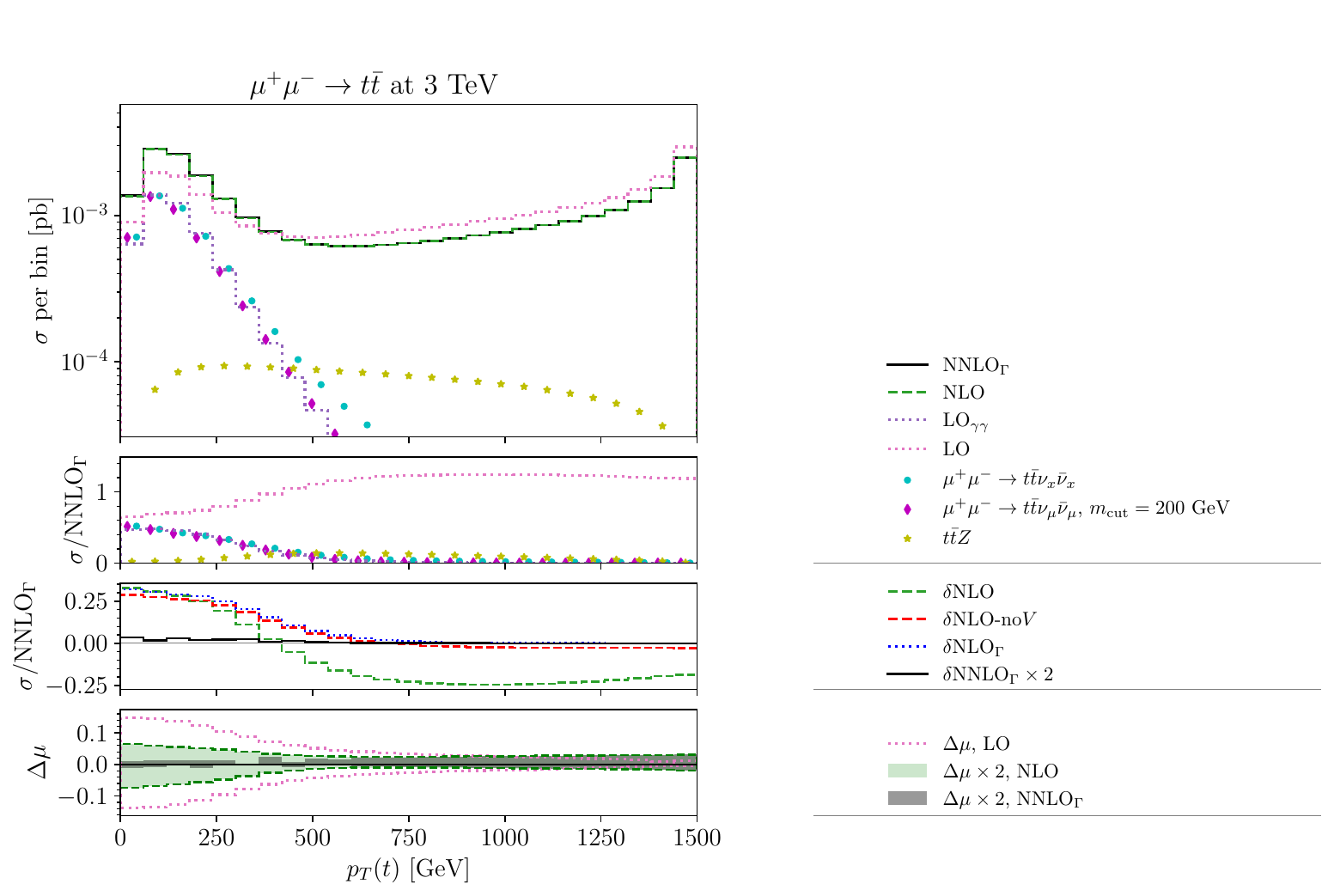}}
        {\includegraphics[height=0.5\textwidth, angle=0]{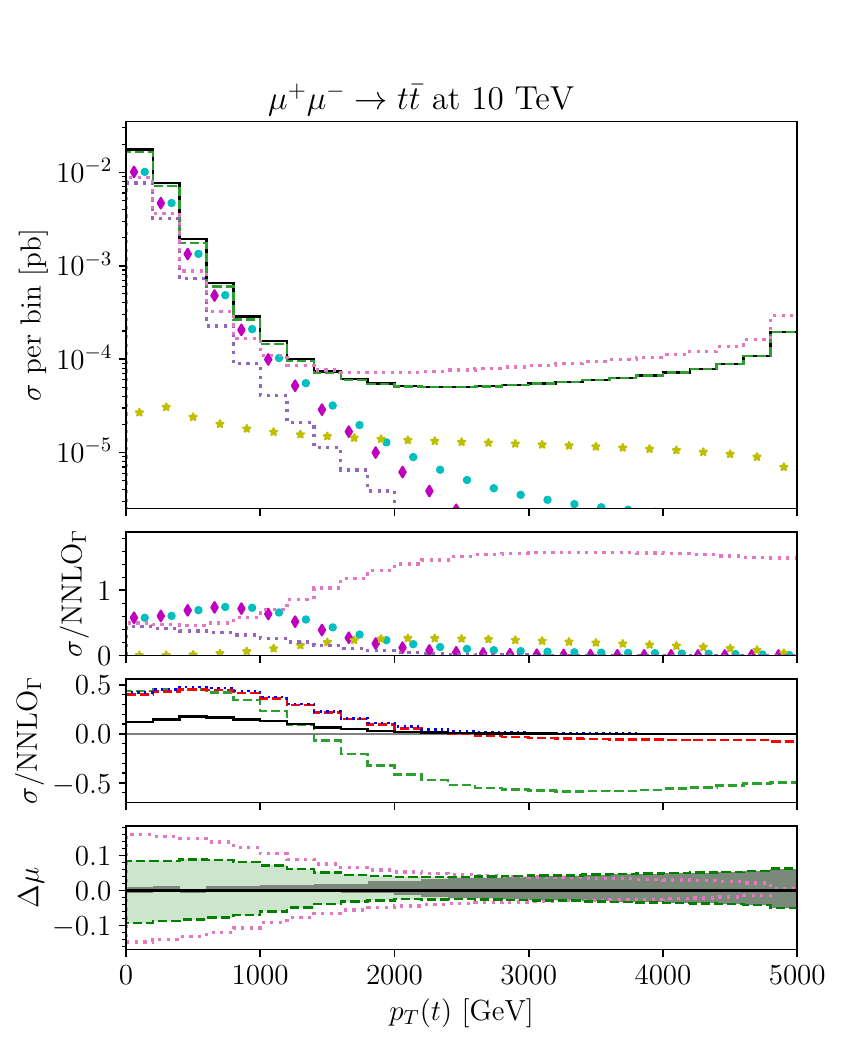}}	
        {\includegraphics[height=0.5\textwidth, angle=0]{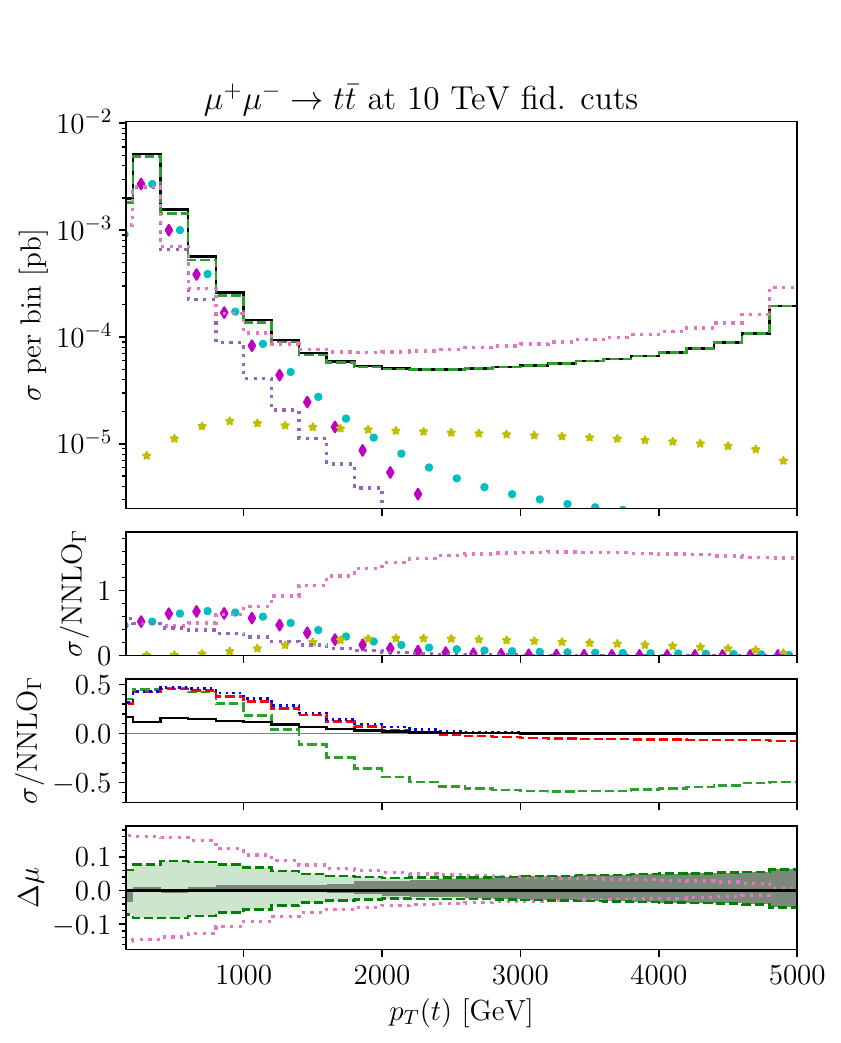}}	
	\caption{Same as in fig.~\ref{fig:mtt}, for the top-quark transverse-momentum distribution, $p_T(t)$. } 
	\label{fig:ptt}%
\end{figure*}

\begin{figure*}
    \centering 
    \figtable
        {\includegraphics[height=0.5\textwidth, angle=0]{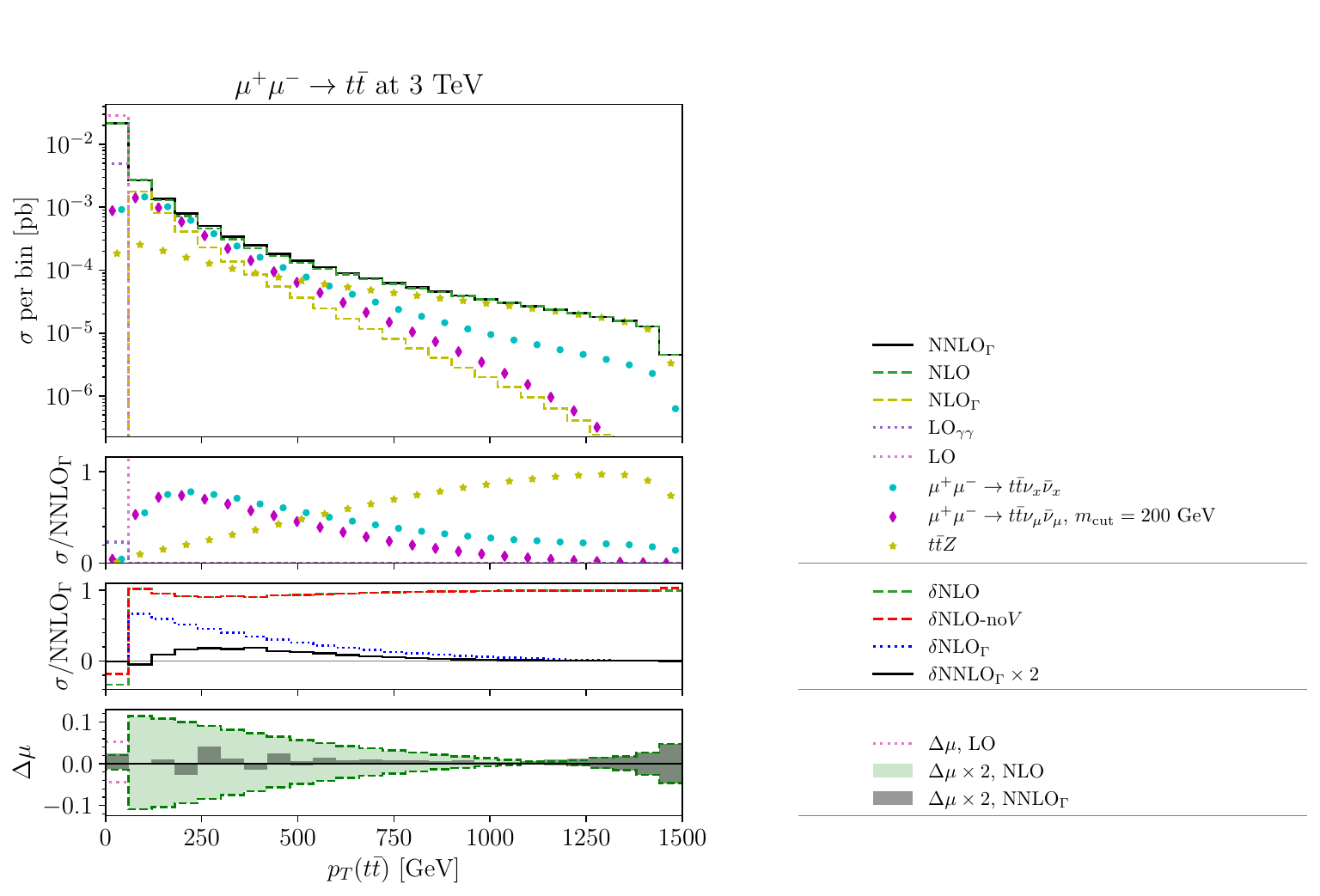}}	
        {\includegraphics[height=0.5\textwidth, angle=0]{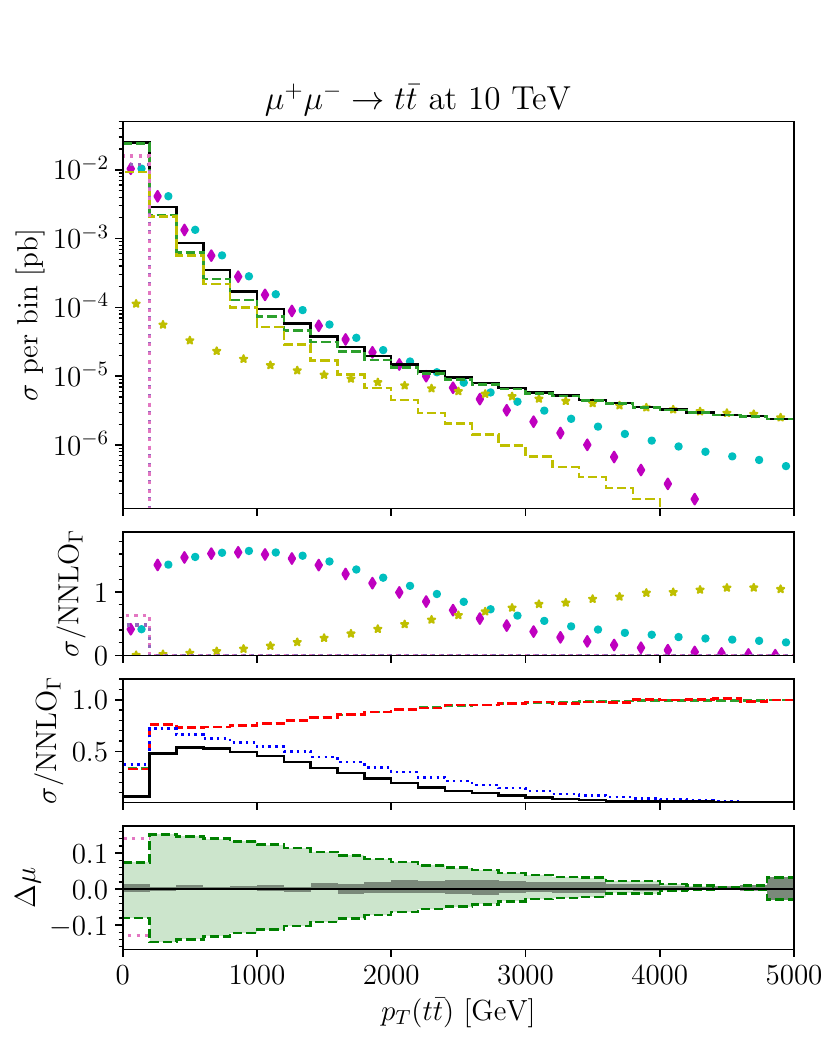}}	
        {\includegraphics[height=0.5\textwidth, angle=0]{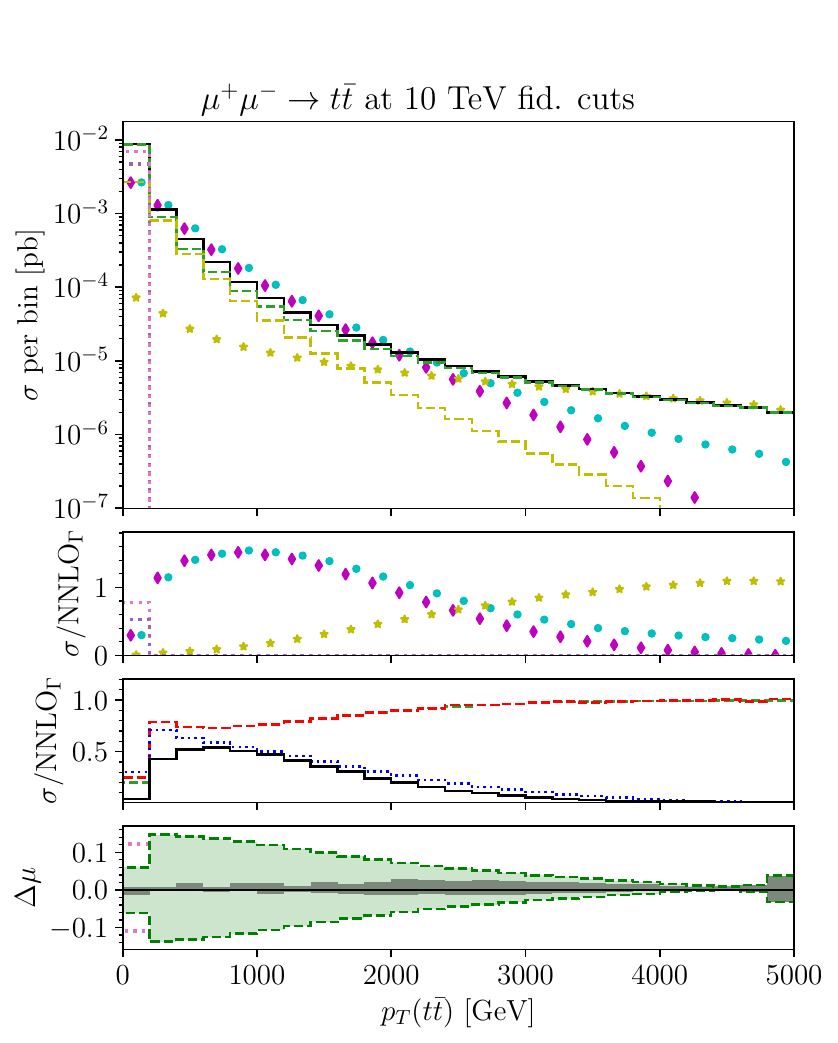}}
	\caption{Same as in fig.~\ref{fig:mtt}, for the top-antitop pair transverse-momentum distribution, $p_T(t\bar t)$. } 
	\label{fig:ptttx}%
\end{figure*}

\begin{figure*}
    \centering 
    \figtable
        {\includegraphics[height=0.5\textwidth, angle=0]{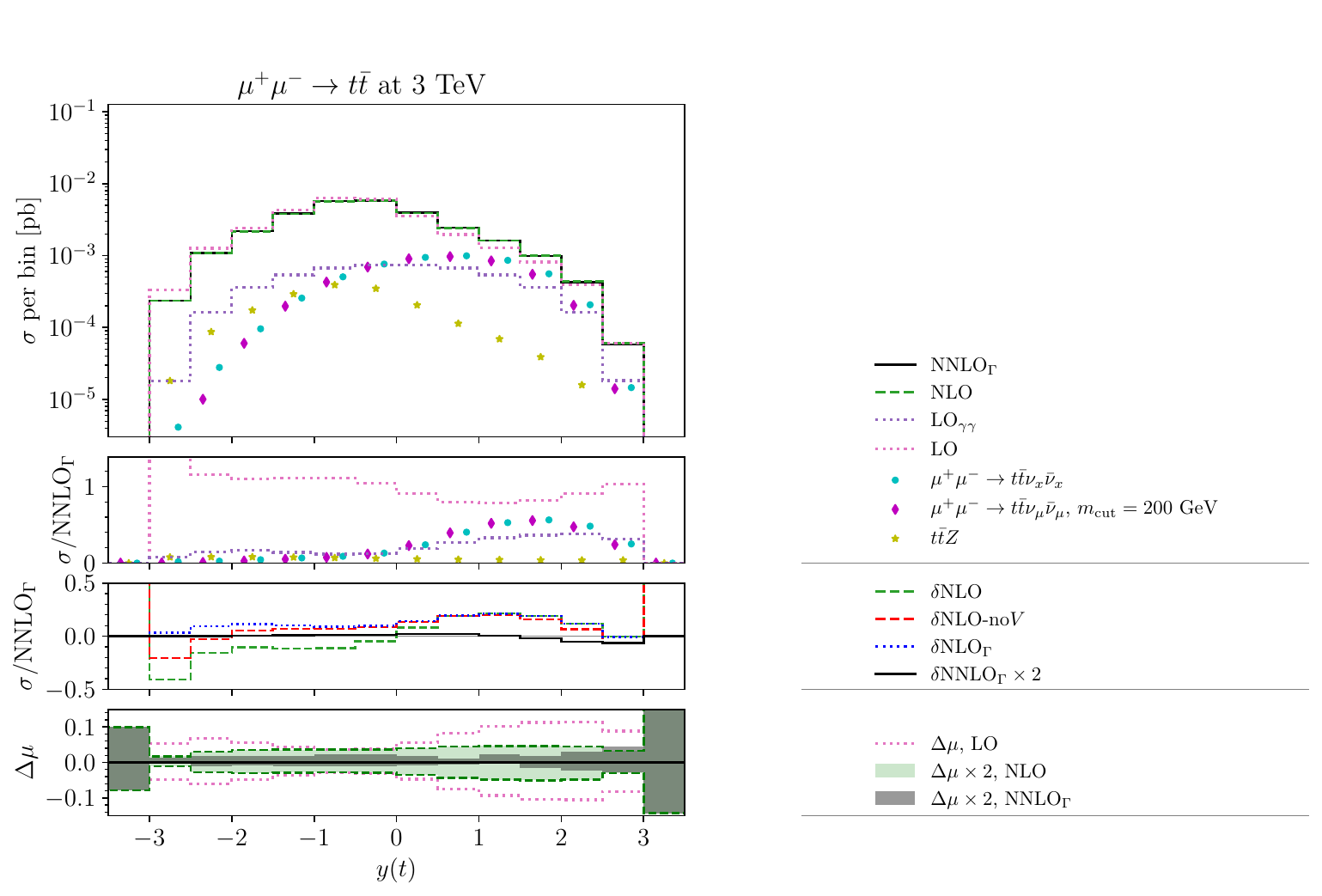}}	
        {\includegraphics[height=0.5\textwidth, angle=0]{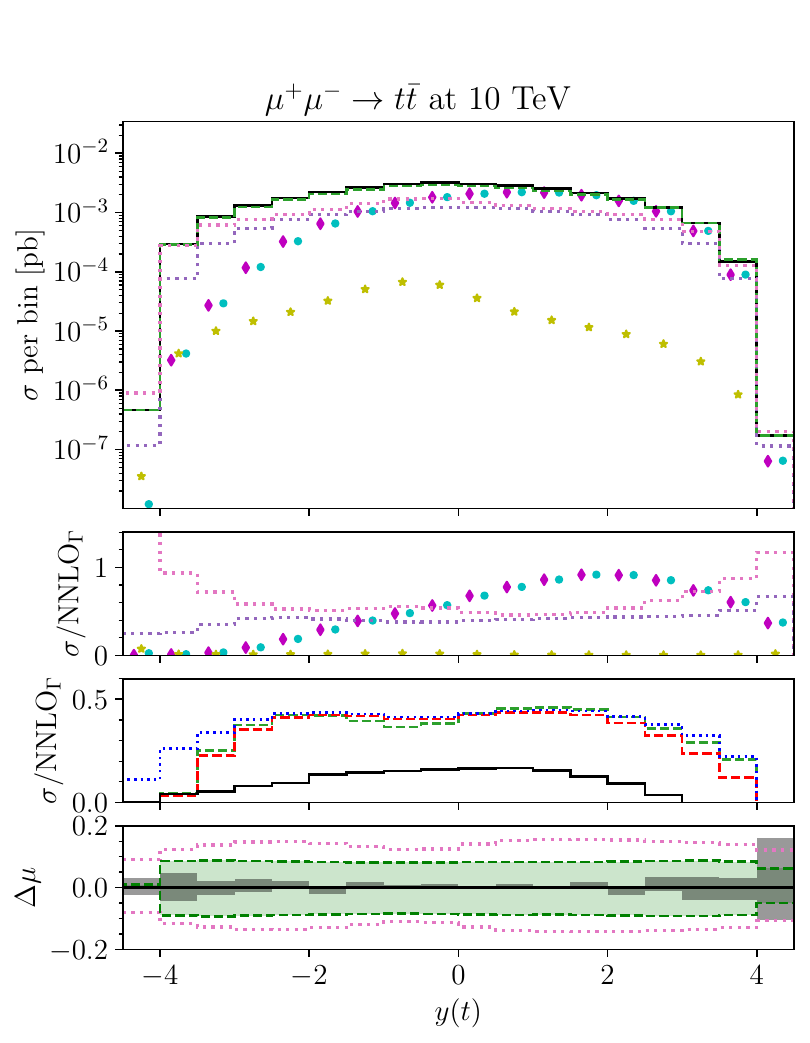}}	
        {\includegraphics[height=0.5\textwidth, angle=0]{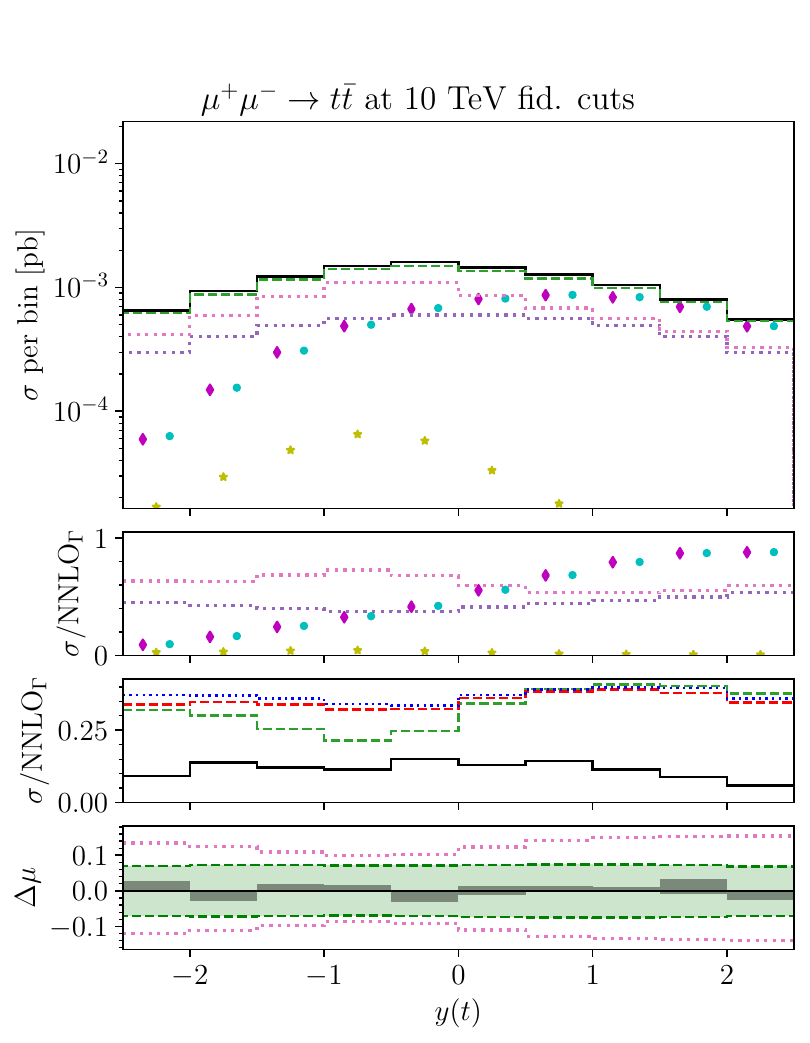}}	
	\caption{Same as in fig.~\ref{fig:mtt}, for the top-quark rapidity distribution, $y(t)$. } 
	\label{fig:yt}%
\end{figure*}

\begin{figure*}
    \centering 
    \figtable
        {\includegraphics[height=0.5\textwidth, angle=0]{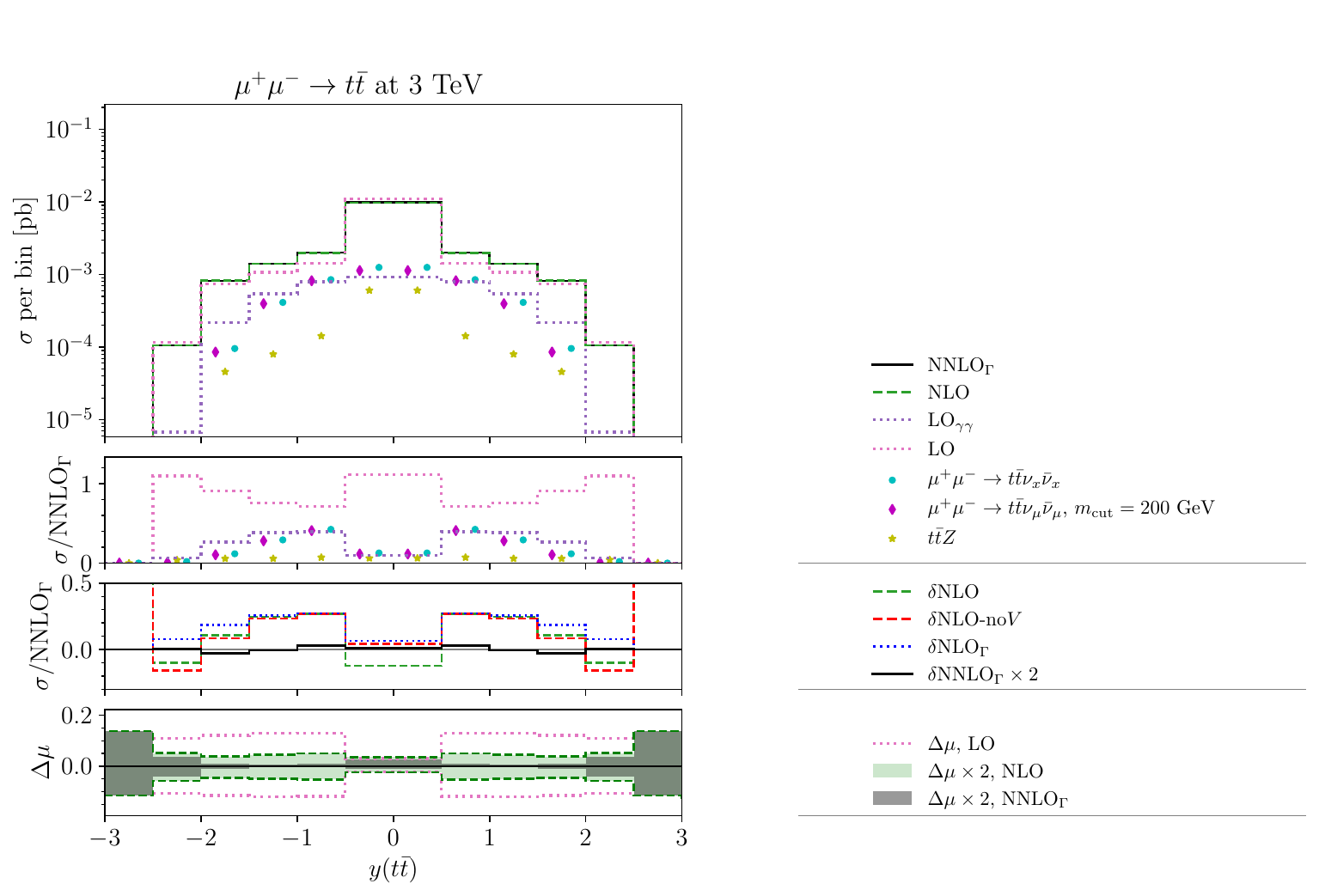}}
        {\includegraphics[height=0.5\textwidth, angle=0]{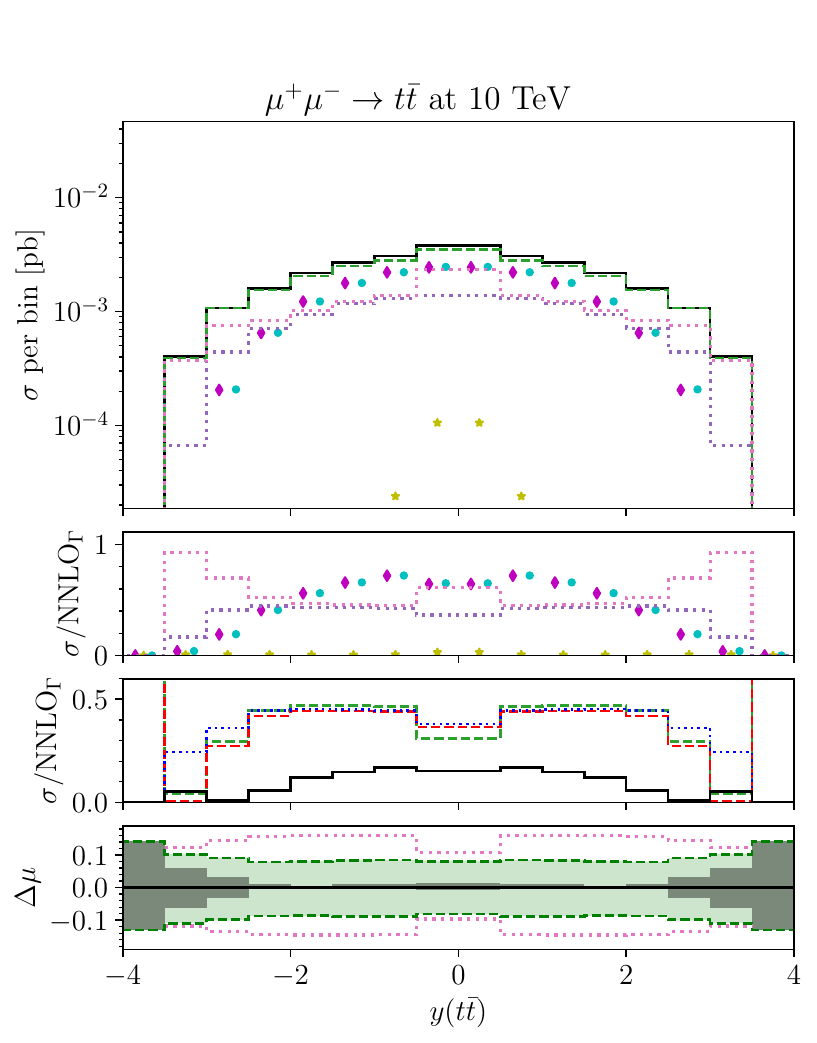}}
        {\includegraphics[height=0.5\textwidth, angle=0]{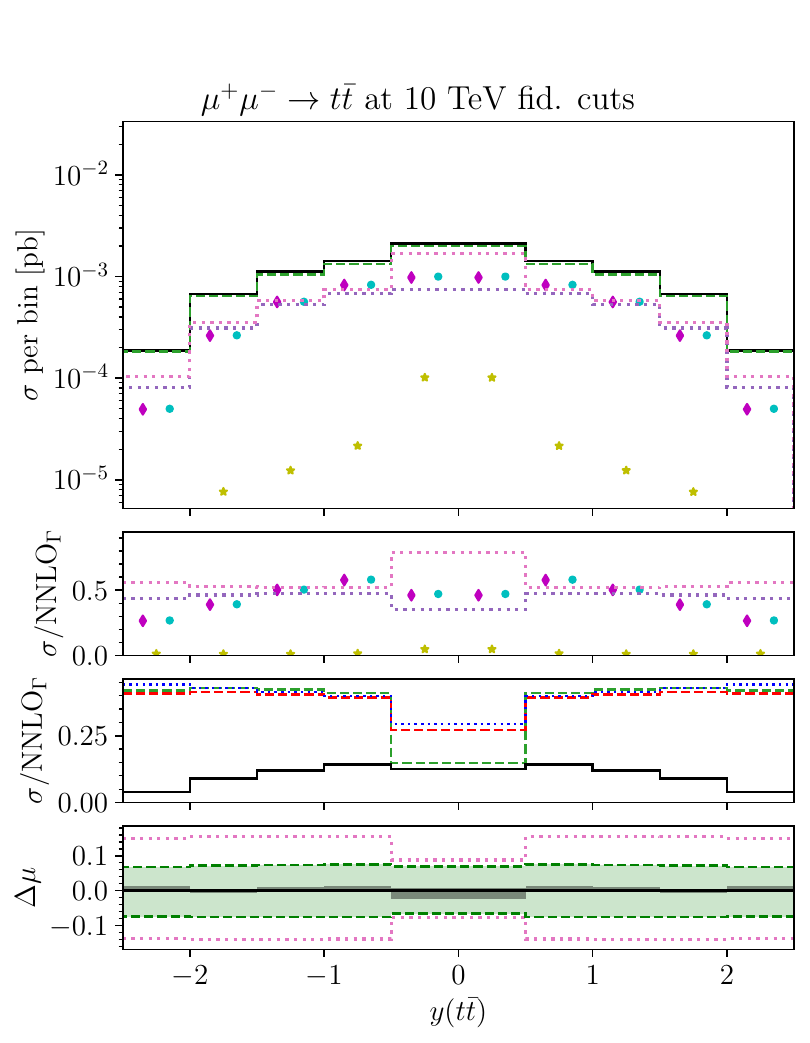}}	
	\caption{Same as in fig.~\ref{fig:mtt}, for the top-antitop pair rapidity distribution, $y(t\bar t)$. } 
	\label{fig:yttx}%
\end{figure*}

As was anticipated, this section collects our predictions for 
several differential distributions stemming from the production of
a $t\bt$ pair. We start with the invariant mass of the pair 
(fig.~\ref{fig:mtt}); we then consider the transverse momentum of the 
top quark (fig.~\ref{fig:ptt}) and that of the pair 
(fig.~\ref{fig:ptttx}), and finally the rapidity of the top quark 
(fig.~\ref{fig:yt}) and that of the pair (fig.~\ref{fig:yttx}).

\medskip
The $t\bar t$ invariant mass distribution, shown in fig.~\ref{fig:mtt}, is
crucial in order to understand many key features of this production process. 
In general, the $\gamma\gamma$-initiated contribution is peaked at small 
values of this observable, while that emerging from $\mu^+\mu^-$ annihilation
peaks at large invariant masses (in fact, at the kinematic limit), a
behaviour which is already evident at the LO (the purple-dotted histogram
rapidly approaches the pink-dotted one when moving towards the threshold).
This is a consequence of the fact that the photon PDF grows with
decreasing Bjorken $x$, while the opposite happens with the muon PDF -- 
it grows with the Bjorken $x$, and has an integrable singularity at 
$x\to 1$\footnote{Without ISR effects, i.e.~by truncating the muon PDF 
at $\mathcal O (\alpha^0)$, the muon PDF is equal to a Dirac delta,
$\delta(1-x)$. Therefore, at the LO the entirety of the invariant-mass
distribution would be contained in the rightmost bin. Note, however,
that higher-order corrections, starting at the NLO, would contribute to
the whole invariant-mass range even in the absence of PDF effects, owing
to real radiation.}. Furthermore, the relative impact of the 
$\gamma\gamma$-induced contribution grows with the c.m.~energy, since
smaller values of $x$ become accessible, and at a 10-TeV collider
LO$_{\gamma\gamma}$ essentially coincides with the total LO cross section 
from the threshold up to \mbox{$m(t\bar t)=2$~TeV}, which in turn is about
equal to 50\% of the NNLO$_\Gamma$ result (this is best seen in the top
inset, where the agreement between the pink- and purple-dotted histograms
improves with the c.m.~energy, with their absolute values equal to about
$1/2$). Before discussing the NNLO$_\Gamma$ prediction in detail, 
it is appropriate to comment on the features of the associated 
production modes, \mbox{$t\bar t \nu\bar\nu$} and \mbox{$t\bar t Z$}. 
Specifically, we see from the main frame and the top inset that the 
charged-current contribution is dominant at the threshold, i.e.~in the 
same region as LO$_{\gamma\gamma}$, and it is of a comparable size 
(twice as large at its peak) as the latter at 3~TeV (10~TeV). 
It is worth observing that, even at a c.m.~energy of 10~TeV, 
from $m(t\bar t)=5$~TeV onwards (i.e.~in the region where the ratio
of the top mass over the relevant kinematic quantity, the pair invariant
mass, is less than $10\%$) the charged-current channel gives a contribution 
which is less than $10\%$ of the NNLO$_\Gamma$  differential cross section.
Therefore, in the region where it gives a sizeable contribution (i.e.~for 
$m(t\bar t)\lesssim 5~$TeV), $Z$-mass effects cannot be neglected 
($m_Z/m(t\bar t)\gtrsim 2\%$). As far as the neutral-current
channel is concerned, see footnote~\ref{foot:zztt}.
Conversely, $t\bar t Z$ production induces a flatter spectrum in absolute 
value, and gives a relative contribution which at its peak is the largest 
among the associated-production channels. However, while this amounts to 
about 40\% of the NNLO$_\Gamma$ cross section at 
\mbox{$m(t\bar t) \simeq 0.7 \sqrt{s}$}, there one finds oneself
at the global minimum of the cross section.

As far as higher-order EW effects are concerned, they are unusually large: 
$\delta\textrm{NLO}$ grows in absolute value
as one approaches both the small- and large-invariant-mass regions
(as is best seen from the green histogram in the middle inset),
but this happens owing to two different mechanisms. Namely, at the 
high-end of the spectrum the growth is driven by the Sudakov negative
enhancement of the virtual matrix elements, (this will be discussed in 
more detail in sect.~\ref{sec:results_sdk}) which is a relative effect 
ranging from $-10\%$ at 3~TeV to $-40\%$ at 10~TeV (similar effects have 
been observed in ref.~\cite{Ma:2024ayr}). Moreover, the rightmost bin of 
the distribution, owing to its sensitivity to soft 
emissions and to the steepness of the spectrum, shows 
even larger effects, almost twice as large as those in the neighbouring bins. 
Conversely, at small invariant masses, a positive enhancement of the cross 
section is due to the $\mu^+\gamma$ and $\gamma \mu^-$ partonic channels 
which open up at the NLO, as real-emission corrections. Indeed, an indirect
evidence of this fact is given by the excellent agreement among the
$\delta\textrm{NLO}$, $\delta\textrm{NLO}_\Gamma$, and
$\delta\textrm{NLO}$-no$V$ predictions (see the green, blue, and red
histograms in the middle inset), which can only happen if one is
dominated by $\gamma$-initiated real-emission processes. More in detail,
since the muon PDF is peaked at $x\simeq 1$, the $\mu\gamma$ channel probes 
the photon PDF at smaller Bjorken-$x$ values w.r.t.~those relevant to
the $\gamma\gamma$ one. Because of this, the NLO correction alone amounts
to 40-50\% (and grows with the collider energy) of the NNLO$_\Gamma$
prediction. For what concerns $\delta\textrm{NNLO}_\Gamma$, at small
invariant masses it has a similar shape as $\delta \textrm{NLO}$
(compare the green and black histograms in the middle inset),
and is smaller than the latter by roughly a factor of $5$, while 
it vanishes by construction at large invariant masses. At the threshold,
it increases the NLO cross section by $\sim 2\%$ and $\sim 10\%$ at 3~TeV 
and 10~TeV, respectively. These effects are unusually large for (EW) NNLO 
corrections, and this happens for the same reasons which we have discussed 
for NLO$_\Gamma$: in the case of NNLO$_\Gamma$, one predominantly probes 
both $\mu^+$ and $\mu^-$ at large $x$'s; both of these radiate a photon 
or $Z$ boson, which eventually fuse and produce the observed $t\bar t$ pair.
As far as the factorisation-scale dependence of the cross section is
concerned, the bottom inset shows clearly how such a dependence is
dramatically reduced when one includes higher-order corrections,
from tens of percents at the LO (the band encompassed by the pink-dotted
histograms) down to a few percents in NNLO$_\Gamma$ (the grey band --
bear in mind that such a band is multiplied by two in order to improve its
visibility). Having said that, we stress that $\delta$NNLO$_\Gamma$ 
constitutes an improvement of the NLO results only in the small-mass 
region, which explains why at large values of $m(t\bar t)$ the NLO 
and NNLO$_\Gamma$ bands coincide. 
We conclude by noting that acceptance cuts have a mild impact on this 
observable, and on the size of the various contributions. 

\medskip
The top-quark transverse momentum distribution, shown in fig.~\ref{fig:ptt}, 
retains most of the features of the pair invariant mass, given that these 
two observables are closely correlated at the LO. For example, the $\pt$ 
spectrum has also two peaks at or near the kinematics edges of its range,
as was the case for $m(t\bt)$. At the LO, small values of $\pt$ correspond 
to the region where LO$_{\gamma\gamma}$ is at its largest, while as the $\pt$ 
grows, one is increasingly dominated by the muon-annihilation contribution --
this is best seen from the top inset.
For small values of $\pt$, the size and hierarchy of the various contributions
largely follow the pattern observed for $m(t\bar t)$. Conversely, the
large-$\pt$ region, starting from about $\pt=\sqrt s/4$, is mostly 
correlated with the rightmost bin of the $m(t\bar t)$ distribution,
since  the large-$\pt$ spectrum is much flatter w.r.t.~its
large-$m(t\bt)$ counterpart. An evidence of this fact can be
obtained from the middle inset, where the NLO corrections are seen to
reach their asymptotic values, dominated by Sudakov logarithms, already 
halfway through the kinematics range.\footnote{The large impact of the virtual contribution, which features
a LO-like factorisation scale dependence, is such that for large $p_T$ values the LO and NLO scale 
uncertainty bands are comparable in size.} 

\medskip
The transverse momentum of the top-antitop pair, shown in fig.~\ref{fig:ptttx},
is non-trivial starting only at the NLO, since at the LO its value is equal 
to zero (the top and the antitop are back-to-back in the transverse plane). 
In view of this, namely because of the outsize importance of real-emission 
contributions for this observable, in the main frame we also show the 
NLO$_\Gamma$ prediction\footnote{This is the NLO counterpart of the NNLO
term of eq.~(\ref{final00NNLO2rep}), and it is basically the $\mu \gamma$-initiated
NLO cross section of real-emission origin -- see ref.~\cite{Frixione:2025wsv}
for further details.} (yellow-dashed histogram). As an aside, we stress
that since larger-than-zero $\pt(t\bt)$ values can only be obtained if
there are final-state particles which the top pair can recoil against,
it is impossible to obtain a non-trivial prediction for this observable
in an approach based on EW PDFs (since that would be limited to a 
LO-accurate kinematic description).
The $\pt(t\bar t)$ spectrum strongly peaks towards small values of this
observable, owing to the enhancement associated with the recoil of the
$t\bt$ pair against a soft/collinear parton. Indeed, as was discussed
above, whenever $\pt(t\bar t)>0$, NLO corrections are entirely
due to real-emission processes, which diverge\footnote{Any IR-sensitive
theory such as QED or QCD cannot have infinite-resolution power; the
cross section diverges with the logarithm of the bin size at $\pt(t\bt)=0$,
but it is finite for any fixed non-zero bin size.} for $\pt(t\bar t)\to 0$. 
As is well known, predictions in this region could be further improved 
by resumming soft/collinear logarithms, by matching fixed-order results
with either a parton shower or an analytical prediction. That said,
the soft-photon-emission dominance in the small-$\pt(t\bt)$ region is
relatively short-lived, as it can be seen by the steep decrease of the 
NLO$_\Gamma$ result. Conversely, the relative size of 
the $\delta$NNLO$_\Gamma$ correction is more significant for this observable
w.r.t.~to the other cases we have dealt with so far, and it reaches $30\%$ 
at 10~TeV -- this is in keeping with the fact that NNLO contributions
to all non-zero values for this observable are in fact the second
(as opposed to the third) non-trivial ones. At small pair transverse
momentum, the charged-current contribution is comparable to (at 3~TeV), 
or even larger than (by a factor of 2, at 10~TeV) the NNLO$_\Gamma$ 
prediction. At the opposite end of the spectrum, up to the kinematic 
edge $\pt(t\bar t) = \sqrt s/2$, one finds a remarkably similar NNLO rate 
as for the $(t\bar t Z)$-induced one. This is of course accidental,
and merely tells one that associated-production channels must be
accounted for for a sensible comparison with experimental data.
Finally, the effective-LO nature of the non-zero $\pt(t\bt)$ bins
is evident when looking at the factorisation scale dependence
(bottom inset), where one sees a very significant reduction when
NNLO$_\Gamma$ corrections are included -- clearly, this should and
does happen only in regions dominated by initial-state photons,
i.e.~at small $\pt(t\bt)$; elsewhere, NLO and NNLO$_\Gamma$ predictions
behave similarly.

\medskip
We now turn to longitudinal distributions, starting with the rapidity of the 
top quark, shown in fig.~\ref{fig:yt}. The most prominent feature of this 
observable is its asymmetry w.r.t.~the reflection around zero, 
\mbox{$y(t)\to -y(t)$}, which is present in all of the results
but LO$_{\gamma \gamma}$. This is due to the fact that, in the 
$\mu^+\mu^-$-annihilation channels, the interference between vector and axial-vector currents results in an asymmetric distribution, 
with the top quark emitted preferably in the same direction as the 
incoming $\mu^-$, i.e.~towards $y(t)<0$ in our simulations; 
hence, the relative contribution of the $\mu^+\mu^-$-initiated process 
grows with $-y(t)$.
The asymmetry pattern is non-trivial, since one finds asymmetric histograms 
also in the top inset (which, we recall, shows contributions relative to 
the NNLO$_\Gamma$ prediction) for \emph{both} LO and LO$_{\gamma \gamma}$;
in the former case, the asymmetry of the numerator is not the same as
that in the denominator, while in the latter case only the denominator
is asymmetric. At 3~TeV, the LO spectrum is bound in the region 
$|y(t)|\lesssim 2.85$ by kinematics constraints; these are relaxed
at higher orders, although marginally so, since the corresponding
rates are very suppressed (by moving from the bin that includes the LO
kinematical limit to its nearest neighbour with larger $|y(t)|$ values,
one finds a cross section which is about four orders of magnitude
smaller). At 10~TeV the LO kinematic limit becomes $|y(t)|\lesssim 4.06$
(still in the absence of acceptance cuts). In addition to this enlarged
range, one notes that at 3~TeV the NNLO$_\Gamma $ prediction is dominated 
by the $\mu^+\mu^-$ annihilation process (i.e.~by large $m(t\bar t)$ values),
while at 10~TeV one finds predominantly $\gamma \gamma$-initiated 
contributions (i.e.~small $m(t\bar t)$ values).
Another interesting feature to notice, when looking at the NLO
results, is that the effect of virtual corrections is largest for
$y(t)<0$, where it suppresses the cross section, a fact which is 
particularly visible either at 3~TeV or at 10~TeV with acceptance cuts (see the
difference between the green- and red-dashed histograms in the middle inset). 
Again, this is explained by the dominance of the $\mu^+\mu^-$ annihilation 
channel in this region, which in turn is strictly correlated with large 
$m(t\bar t)$ values; since virtual matrix elements are negative there,
NLO corrections reduce the asymmetry. At 10~TeV the size of the NLO and
the NNLO$_\Gamma$ corrections is strikingly large (see the middle inset), 
and this reflects what happens for the top-pair invariant mass at low 
values of such an observable. Finally, for what concerns the 
associated-production channels, namely $t\bar t \nu\bar \nu$ and 
$t\bar t Z$, the key points are as follows. Firstly, they also underpin
asymmetric spectra, with the asymmetry of $t\bar t Z$ ($t\bar t \nu\bar\nu$)
being in the same (opposite) direction as that of $t\bar t$. 
Secondly, $t\bar t Z$ rates are quite suppressed for this distribution, 
while those for $t\bar t \nu\bar \nu$ production are chiefly
due to the dominant charge-current channel. 
Lastly, the effect of factorisation-scale variations is quite flat, except 
at the LO, in which case the corresponding uncertainty is smaller
at $y(t)<0$, where the competition between the $\gamma\gamma$ and
$\mu^+\mu^-$ contributions leads (accidentally) to a compensating effect. 
NLO corrections reduce the LO uncertainty band by a factor of 2 (3) at 
3 (10)~TeV w.r.t.~its minimum. The inclusion of 
$\delta$NNLO$_\Gamma$ leads again to a reduction of the band width, 
by another factor of 2 (4).

\medskip
The last observable we consider is the rapidity of the top pair, $y(t\bar t)$,
shown in fig.~\ref{fig:yttx}. At variance with $y(t)$, this observable is 
symmetric under reflections, but nonetheless displays a rather peculiar 
shape, in particular at 3~TeV where the muon-annihilation channel has the
largest relative impact. Indeed, the distribution shows a marked
peak at $y(t\bar t)=0$, which stems from the muon-annihilation channel 
at the LO, when both the $\mu^+$ and $\mu^-$ PDFs are probed at large 
Bjorken $x$'s -- since both $x$'s are very close to one, the resulting
$y(t\bt)$ distribution (which is fairly strictly correlated with the
logarithm of the ratio of the two $x$'s) exhibits a narrow peak,
centred at zero. This peak structure is superimposed to the results
of other channels, and is thus diluted by them -- indeed, NLO and NNLO 
corrections have much flatter spectra, thanks to the fact that they 
feature extra final-state particles in addition to the $t\bt$ pair. In
particular, the inclusion of NLO corrections reduces the difference between 
the two central bins and the adjacent ones\footnote{The contributions that do not feature such an enhancement in the central 
region have a dip in the ratio over NNLO$_\Gamma$ plotted in the top and
middle insets. Note that such a dip is due to the denominator 
rather than to the numerator.}. The LO prediction for this observable 
is bound in the range \mbox{$|y|<-\log(2m_t/\sqrt{s})$}, while beyond
LO larger absolute values can be accessed, albeit with much suppressed
rates, and an increased factorisation-scale dependence (as can be seen
in the bottom inset). Similarly to the case of $y(t)$, $y(t\bt)$ is 
dominated by the large invariant-mass muon-induced peak at 3~TeV, and by 
the small invariant-mass photon-induced contribution at 10~TeV.


\subsubsection{$W^+W^-$ inclusive production}
\label{sec:results_ww}

\begin{figure*}	
    \centering
    \figtable
        {\includegraphics[height=0.5\textwidth, angle=0]{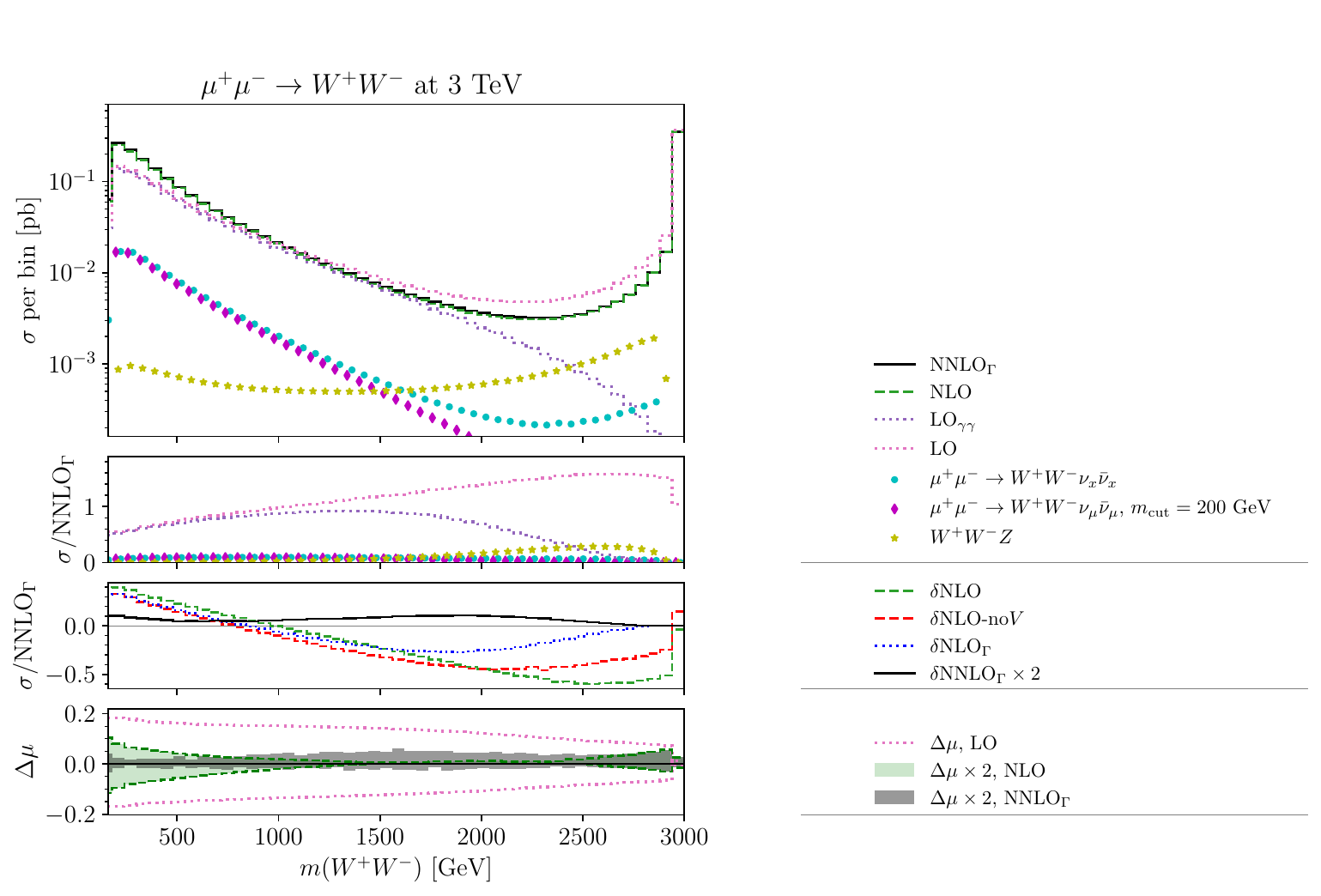}}
        {\includegraphics[height=0.5\textwidth, angle=0]{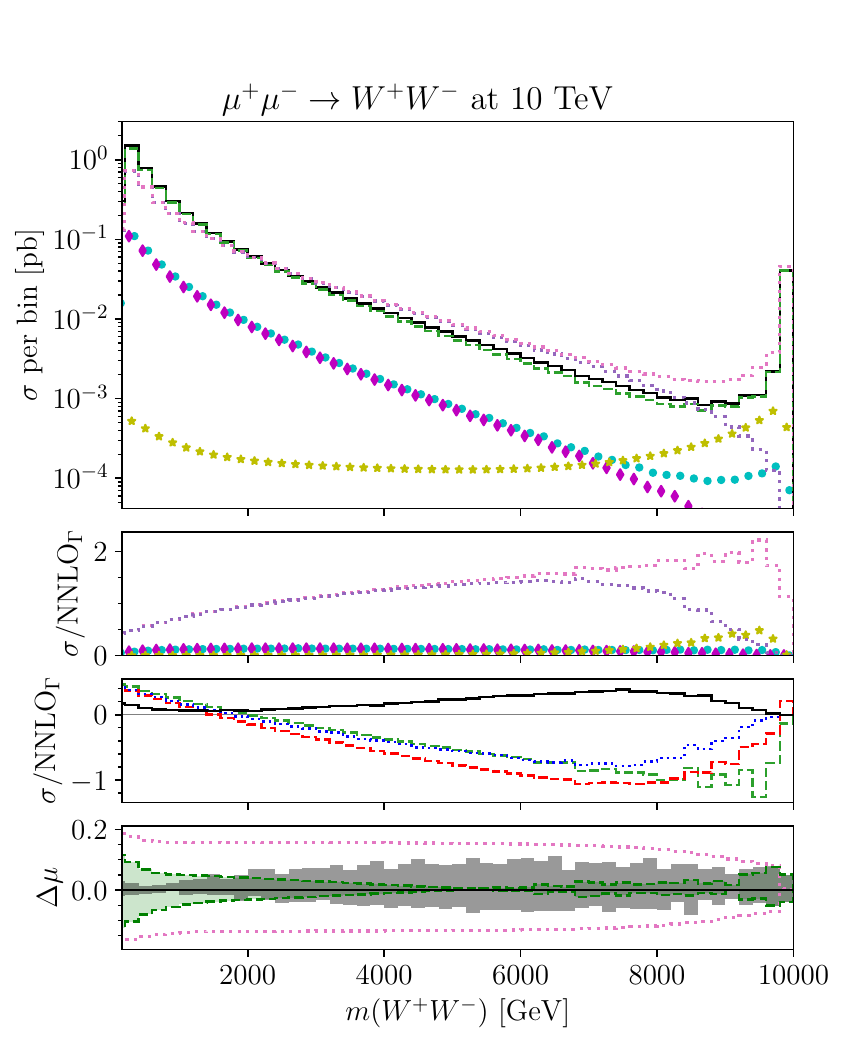}}	
        {\includegraphics[height=0.5\textwidth, angle=0]{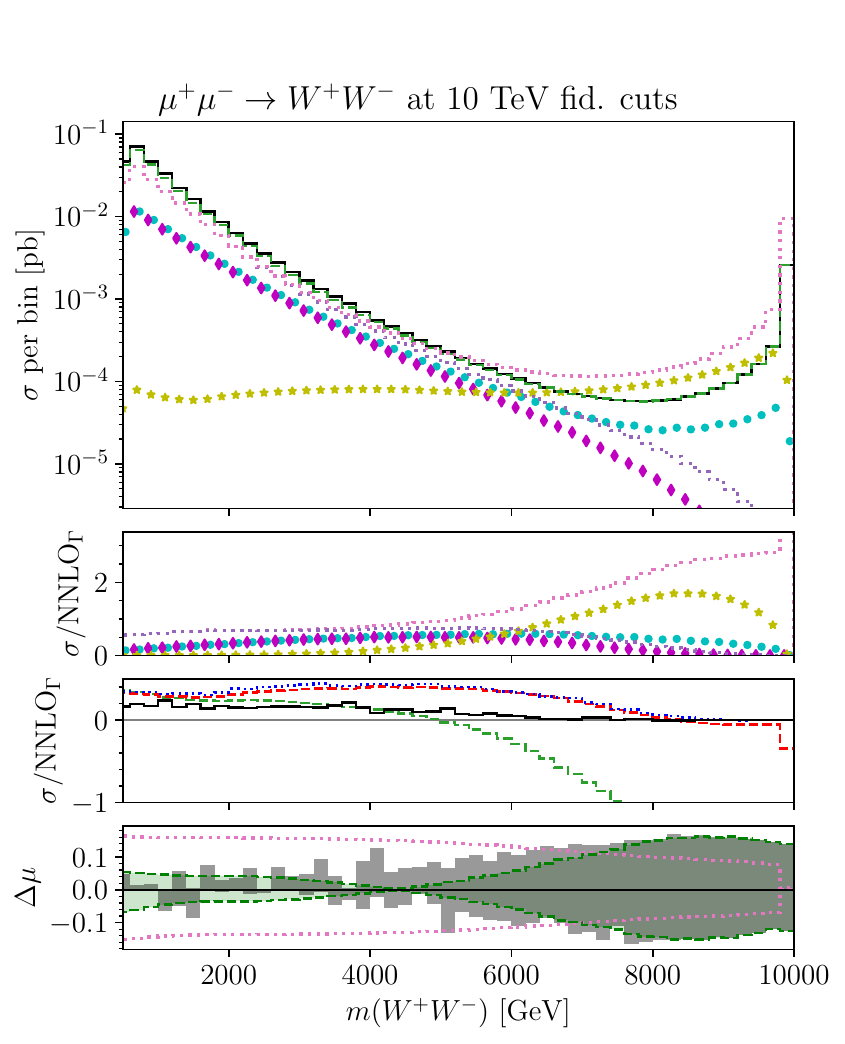}}
	\caption{Same as in fig.~\ref{fig:mtt}, for the $W^+W^-$ pair invariant-mass distribution, $m(W^+W^-)$, in $W^+W^-$ production. } 
	\label{fig:mww}%
\end{figure*}

\begin{figure*}
    \centering 
    \figtable
        {\includegraphics[height=0.5\textwidth, angle=0]{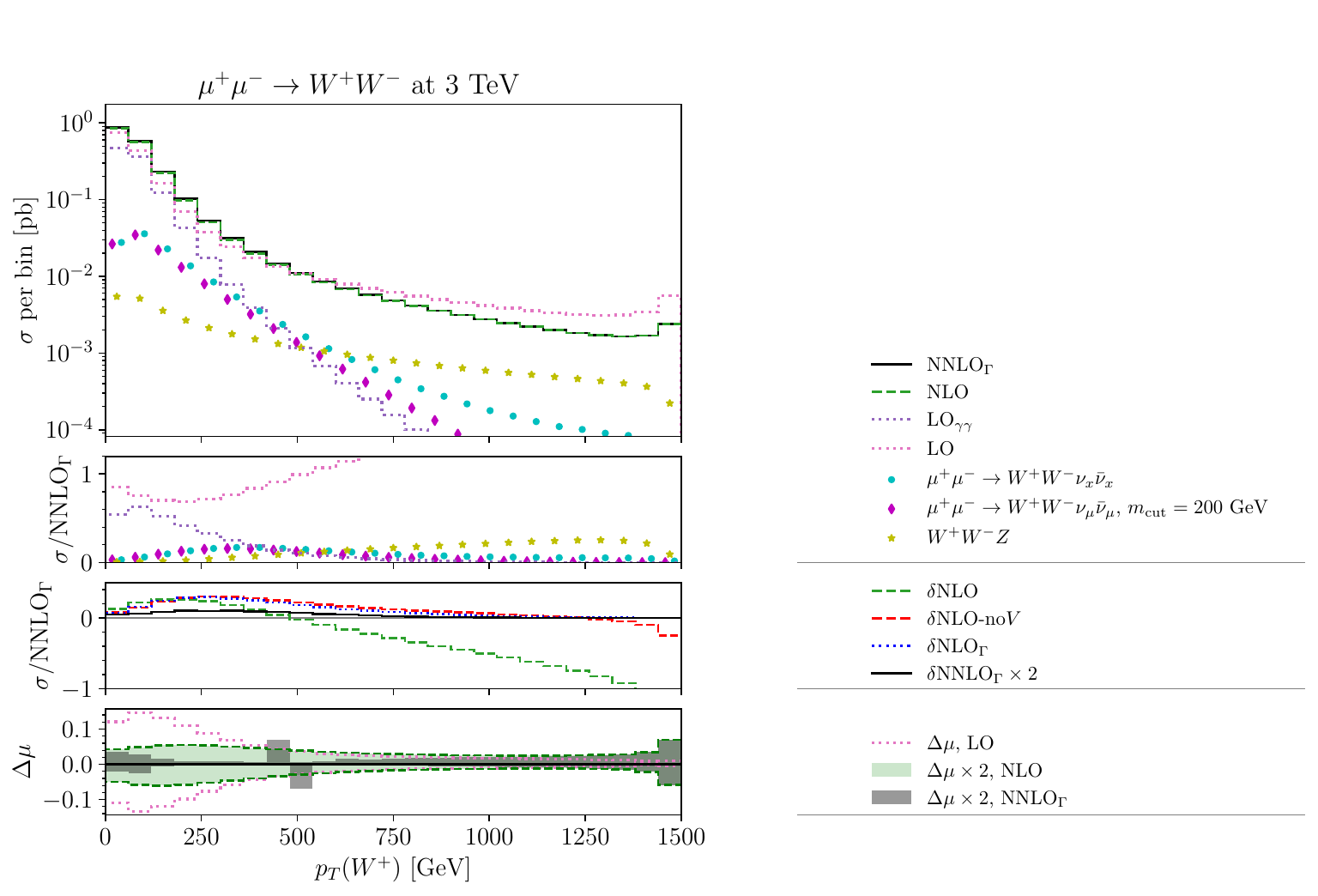}}	
        {\includegraphics[height=0.5\textwidth, angle=0]{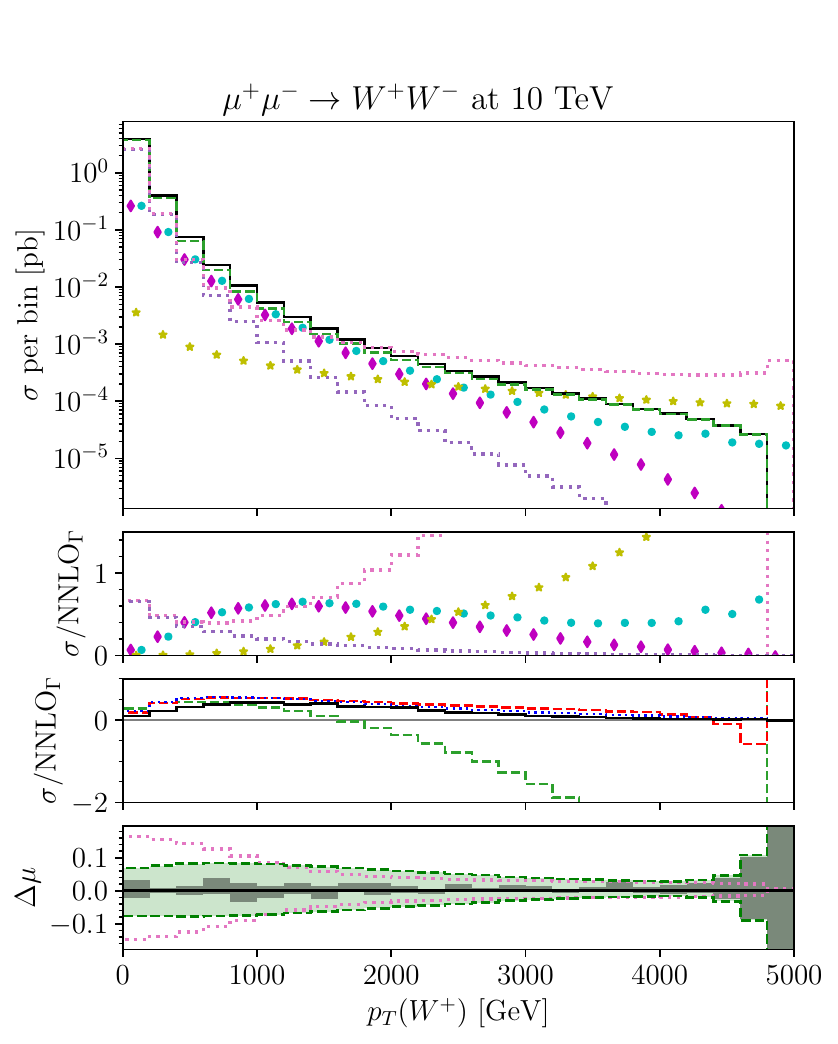}}
        {\includegraphics[height=0.5\textwidth, angle=0]{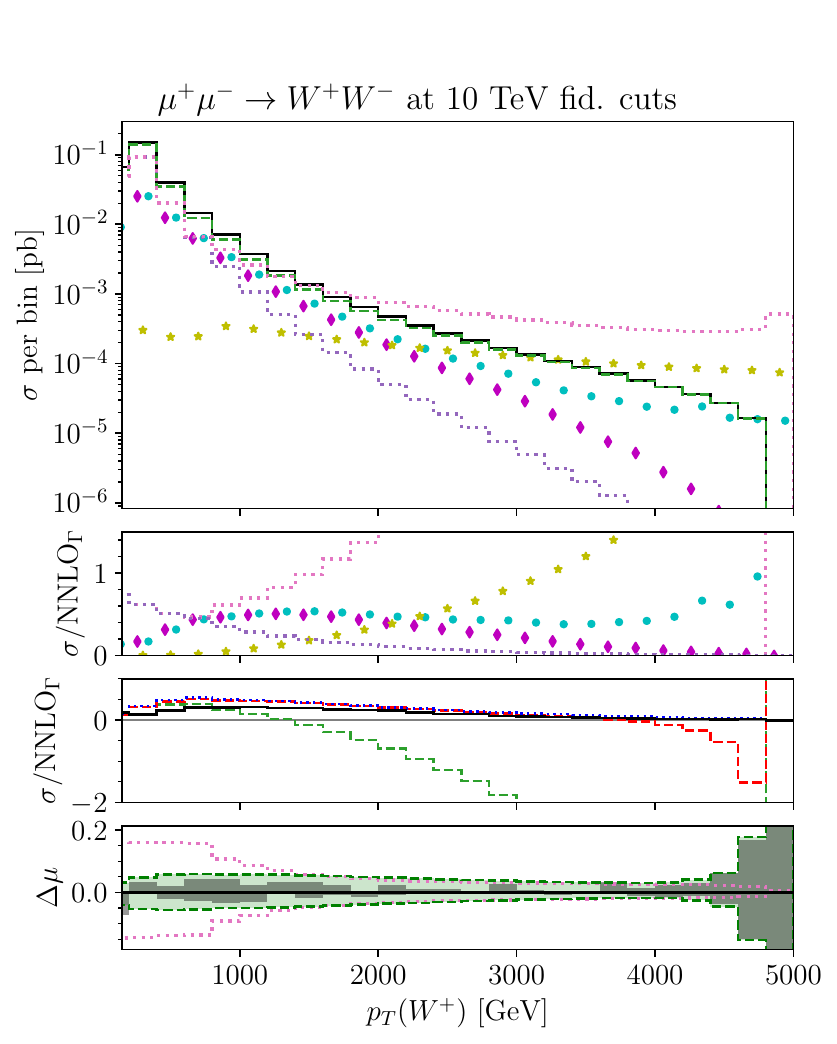}}
	\caption{Same as in fig.~\ref{fig:mww}, for the $W^+$-boson transverse-momentum distribution, $\pt(W^+)$. } 
	\label{fig:ptw}%
\end{figure*}

\begin{figure*}
    \centering 
    \figtable
        {\includegraphics[height=0.5\textwidth, angle=0]{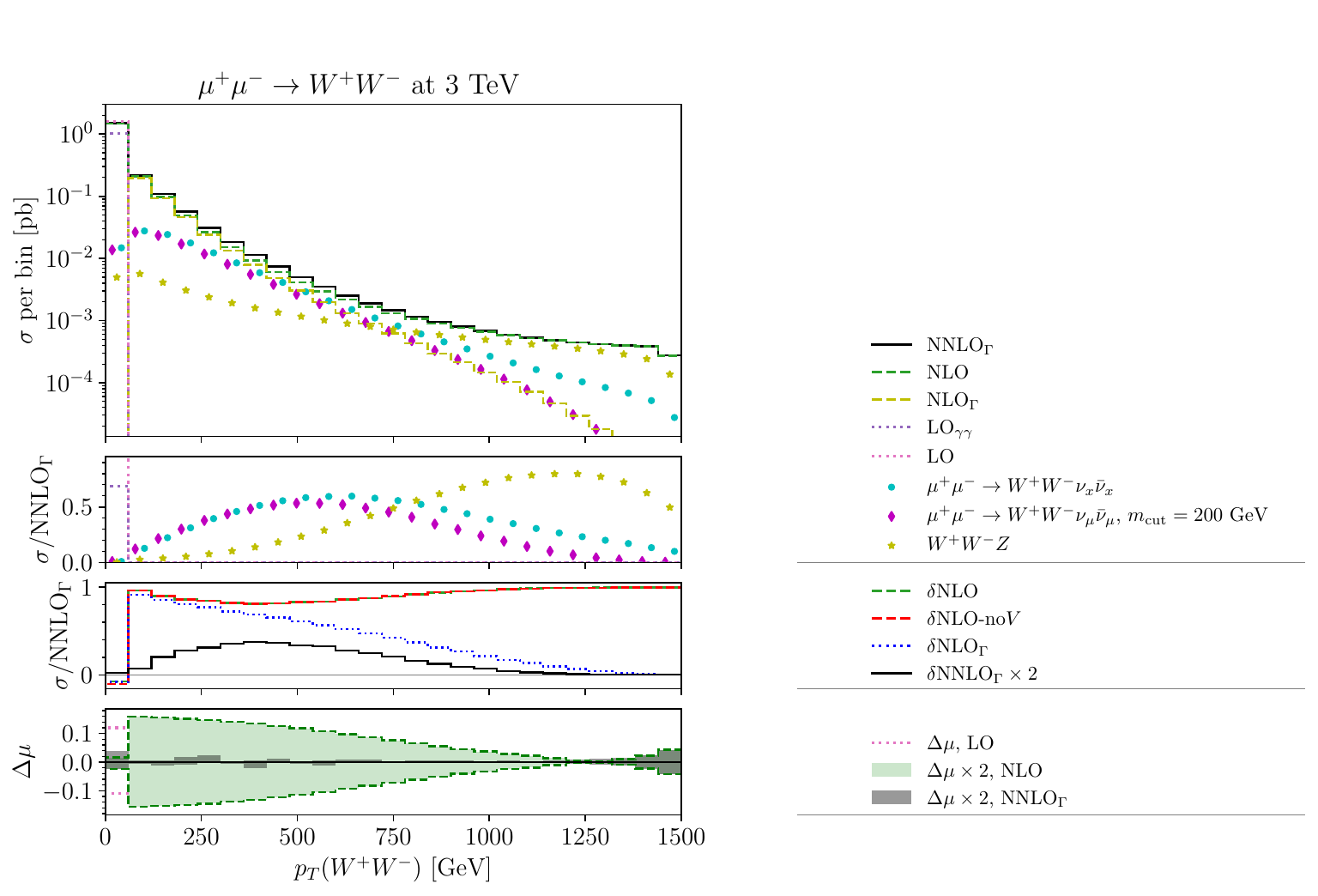}}	
        {\includegraphics[height=0.5\textwidth, angle=0]{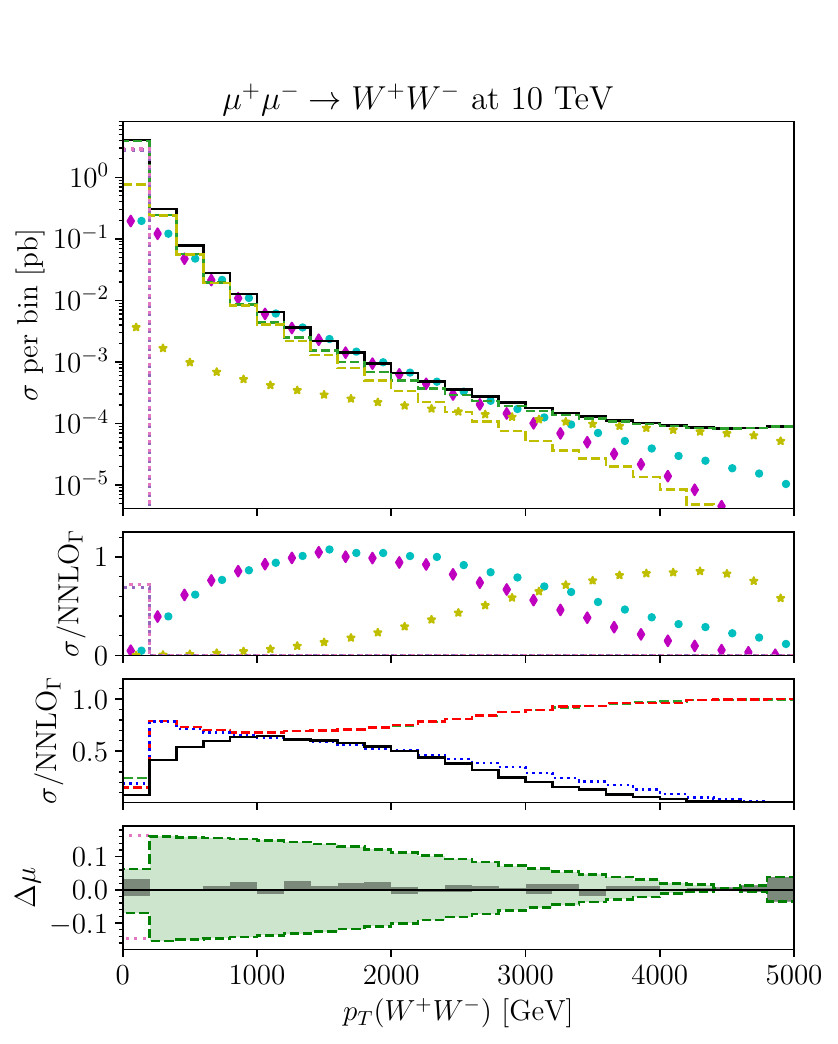}}
        {\includegraphics[height=0.5\textwidth, angle=0]{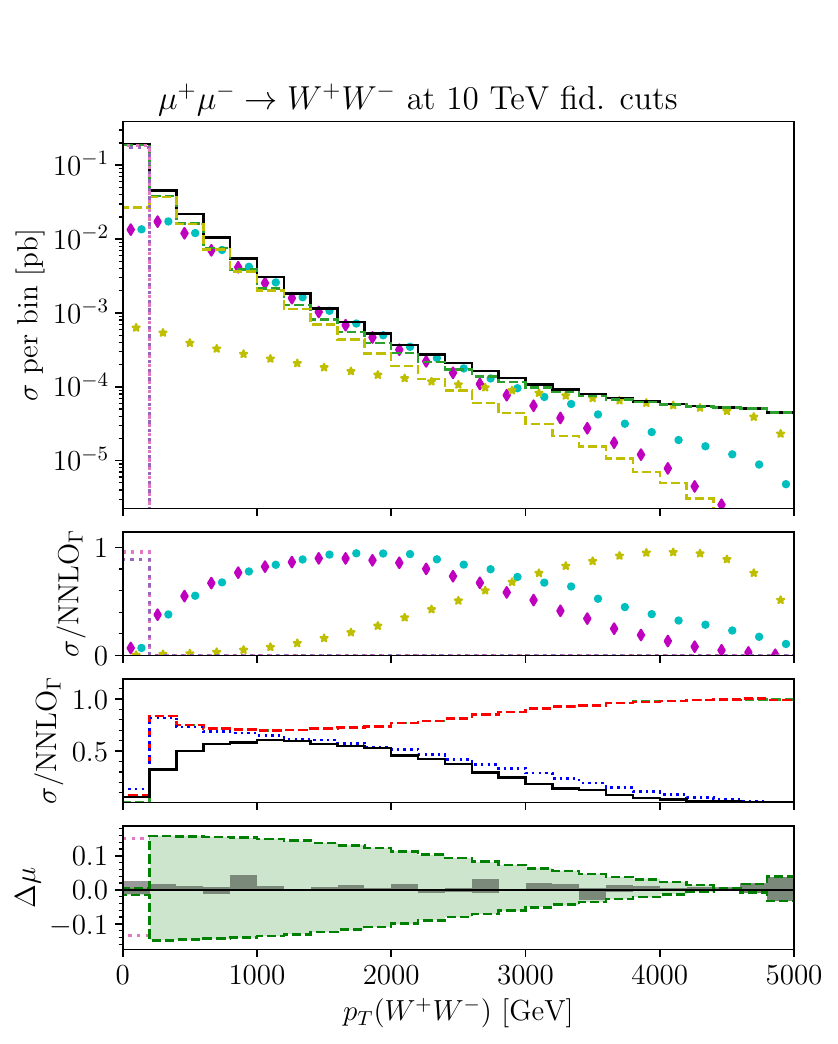}}
	\caption{Same as in fig.~\ref{fig:mww}, for the $W^+W^-$ pair transverse-momentum distribution, $\pt(W^+W^-)$.  } 
	\label{fig:ptww}%
\end{figure*}

\begin{figure*}
    \centering 
    \figtable
        {\includegraphics[height=0.5\textwidth, angle=0]{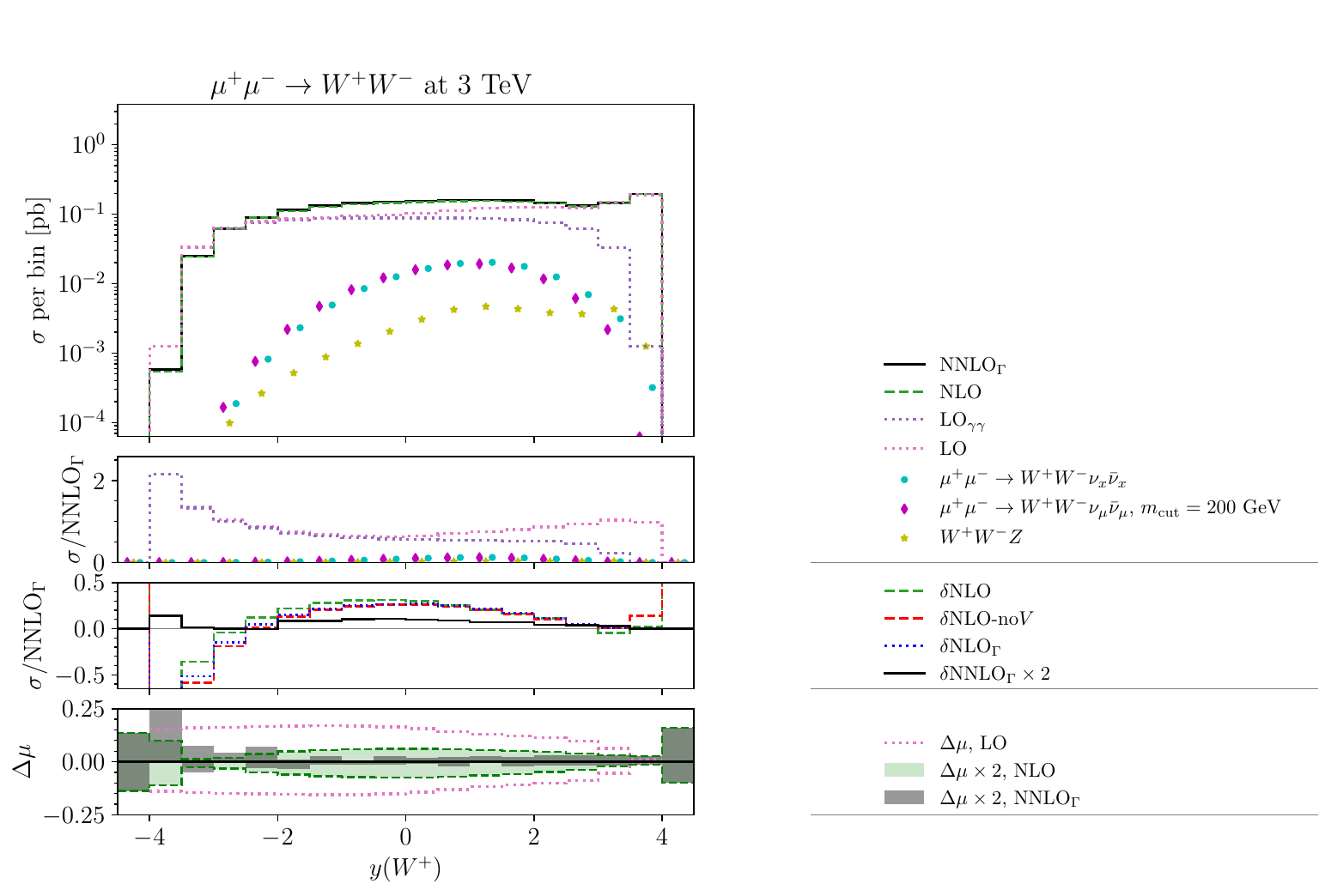}}	
        {\includegraphics[height=0.5\textwidth, angle=0]{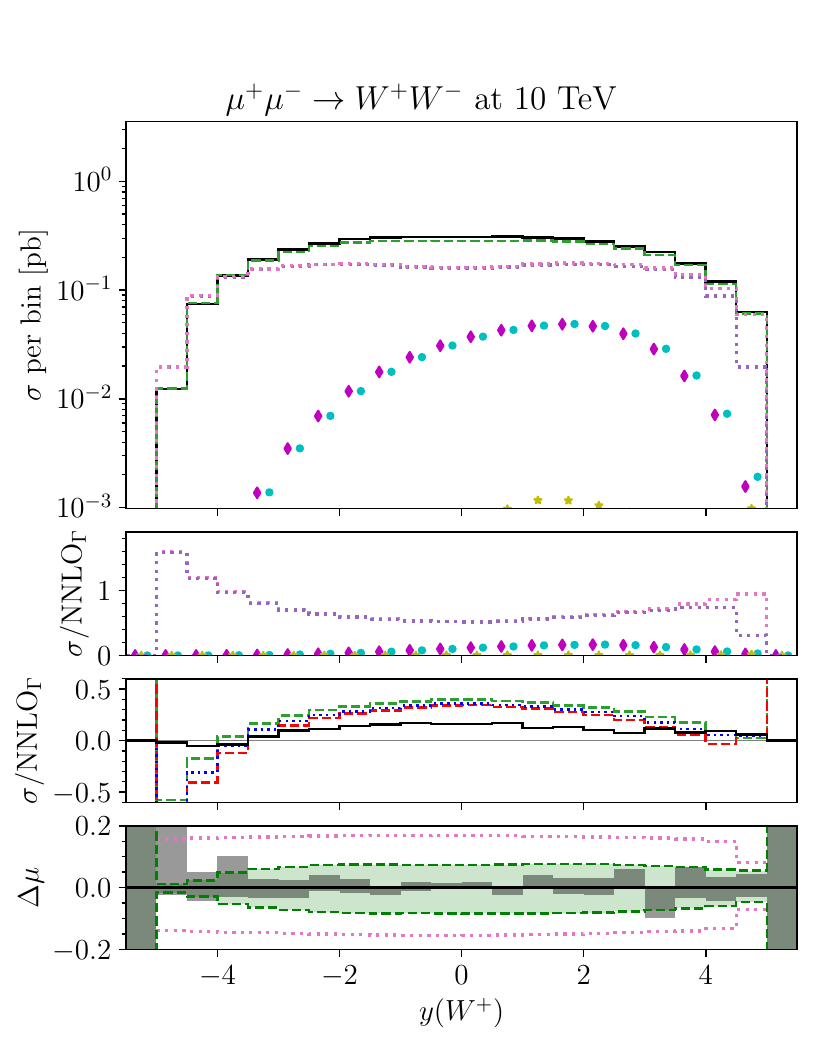}}
        {\includegraphics[height=0.5\textwidth, angle=0]{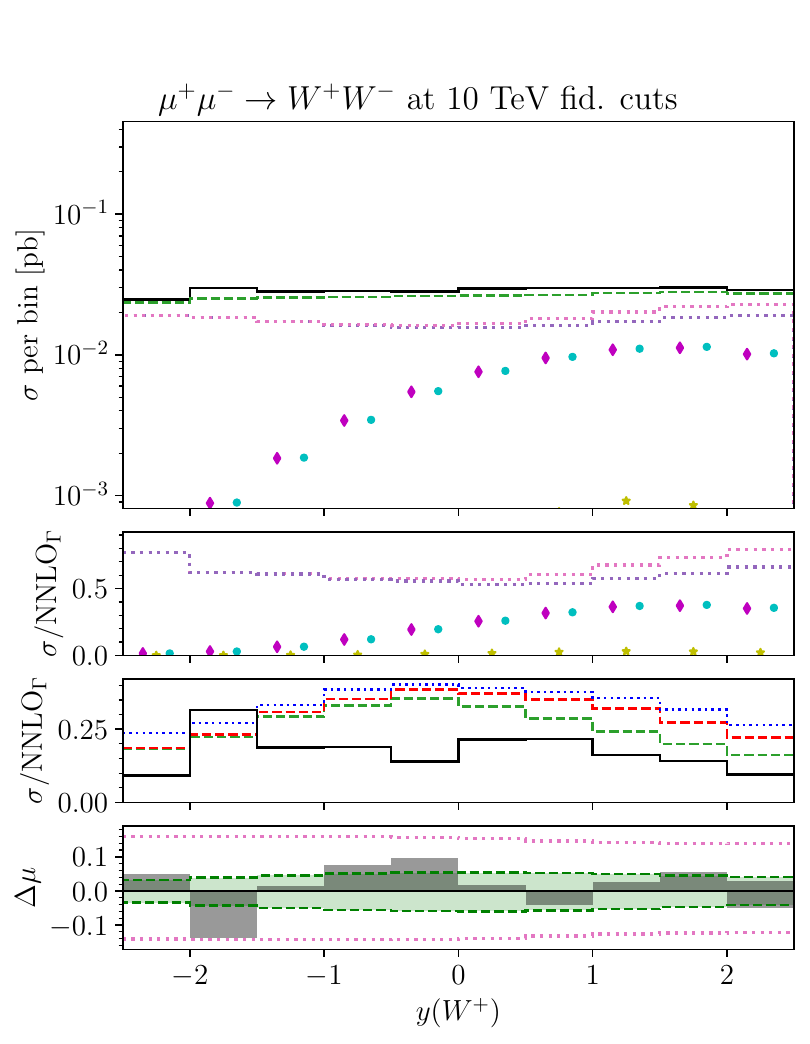}}	
	\caption{Same as in fig.~\ref{fig:mww}, for the $W^+$-boson rapidity distribution, $y(W^+)$. } 
	\label{fig:yw}%
\end{figure*}

\begin{figure*}
    \centering 
    \figtable
        {\includegraphics[height=0.5\textwidth, angle=0]{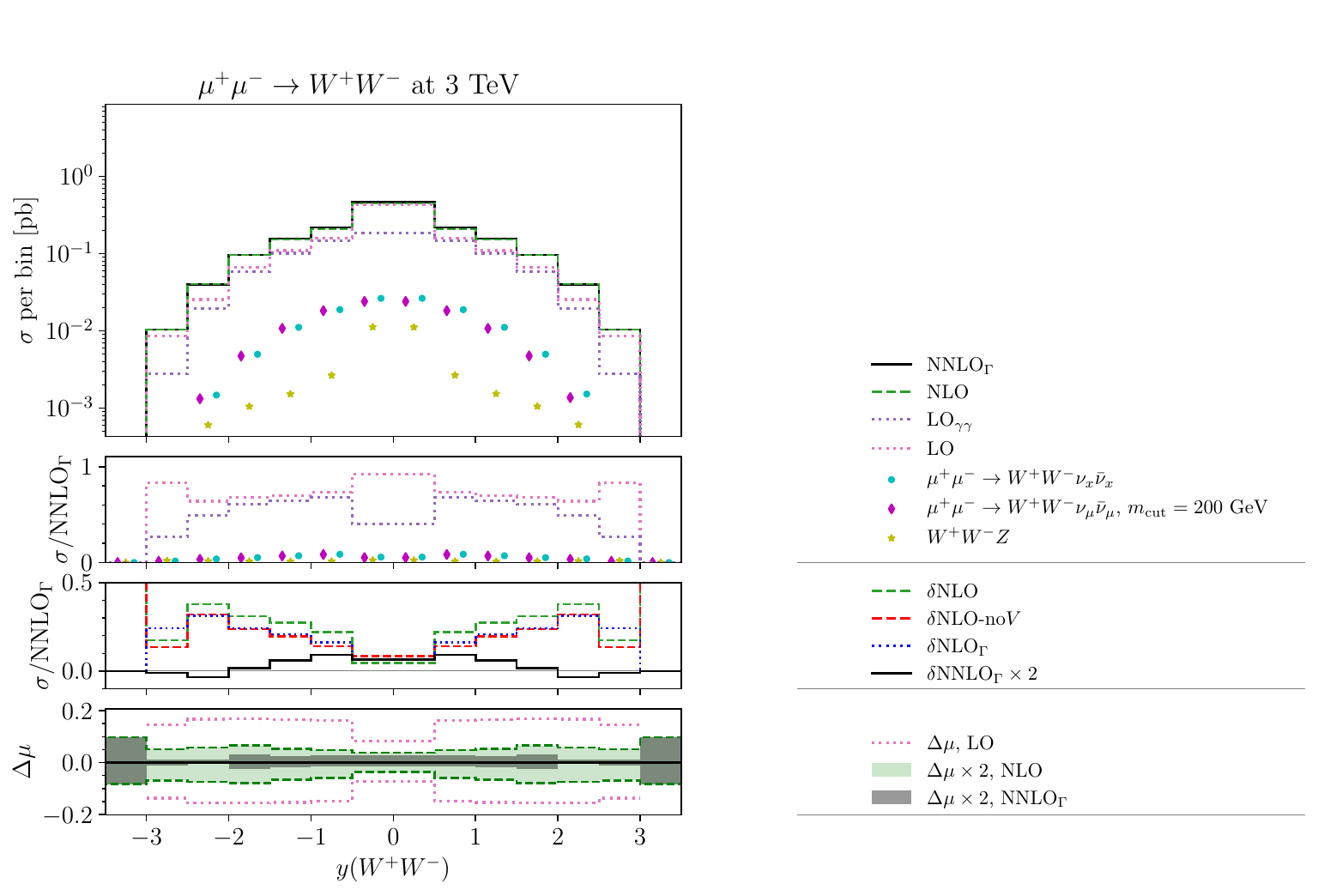}}	
        {\includegraphics[height=0.5\textwidth, angle=0]{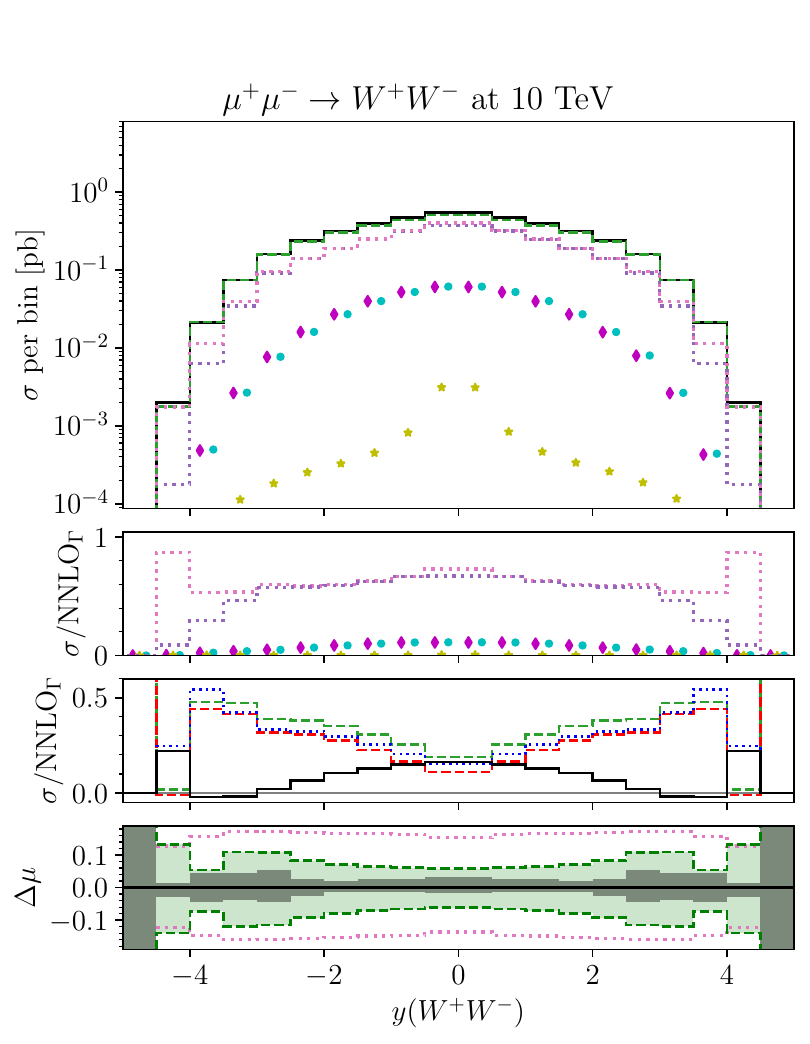}}
        {\includegraphics[height=0.5\textwidth, angle=0]{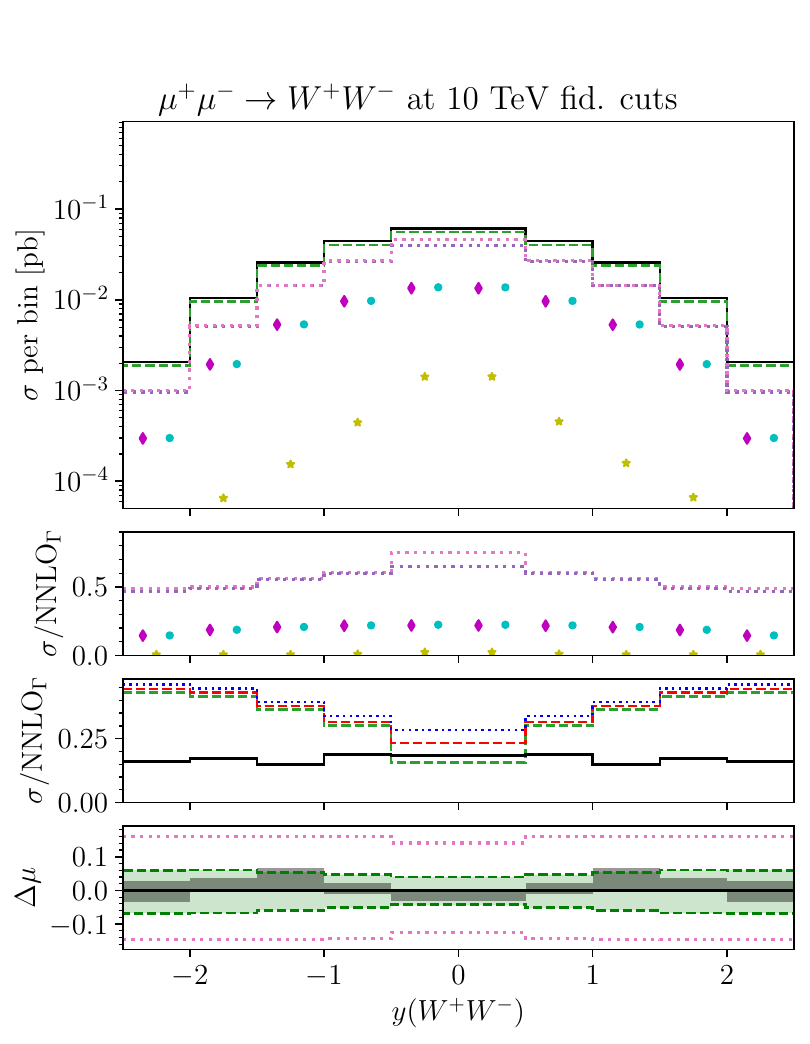}}
	\caption{Same as in fig.~\ref{fig:mww}, for the $W^+W^-$ pair rapidity distribution, $y(W^+W^-)$. } 
	\label{fig:yww}%
\end{figure*}

We now turn to discussing $W^+W^-$ production. The observables which we 
consider are the strict analogues of those analysed in 
sect.~\ref{sec:results_tt}, namely the invariant mass of the $W^+W^-$ pair 
(fig.~\ref{fig:mww}), the transverse momentum of the $W^+$ 
(fig.~\ref{fig:ptw}), the transverse momentum of the
pair (fig.~\ref{fig:ptww}), the rapidity of the $W^+$ (fig.~\ref{fig:yw}), 
and the rapidity of the pair (fig.~\ref{fig:yww}). Having
commented at length on the case of $t\bar t$ production in 
sect.~\ref{sec:results_tt}, here we mostly focus on the {\em differences} 
that one finds in $W^+W^-$ production w.r.t.~the former process. When
looking at the distributions and comparing them to the case of $t\bar t$
production, it may be helpful to bear in mind some general features of
$W^+W^-$ production. Firstly, the relative impact of the photon-initiated
contribution is larger than in $t\bar t$ production, both because of
matrix-element enhancements and because the $W$ boson, being lighter than the
top quark, makes it possible to probe smaller Bjorken $x$'s. Secondly, while 
in $t\bar t$ production the two heavy quarks always belong
to the same fermionic line, in turn connected to the other fermionic
lines of the processes (the muons') by means of internal spin-$1$ propagators,
this is not the case in $W^+W^-$ production, where the bosons can also be
directly radiated by the muon themselves. An immediate consequence of
this is that $W^+W^-$ production features a larger number of topologies
w.r.t.~$t\bt$ production. For example, the muon-induced contribution at the LO for
$t\bar t$ proceeds only via $s$-channel diagrams, while for $W^+W^-$ also
$t$-channels exist. 
Thirdly, virtual corrections, in particular when large (w.r.t.~the EW
scale) energy scales are relevant, become huge in absolute value (almost
as large as the LO results) and negative. This implies that, in order 
to have reliable predictions, effects beyond NLO must be included in the cross
section, e.g.~via resummation of EW Sudakov logarithms~\cite{Denner:2024yut}. 
We shall return to this point in sect.~\ref{sec:results_sdk}.

\medskip
The $W$-pair invariant mass, shown in fig.~\ref{fig:mww}, is a prototypical
observable where the differences w.r.t.~its $t\bt$-analogue, which we have 
just discussed, are apparent. Indeed, already at 3~TeV, the LO prediction is
dominated by the photon-induced contribution up to $m(W^+W^-)\sim 2$~TeV
(i.e., $\gamma$-initiated channels account for more than 50\% 
of the cross section); at a 10-TeV collider, this dominance extends up to
$m(W^+W^-)\sim 8$~TeV. The positive NLO corrections which appear at small 
$m(W^+W^-)$ and are chiefly due to the opening of the $\mu^+\gamma$ and 
$\gamma \mu^-$ channels, reach 40-50\% of $\NNLOG$ at the threshold (see the 
green-dashed histogram in the middle insets) . At variance with $t\bt$
production, their contributions become negative and large in absolute 
value as the invariant mass grows, until one reaches the region dominated 
by the $\mu^+\mu^-$-annihilation channel, where instead the Sudakov 
suppression stemming from the virtual contribution is apparent. 
From the histograms in the middle insets, one can also see that the 
contribution of the virtual matrix elements is positive for the 
$\gamma\gamma$ channel. The impact of the $\delta$NNLO$_\Gamma$ correction 
is also relatively large: it amounts to 5\% of the total $\NNLOG$
predictions at 3~TeV, and exceeds 20\% at 10~TeV when $m(W^+W^-)\sim 8$~TeV. 
At both c.m.~energies, as is expected, such a
correction dies off together with LO$_{\gamma\gamma}$. Finally, the impact of
the associated production channels is rather negligible, with the only 
exception of $W^+W^-Z$ production amounting to some $10-20\%$
of $\NNLOG$ at large invariant masses. 

The presence of acceptance cuts at 10 TeV significantly changes the
pattern we have just discussed. The main effect of such cuts is due to 
the ones in rapidity, as we shall see in detail when considering the 
$W$-rapidity distribution. In particular, the rapidity cut suppresses
 $W$-boson emissions from the $\mu^\pm$ lines in the
forward/backward regions, and thus it reduces the rate across the 
whole $m(W^+W^-)$ range by more than one order of magnitude, with a 
stronger suppression of the small-mass $\gamma\gamma$-induced peak 
than of the large-mass $\mu^+\mu^-$-induced one. In this case, the
dominance of the photon-initiated contribution fades off earlier, at around
$m(W^+W^-)=6$~TeV. The contribution of the virtual matrix elements is 
negative already at 2 TeV, and grows quickly in absolute value with 
the pair-invariant mass\footnote{For better visibility, the range of the 
middle inset has been limited, and does not show the green histogram in
its entirety. A more detailed discussion will follow in 
sect.~\ref{sec:results_sdk}.}. The acceptance cuts have a milder effect 
on $W^+W^-Z$ associated production; therefore, this channel has a larger
relative contribution w.r.t.~the case where no cuts are imposed.
Last but not least, it is counterintuitive (and at variance with all
of the other cases discussed so far) that the $\NNLOG$ factorisation-scale 
variation bands in the bottom insets are larger than their analogues at the
NLO. However, we have explicitly verified that this is due to an accidental 
reduction (stemming from a cancellation) of the scale dependence at the NLO, 
in turn due to the large impact of virtual corrections (which have a LO-like 
scale dependence). If the virtual matrix elements are not included in the
predictions, the $\dNNLOG$ corrections reduce the scale dependence as is
expected.

\medskip
The very large impact of virtual corrections at the NLO is apparent in the
$\pt(W^+)$ distribution, shown in fig.~\ref{fig:ptw}, for large values of 
the transverse momentum (see the difference between the green and red
histograms in the middle panel). This observable displays a much steeper 
decrease w.r.t.~its $t\bar t$-production counterpart which we have discussed
earlier, owing to the larger impact of the photon-induced contribution. While 
at the LO the predictions become less steep once moderate/large values of 
$\pt$ are attained, the inclusion of NLO corrections changes this pattern,
leading to steeper-than-LO results in this kinematic region. It is worth
mentioning that NLO corrections grow so large in absolute value that the 
cross section in the rightmost bin becomes negative. Again, this is a sign 
of the need to resum the large effects due to the EW Sudakov logarithms. 
In the small-$\pt$ range, which is dominated by the photon-induced 
contribution, higher-order corrections are positive, similarly to what 
happens with the analogous observable in $t\bar t$ production. While
at 3~TeV the impact of the associated production processes is small, 
at 10~TeV this is not the case. In particular, the charged-current 
channel gives a contribution which is comparable to the complete 
$\NNLOG$ prediction at around $\pt(W^+)=1$~TeV; acceptance cuts only 
slightly reduce its impact. Conversely, $W^+W^- Z$ production contributes 
mostly towards the large-$\pt$ tail of the spectrum. However, one must take
into account that this associated channel is simulated at the LO accuracy 
only, while NLO corrections are likely to reduce its impact significantly
(see e.g.~ref.~\cite{Ma:2024ayr}).

\medskip
Turning to $\pt(W^+W^-)$, shown in fig.~\ref{fig:ptww}, we remind the
reader that, in full analogy with its $t\bt$ counterpart, this observable 
is trivial in $2\to 2$ kinematic configurations, such as at the LO or within 
an EW-PDF description. Indeed, all processes that do not feature any initial-
or final-state radiation contribute to the $\pt(W^+W^-)=0$ bin. NLO is thus
the lowest order which populates the $\pt(W^+W^-)>0$ region. Similarly
to what happens for $\pt(t\bar t)$, this translates into a very large
impact of the $\delta$NNLO$_\Gamma$ correction, whose contribution reaches 
20\% (35\%) of the total NNLO$_\Gamma$ prediction at 3~TeV (10~TeV) --
see the black histogram in the middle inset. 
The contributions of
the associated-production processes follow a similar pattern as in
the case of $\pt(W^+)$, with $W^+W^-Z$ becoming comparable to NNLO$_\Gamma$ 
at large values of $\pt(W^+W^-)$ at both 3~TeV and 10~TeV.
Conversely, at small and intermediate values of this observable it is
the $W$-annihilation contribution which is comparable to $\NNLOG$.
As in the case of $\pt(t\bar t)$, the accuracy in the neighbourhood of
the first bin can be further improved by resumming soft/collinear 
logarithms. Also, being the first non-trivial order, the NLO prediction 
has a LO-like (in size) factorisation scale dependence, while the 
NNLO$_\Gamma$ one significantly reduces it, at least for small values 
of this observable.

\medskip
As far as rapidity distributions are concerned, we begin by looking at 
$y(W^+)$, shown in fig.~\ref{fig:yw}. Similarly to its $t\bt$ counterpart, 
this observable is asymmetric, with a much more pronounced asymmetry
w.r.t.~to the former case. Indeed, because of the presence of the 
$t$-channel diagrams that feature $\mu^+\mu^-\to W^+ W^-$ subgraphs, which 
would diverge at zero angle if the $W$ were massless, the $W^\pm$ boson is 
emitted preferably along the direction of the incoming $\mu^\pm$. 
At a c.m.~of 3 TeV, this translates into a strongly-enhanced rate
for the $W^+$ boson at large, positive rapidity, with the distribution
featuring a rather curious shape with two maxima. Such a feature is absent at
10 TeV, due to the dominance of the symmetric $\gamma\gamma$ channel. The
associated-production processes display an asymmetry with the same sign as
that of the $\NNLOG$ cross section. However, their rate is negligible, 
unless cuts are applied. In that case, the $W$-fusion channel is roughly 
40\% of the NNLO$_\Gamma$ prediction for $y(W^+)>0$.  The pattern of
higher-order corrections is quite non-trivial: towards the negative edge of
the LO spectrum (dominated by $\gamma\gamma$ fusion), NLO corrections start with being 
negative, then become positive around $y(W^+)\simeq -3$ ($y(W^+)\simeq -4$ )
at 3 (10) TeV. They reach the maximum (about 30\% of NNLO$_\Gamma$ for either
energy) around $y(W^+)=0$ and start to decrease, approaching zero at the
high-end of the distribution. Remarkably, and consistently with what has
been observed for the $m(W^+W^-)$ distribution, there is an overall good
agreement between NLO$_\Gamma$ and NLO without virtual matrix elements. 
The impact from the virtuals is the largest (and positive) for $y(W^+)<0$, 
where the $\gamma\gamma$ initial state is dominant. The $\dNNLOG$ correction is
generally positive, more symmetric, and it grows as one approaches the central
region, up to 10\% or 20\% respectively at 3 and 10 TeV. At 10 TeV, when
acceptance cuts are applied, a large part of the cross section is discarded,
and one obtains a very flat distribution. In this case, as was mentioned when
discussing $m(W^+W^-)$, cuts have a less severe impact on the $\mu^+\mu^-$ 
contribution than on the $\gamma\gamma$ one. This enhances the 
$\mu^+\mu^-$-initiated fraction of the cross section at positive rapidity, 
for which NLO corrections receive a negative virtual contribution. Conversely,
at negative rapidity, the LO cross section is entirely due to $\gamma\gamma$
fusion. This fact leads to an asymmetric impact of the NLO corrections.

\medskip
Finally, we look at the rapidity of the $W^+W^-$ pair, shown in 
fig.~\ref{fig:yww}. The kinematics bounds for this observable are  
$|y(W^+W^-)|<2.9$ and $|y(W^+W^-)|<4.1$ at 3 and 10~TeV, respectively. 
The central peak, due to the $\mu^+\mu^-$-annihilation channel, is less 
apparent than for $t\bar t$ production, again consistently
with the smaller impact of this channel on the cross section. Higher-order
corrections are all positive: at the NLO, they grow when moving from 
the central towards the forward/backward regions, starting from 5\%(20\%) 
at 3~TeV (10~TeV), and growing up to 40\% (50\% at 10~TeV). We note that
the $\dNNLOG$ corrections display the opposite behaviour: they are very 
small in the forward/backward region, and grow up to 5\% (10\%) at 
$y(W^+W^-)=0$. The impact of the associated-production processes
is similar to the case of $y(W^+)$.

\subsubsection{Summary of key phenomenological features}
In general, our results for $W^+W^-$ production reinforce the conclusions
drawn from the $t\bar{t}$ analysis; at the same time, they also highlight 
a few differences due to some underlying physics characteristics, as
follows.
\begin{itemize}
\item For both $t\bar t$ and $W^+W^-$ production, the uncertainties 
associated with missing higher orders, which we have estimated by studying 
the behaviour of the perturbative series and its scale (and scheme, as 
we shall see later) dependence, appear to be under control.

\item The number of subprocesses and topologies, in particular those
which are not VBF-like, that lead to a $W^+W^-$ final state is larger 
than that relevant to $t\bt$ production.
This results in larger corrections across all observables.

\item The shape and scale of $W^+W^-$ differential distributions are strongly
affected by double-emissions from the muonic lines, which are absent 
in $t\bar{t}$ production.

\item $W^+W^-$ rapidity distributions show a stronger forward shift, 
consistently with the higher probability of independent boson emissions.

\end{itemize}

Taken together, the $W^+W^-$ and $t\bar{t}$ production processes provide
complementary tests of the robustness of the QED-resummed and fixed-order
electroweak framework we have introduced. They both show that our approach is
not only consistent with our stated goals, but also that it can reliably
account for process-dependent behaviours, and provide a systematic way to
evaluate uncertainties, which appear to be under control.


\subsection{Inclusion of the Sudakov effects}
\label{sec:results_sdk}
In inclusive-production processes at multi-TeV muon colliders, the
kinematic configuration of high energies {\em and} large scales is 
associated with a large fraction of events, at variance with what
happens in hadronic collisions -- this behaviour is summarised in
the sketch of fig.~\ref{fig:flag}, and is underscored by the
explicit results of sect.~\ref{sec:results_gen}. In such a configuration,
EW corrections are dominated by large logarithmic terms of the form 
\mbox{$\alpha^n\log^k(Q^2/M_W^2)$} with $1\le k\le 2n$; at the NLO, 
in particular, there are double ($k=2$) and single ($k=1$) logarithms. 
By $Q^2$ we have denoted any kinematic invariant that can be constructed 
with a pair of external momenta, and which satisfies the condition 
\mbox{$|Q^2|\gg M_W^2$} (in other words, $Q^2$ is equal to either $s$,
$t$, or $u$ in a $2\to 2$ process, away from the collinear region). These 
logarithmic terms
are commonly referred to as Sudakov logarithms~\cite{Sudakov:1954sw}, and 
stem from the calculation of the virtual contribution to EW corrections,
where they emerge from mechanisms of both UV origin (associated with
the running of the SM parameters) and IR origin (soft and/or collinear).
If one would include their real-emission counterparts, due to the weak 
Heavy-Boson-Radiation (HBR)\footnote{While this inclusion is mandatory in 
massless theories such as QCD and QED, lest the cross section diverge, 
in the weak sector one {\em can} treat HBR as a separate process, owing to
the fact that the heavy bosons are (in principle) taggable. Whether one 
{\em should} do it is another matter, which is besides the point here.
More detailed discussions can be found in refs.~\cite{Frixione:2014qaa,
Czakon:2017wor} and in ref.~\cite{Ma:2024ayr}, for the LHC and the 
muon-collider cases, respectively.}, there would be cancellations among 
these logarithms, but such cancellations would in any case be incomplete, 
and thus still lead to negative corrections which grow with 
energy~\cite{Ciafaloni:1998xg,Ciafaloni:1999ub,Ciafaloni:2000df, Bell:2010gi,
Manohar:2014vxa}\footnote{
The consequences of the incomplete cancellation have been exploited in ref.~\cite{Chen:2022msz} in order to
enhance the sensitivity to new short-distance physics via EW
corrections; this is an intriguing strategy which has also been considered in
ref.~\cite{ElFaham:2024egs}.}.

Therefore, irrespective of whether one includes HBR contributions in inclusive predictions, these electroweak Sudakov logarithms (EWSL henceforth)
become sizeable in the multi-TeV regime, and show up in particular in the 
large-scale regions of kinematic distributions such as invariant masses and transverse 
momenta, which are central to many new-physics searches. In fact, EWSL
can be so large that they can lead to negative (thus unphysical) cross 
sections at any fixed order in perturbation theory, as is also discussed in
ref.~\cite{Ma:2024ayr}. This is precisely the effect observed 
in the large-$\pt(W^+)$ tail of the 10~TeV result of fig.~\ref{fig:ptw}.
Similar issues, although not as dramatic, are also seen in the
$m(W^+W^-)$ distribution, see fig.~\ref{fig:mww}.

Our framework, which combines QED-resummed collinear photon radiation 
with fixed-order NLO EW corrections and VBF-like contributions of NNLO,
automatically includes the EWSL that are part of the NLO EW corrections,
$\aem^{b+n}\log^k(Q^2/M_W^2)$ with $n=1$, and $k=2$ and $k=1$; $b$ is
the power of $\aem$ associated with the Born (i.e.~$b=2$ for $t\bt$ and
$W^+W^-$ production). In addition to that, these double and single logarithms 
of relative order $\alpha$ can also be calculated, within the \aNLOs\ 
framework, by means of an automation and extension~\cite{Pagani:2021vyk} 
of the Denner--Pozzorini algorithm~\cite{Denner:2000jv,Denner:2001gw}. 
Such an algorithm is based on the factorisation of universal high-energy 
logarithmic corrections at the amplitude level in a process-independent 
fashion.  

The advantage of having the possibility of computing in a standalone
manner the EWSL contribution to NLO EW corrections is that it is 
not only possible to confirm that the origin of any large effects 
observed (such as that in fig.~\ref{fig:ptw}, which we have already 
mentioned) is indeed due to EWSL, but also to estimate the impact 
of EWSL from higher orders in a fixed-order expansion. The study of
ref.~\cite{Ma:2024ayr} does exactly that, for several processes and 
observables;
however, in that paper, the focus was exclusively on the direct 
$\mu^+\mu^-$-annihilation process; furthermore, a cut 
\mbox{$m(W^+W^-)>0.8\,\sqrt{s}$} was systematically imposed. 
By relaxing both of these conditions we carry out here an analogous
study, in order to prove that the negative cross section observed at 
large $\pt(W^+)$ originates from EWSL. By doing so we also show that, 
{\em for this process}, effects due to the EWSL present 
in the NNLO cross section must be taken into account in order to achieve 
positive, and phenomenologically sensible, results. Conversely, if one 
aimed at precise (say, of relative $10^{-2}$ or smaller) predictions,
the resummation of the whole EWSL tower would be necessary.

In order to unambiguously identify the EWSL contributions to the NLO EW 
results for phenomenological observables, we use the so-called 
${\rm SDK}_{\rm weak}$ scheme, which has been introduced in 
ref.~\cite{Pagani:2021vyk}, and is further discussed in detail in 
ref.~\cite{Pagani:2023wgc}. This scheme accounts for the soft and/or 
collinear logarithms of only weak origin, while UV logarithms of both weak and QED origin are retained. The logic behind it 
is that, in physical observables, soft and collinear logarithms of QED origin completely cancel 
when the real radiation of photons is clustered (a.k.a.~recombined)  
with any charged particle (massive or massless, thus including the top quark 
and the $W$ boson) -- this corresponds to defining dressed particles. This 
is the reason why in this section we also show predictions with photon 
recombination for charged heavy final states. We consider both the 
case of $t \bar t$ and $W^+W^-$ production at 10~TeV, where the acceptance 
cuts are imposed, and we address the cases of the $\pt(t)$ and $m(t \bar t)$
distributions, and of the $\pt(W^+)$ and $m(W^+W^-)$ ones, 
respectively\footnote{Without any cuts only the $\pt$ distributions 
can be studied in this context, since for large invariant masses transverse
momenta could be small and therefore $|t|\sim M_W^2$, thereby invalidating 
the Sudakov approximation.}.

\begin{figure*}[!t]
	\centering 
    \includegraphics[width=0.49\textwidth, angle=0]{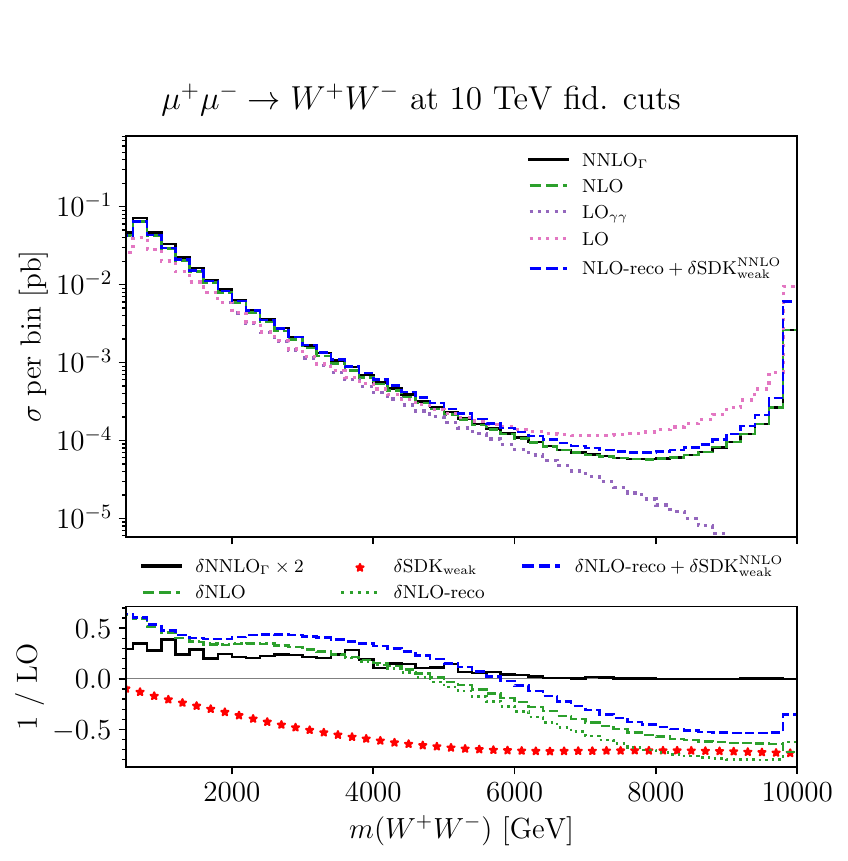}	
    \includegraphics[width=0.49\textwidth, angle=0]{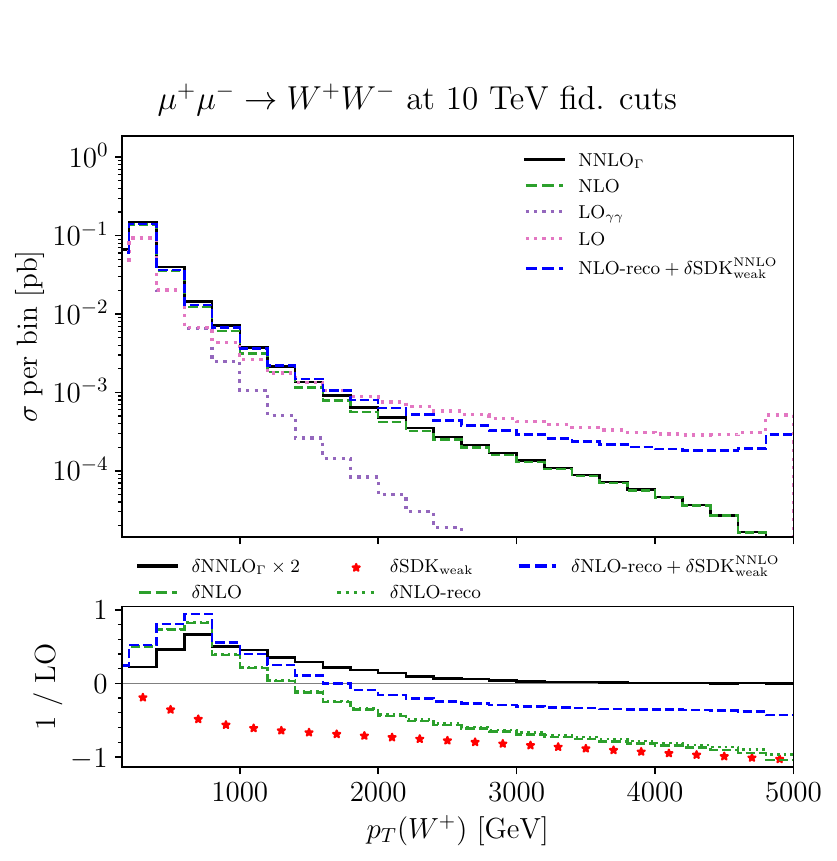}	\\
    \includegraphics[width=0.49\textwidth, angle=0]{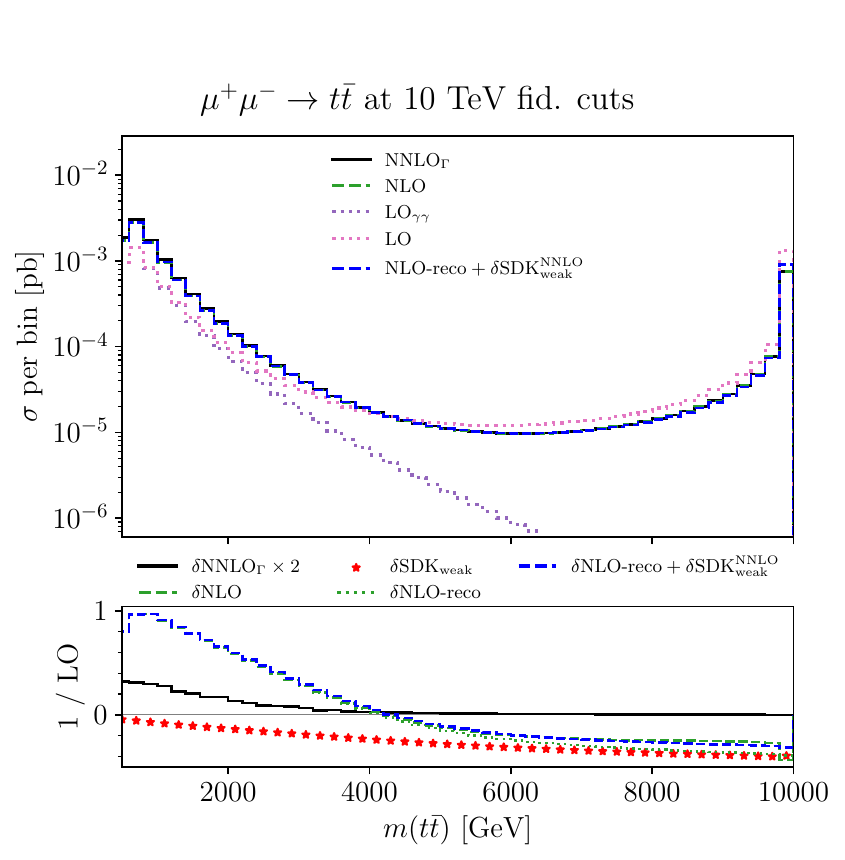}	
    \includegraphics[width=0.49\textwidth, angle=0]{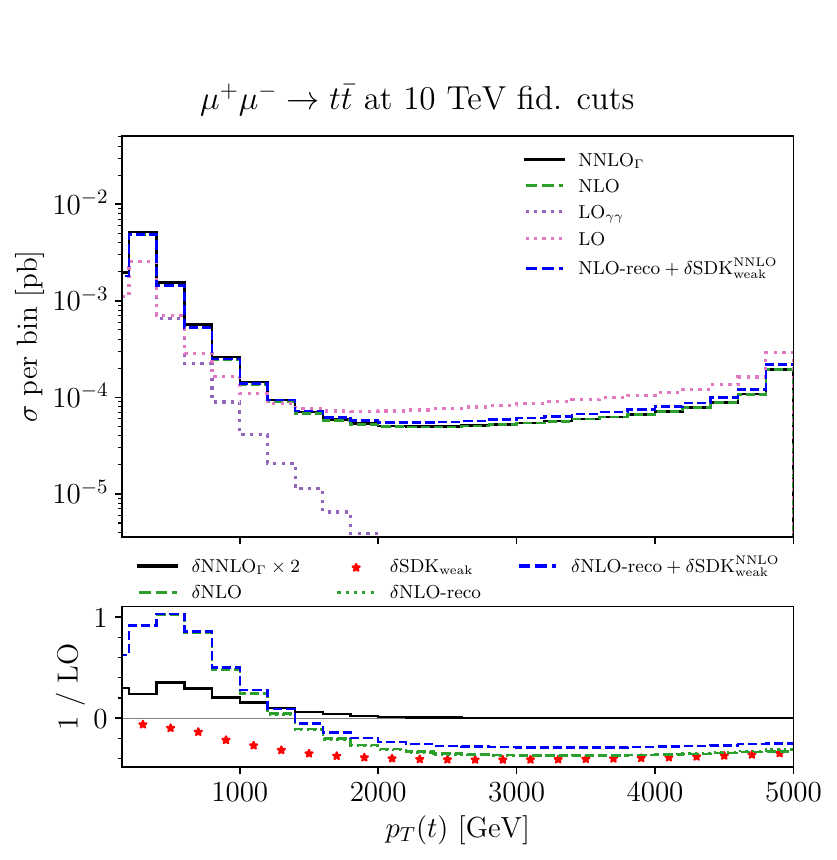}	
  	\caption{Study of the effects coming from the presence of large Sudakov logarithms in representative distributions for $t\bar t$ and $W^+W^-$ final states. See text for details.} 
	\label{fig:sdk}
\end{figure*}

The effects of the EWSL are illustrated in fig.~\ref{fig:sdk}, where we also
show the potential impact of their resummation. In the main panel, in
addition to the results already shown in the bottom-right panels of
figs.~\ref{fig:mtt}, \ref{fig:ptt}, \ref{fig:mww}, and~\ref{fig:ptw},
we present the NLO EW predictions with photon recombination, obtained by 
using a clustering radius $R=1$\footnote{Any sensible value of $R$ may be
employed. Here we choose $R=1$ in order to clearly isolate the effects of
EWSL, since in this way no $\log R$ term affects the cross section.},
supplemented by the approximate EWSL two-loop results. This is called
NLO-reco+$\delta\rm SDK_{weak}^{NNLO}$ (blue-dashed histograms), 
with:
\begin{equation} 
\delta{\rm SDK_{weak}^{NNLO}} \equiv {\rm LO} \,
\frac{1}{2!} \left(\frac{\delta{\rm SDK_{weak}}}{\rm LO}\right)^2
=\frac{\delta{\rm SDK_{weak}^2}}{2\, \rm LO}\,.  
\end{equation} 
In other words, $\delta{\rm SDK_{weak}^{NNLO}}$ originates from the 
Taylor expansion of the exponentiation of the corrections induced 
by EWSL at the NLO, that is:
\begin{equation} 
{\rm LO}\times\exp\left(\frac{\delta{\rm SDK_{weak}}}{\rm LO}\right)=
{\rm LO}+\delta{\rm SDK_{weak}}+\delta{\rm SDK_{weak}^{NNLO}}+\ldots\,.
\end{equation} 
In particular, $\delta{\rm SDK_{weak}}$ is the EWSL component of
$\delta$NLO computed in the SDK$_{\rm weak}$ scheme, and 
$\delta{\rm SDK_{weak}^{NNLO}}$ approximates the effects of EWSL at the
NNLO. The latter takes into account in an exact manner the dominant 
EWSL effects at high energy in the NNLO EW corrections, namely those of order
\mbox{$\alpha^{b+2}\log^4(Q^2/M_W^2)$}, and estimates by means of a naive
exponentiation those of order \mbox{$\alpha^{b+2}\log^n(Q^2/M_W^2)$}, with
$n=3,2$\footnote{This prediction coincides with what is called $\rm EXP_{EW}$
in ref.~\cite{Ma:2024ayr}, truncated at relative order $\alpha^2$ 
w.r.t.~the LO. For a proper treatment of resummation effects, one should 
consider studies as, e.g., that of ref.~\cite{Denner:2024yut}.}.
In the inset of fig.~\ref{fig:sdk} some of the same results as in the main 
panel are shown as relative with respect to the LO\footnote{Note that this 
is different w.r.t. what is done in sect.~\ref{sec:results_gen}, where the 
ratios over $\NNLOG$ predictions are employed. This is so in view of the 
fact that potentially large cancellations occurring beyond LO would create 
artefacts in the ratios, and thus obscure the conclusions here.}. In addition
to those, we also present the $\delta\rm NLO$-reco (i.e.~the NLO corrections
with photon recombination, green dots) and the $\delta{\rm SDK_{weak}}$ (red
stars) predictions.

We start by discussing the case of $W^+W^-$ production, where the EWSL
effects are larger, and later briefly move on to the case of $t\bt$
production.  For large values of $m(W^+W^-)$, and especially of $\pt(W^+)$,
NLO EW corrections are very large in absolute value and negative, as was
already noted. Here we also see that, by recombining photons, the NLO EW
corrections very slightly decrease in the tail of the $\pt(W^+)$
distribution. This phase-space region is mostly correlated with the rightmost
bin of $m(W^+W^-)$, which also displays the same behaviour. The key message to
be taken out of that figure is that the EWSL from $\delta{\rm SDK}$ capture a
large part of the \mbox{$\delta\rm NLO$-$\rm reco$} result, and that the 
pathological behaviour observed at fixed order is cured by adding the 
contribution of $\delta \rm SDK_{weak}^{NNLO}$.

Below $p_T(W^+)\simeq 2~\rm TeV$, we notice how the difference between 
$\delta\rm NLO$-$\rm reco$ and $\delta{\rm SDK_{weak}}$ is not a constant, 
a fact that signals the presence of other kinematic-dependent and 
non-negligible effects unrelated to the EWSL. These effects are indeed 
those already discussed in sect.~\ref{sec:results_gen}, stemming from
the contributions of the $\mu^+\gamma$- and $\gamma\mu^-$-initiated processes, 
which have a significant impact on $\NNLOG$. The same observation applies 
to the $m(W^+W^-)$ distribution for \mbox{$m(W^+W^-)\lesssim 8~\rm TeV$}.

In summary, our analysis shows that in $W^+W^-$ production
we understand the origin of the pathological behaviours of the
$m(W^+W^-)$ and $\pt(W^+)$ distributions at high energy. 
The implication is that at least the NNLO EWSL contributions 
(i.e.~not only those present in $\NNLOG$), in addition to the full NLO EW 
corrections, must be taken into account in order to obtain sensible 
predictions (i.e.~positive-defined cross sections) in the presence of
corrections as large as $-50\%$ w.r.t.~the LO. Having said that, further
effects of $\ord(10\%)$ can still appear due to EWSL of higher orders 
(e.g., $\alpha^{b+3}\log^6(Q^2/M_W^2)$)\footnote{One can
easily see that when the impact of EWSL at the NLO is of order $-100\%$, 
at the NNLO one is liable to see order $(-100\%)^2/2!\simeq +50\%$ effects, 
which may become of order $(-100\%)^3/3!\simeq -15\%$ at the N$^3$LO,
of order $(-100\%)^2/4!\simeq +2\%$ at the N$^4$LO, so forth.\label{ft:bla}},
and therefore the resummation of the full EWSL tower would be necessary 
for precision physics, which in turn would require calculations such as 
the one of ref.~\cite{Denner:2024yut}.  However, it is important
to bear in mind that this kind of conclusions are both observable- and 
process-dependent. Indeed, by simply looking at the analogous plots for 
$t\bt$ production (in turn compatible with the results presented in 
ref.~\cite{Ma:2024ayr,ElFaham:2024egs}), one ends up with a different
overarching message. Namely, that the EWSL NNLO contributions are not 
necessary for obtaining positive-definite, and generally sensible, 
predictions. Moreover, they are sufficient for reducing the uncertainty 
due to higher-order EWSL at the percent level\footnote{By 
employing the same argument as in footnote~\ref{ft:bla}, one can easily 
see that since EWSL effects at the NLO are at most $-40\%$ in this case, 
those are the NNLO are of order $(-40\%)^2/2!\simeq +8\%$, and therefore 
important only in order to achieve percent-level accuracy. However, at the
N$^3$LO they are expected to be of order $(-4\%)^3/3!\simeq -1\%$.}. Thus, in $t\bt$ production
the resummation of EWSL appears not to be necessary even if one focuses
on precision physics at the percent level.

In conclusion, the impact of EWSL must be carefully assessed 
process-by-process and even observable-by-observable, and it is therefore
important to have tools that allow one to do that in a flexible way.

\subsection{Factorisation scheme dependence}
\label{sec:results_scheme}

A potentially important source of theoretical uncertainty in predictions that
combine QED resummation with fixed-order electroweak corrections is the choice
of the factorisation scheme used to define lepton PDFs. In this section we
explicitly show how such a choice (exemplified by the differences between
the results obtained with the $\MSb$~\cite{Bardeen:1978yd} and the 
$\Delta$~\cite{Frixione:2012wtz} schemes) has a huge impact on LO 
predictions (e.g., inducing $\ord(50\%)$ systematics at the threshold of 
the $m(W^+W^-)$ distribution in 10-TeV muon collisions). Fortunately, 
we also show how the factorisation-scheme dependence is reduced to the
percent level when both NLO EW corrections and the $\NNLOG$ contribution
are taken into account -- such a dramatic reduction is particularly
effective in the threshold region, and more in general where one is
dominated by $\gamma\gamma$-induced processes. In the following, we 
briefly discuss some of the technical details underpinning the scheme 
dependence, and show the improvements that can be obtained by including 
higher orders.

\medskip
While the 
$\overline{\rm MS}$ scheme is generally assumed to be the default choice
in massless theories such as QCD and QED, and indeed it is nowadays
universally adopted in hadron-collision simulations, the possibility
of calculating analytically lepton PDFs in QED allows one to show that
double soft logarithms appear beyond the LO in both the PDF initial
conditions and the short-distance cross sections, only to eventually
cancel at the level of physical observables. At the NLO, the initial
conditions can be written as follows~\cite{Frixione:2019lga}:
\begin{eqnarray}
\Gamma^{[1]}_{\mu/\upmu}(z, \mu_0) &=&  \left[\frac{1+z^2}{1-z} 
\left(\log \frac{\mu_0^2}{m^2} - 2 \log (1-z) - 1\right) \right]_+ + 
K_{\mu\upmu}(z),
\label{inilep}
\\
\Gamma^{[1]}_{\gamma/\upmu}(z, \mu_0) &=& \frac{1 + (1 - z)^2}{z} 
\left( \log \frac{\mu_0^2}{m^2} - 2 \log z - 1 \right) + K_{\gamma\upmu}(z),
\label{inigam}
\end{eqnarray}
where the functions $K_{a\upmu}(z)$ encode all information about the
factorisation scheme. Conventionally, in $\overline{\rm MS}$ one
has $K_{\mu\upmu}^{\overline{\rm MS}}(z)=K^{\overline{\rm MS}}_{\gamma\upmu}(z)=0$,
whereby the double logarithmic term in eq.~(\ref{inilep}) is manifest
(stemming from the second term within round brackets).
The scheme independence of observables becomes intuitively clear when
bearing in mind that contributions which feature the $K$ functions must 
be included (importantly, with a minus sign) in the short-distance cross 
sections. In the FKS formalism~\cite{Frixione:1995ms,Frixione:1997np,
Frederix:2009yq}, this is done in the so-called $(n+1)$-body degenerate 
contributions. A residual dependence on the factorisation scheme remains 
after the convolution with PDFs, but such a dependence is beyond the 
perturbative order at which the cross section is computed. In this
section, we shall present a few explicit examples of 
factorisation-scheme (in)dependence -- the interested reader can find
further results in ref.~\cite{Frixione:2025wsv}.

The $\Delta$ scheme~\cite{Frixione:2012wtz} has been devised to eliminate
the double logarithms from the beginning, by defining:
\begin{eqnarray}
K_{\mu\upmu}^{\Delta}(z)&=&
\left[\frac{1+z^2}{1-z} \left( 2 \log (1-z) + 1\right)\right]_+,
\\
K_{\mu\upmu}^{\Delta}(z)&=&
\frac{1 + (1 - z)^2}{z} \Big( 2 \log z + 1 \Big)\,, 
\end{eqnarray}
so that in the lepton-PDF initial condition the only surviving term
is a single soft logarithm, with coefficient proportional to the logarithm
of the scale ($\mu_0$) where the initial conditions are imposed, and which
thus vanishes with the standard choice $\mu_0=m$\footnote{It has to be 
stressed that, while the $\Delta$ scheme is unambiguously defined in the 
large-$z$ region, more freedom exists at small $z$, so that it can effectively 
 be considered as a \emph{class} of different 
schemes~\cite{Bonvini:2025xxx}.}.

A very convenient by-product of the usage of the $\Delta$ scheme is that
observables computed in such a scheme behave much better from the numerical
point of view. For example, the muon density in $\overline {\rm MS}$ can
become negative as $z\to 1$\footnote{See e.g.~eq.~(5.63) in
ref.~\cite{Bertone:2019hks} for the asymptotic behaviour in the
$\overline{\rm MS}$ scheme.} (this happens for $1-z=\mathcal
O(10^{-20})-\mathcal O(10^{-30})$), while it remains positive in the $\Delta$
scheme. This results in significantly faster integration times. To give
just one example, the accumulated running time for the integration of $W^+W^-$
production at the NLO with \aNLOs, at 3~TeV and with a
target relative accuracy of $10^{-4}$ on the total integral is 190 days in 
$\overline {\rm MS}$, and 35 days in $\Delta$. Owing to the smaller 
contribution of the muon-initiated channels, the gain is reduced down 
to ``just'' a factor 2 at 10 TeV.

\begin{figure*}
	\centering 
    \includegraphics[width=0.49\textwidth, angle=0]{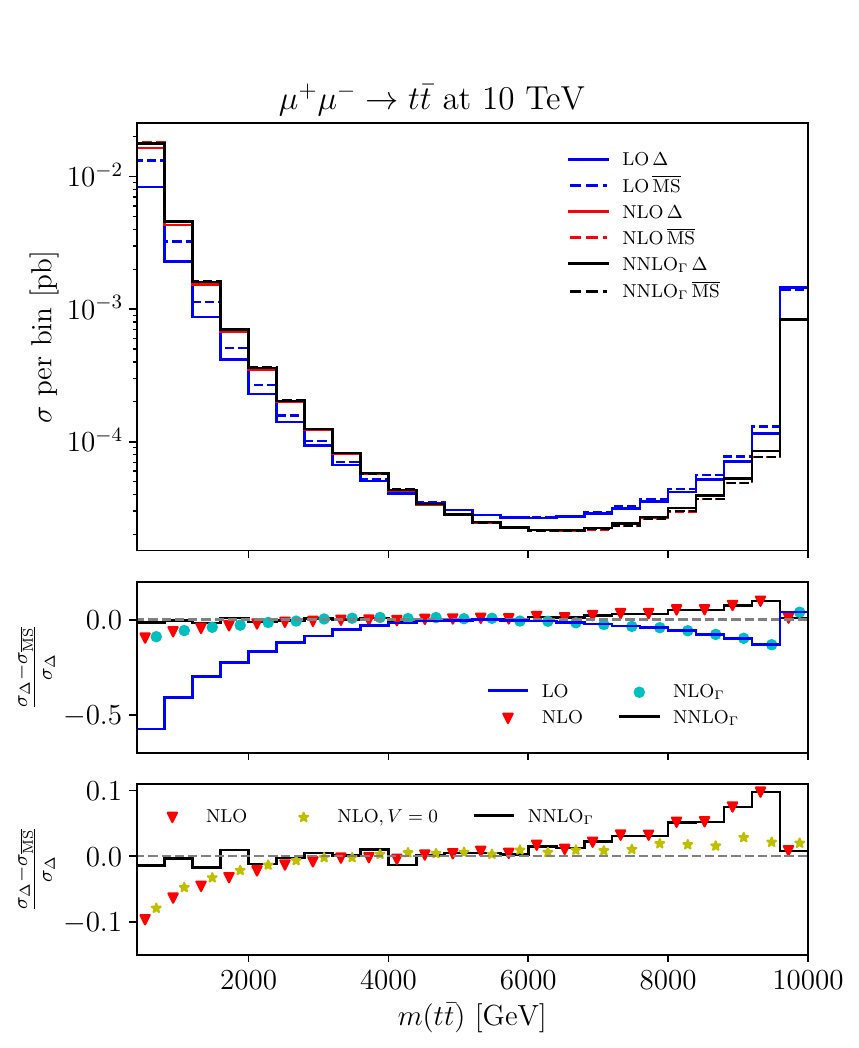}
    \includegraphics[width=0.49\textwidth, angle=0]{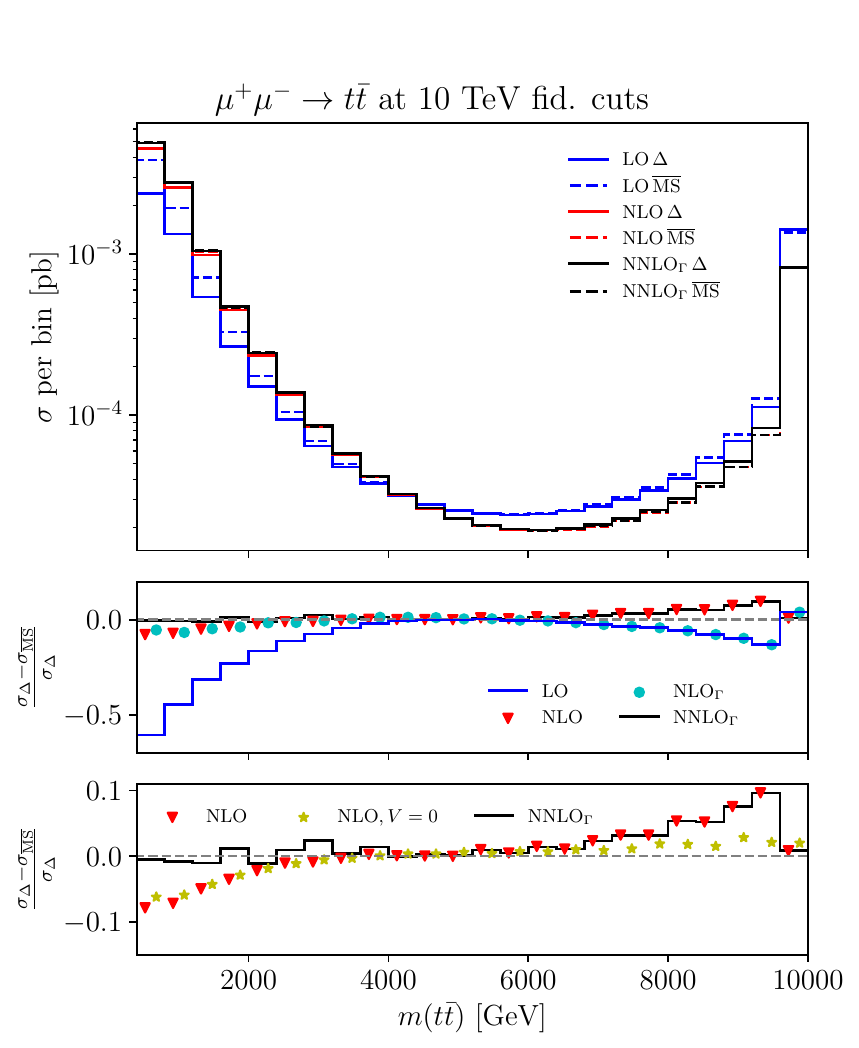}\\	
    \includegraphics[width=0.49\textwidth, angle=0]{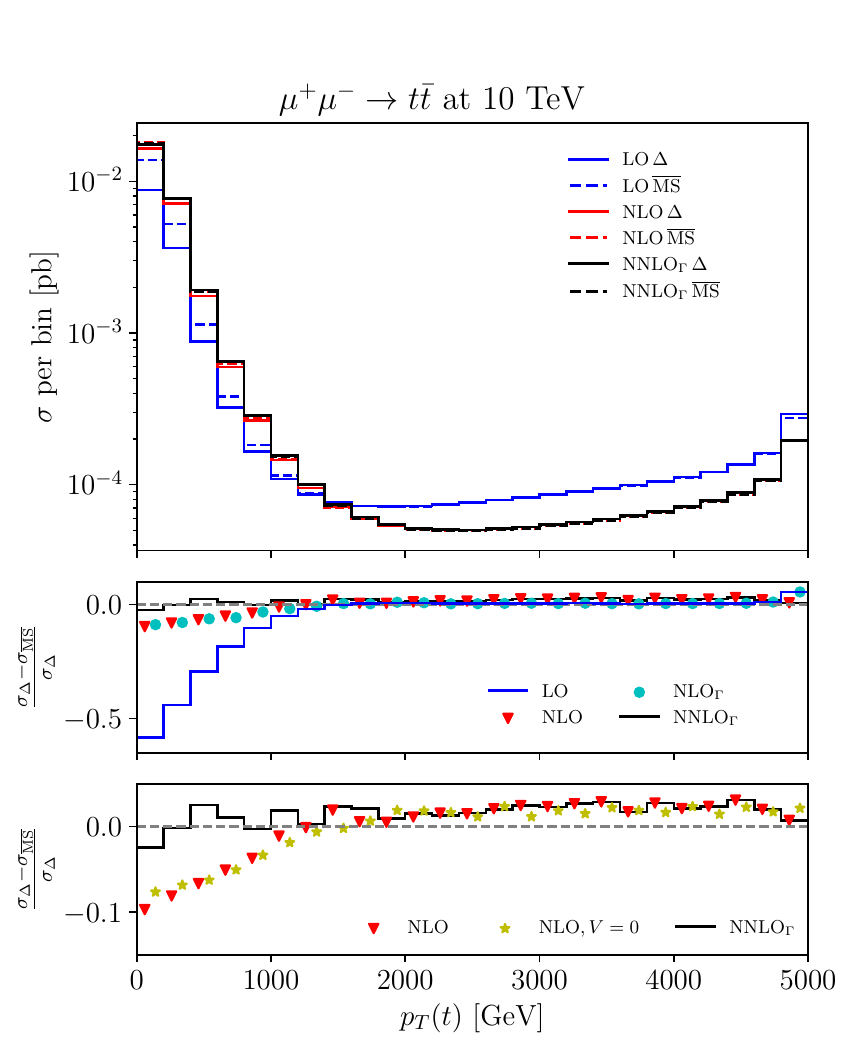}
    \includegraphics[width=0.49\textwidth, angle=0]{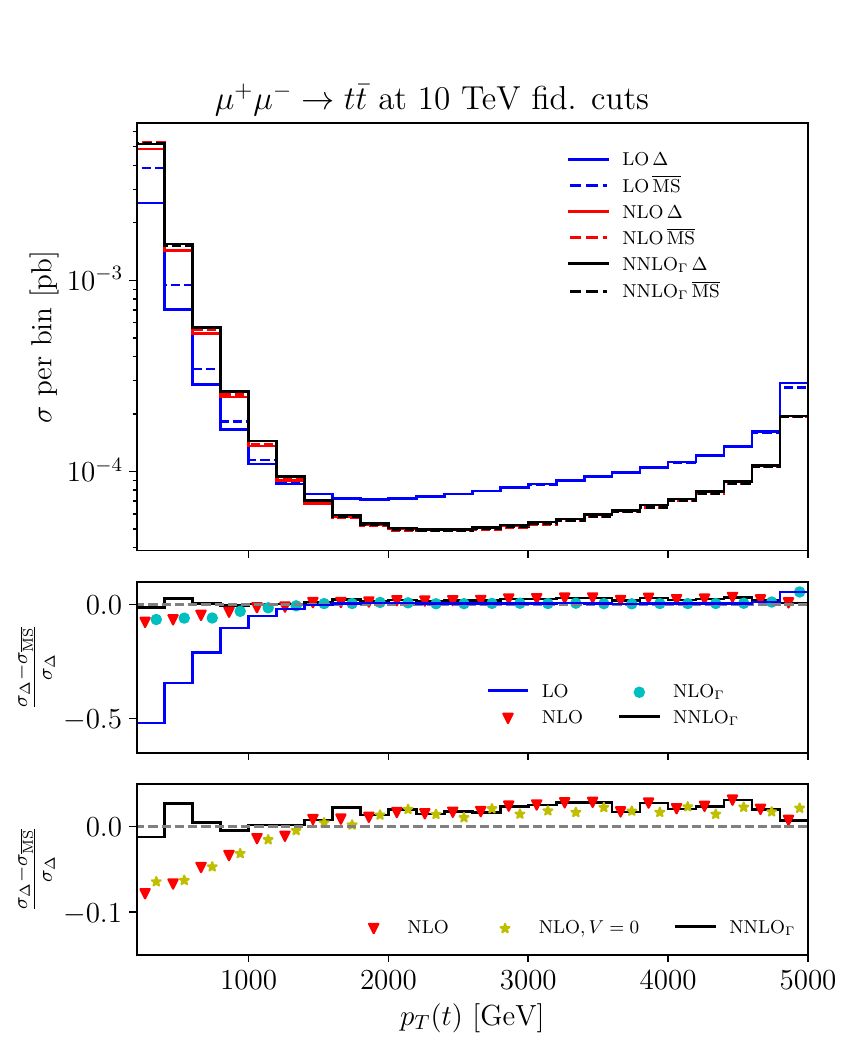}	
	\caption{Factorisation-scheme dependence of the invariant-mass and transverse-momentum distributions in  $t \bar t $ production. See the text for details.  } 
	\label{fig:scheme_tt}%
\end{figure*}

\begin{figure*}
	\centering 
    \includegraphics[width=0.49\textwidth, angle=0]{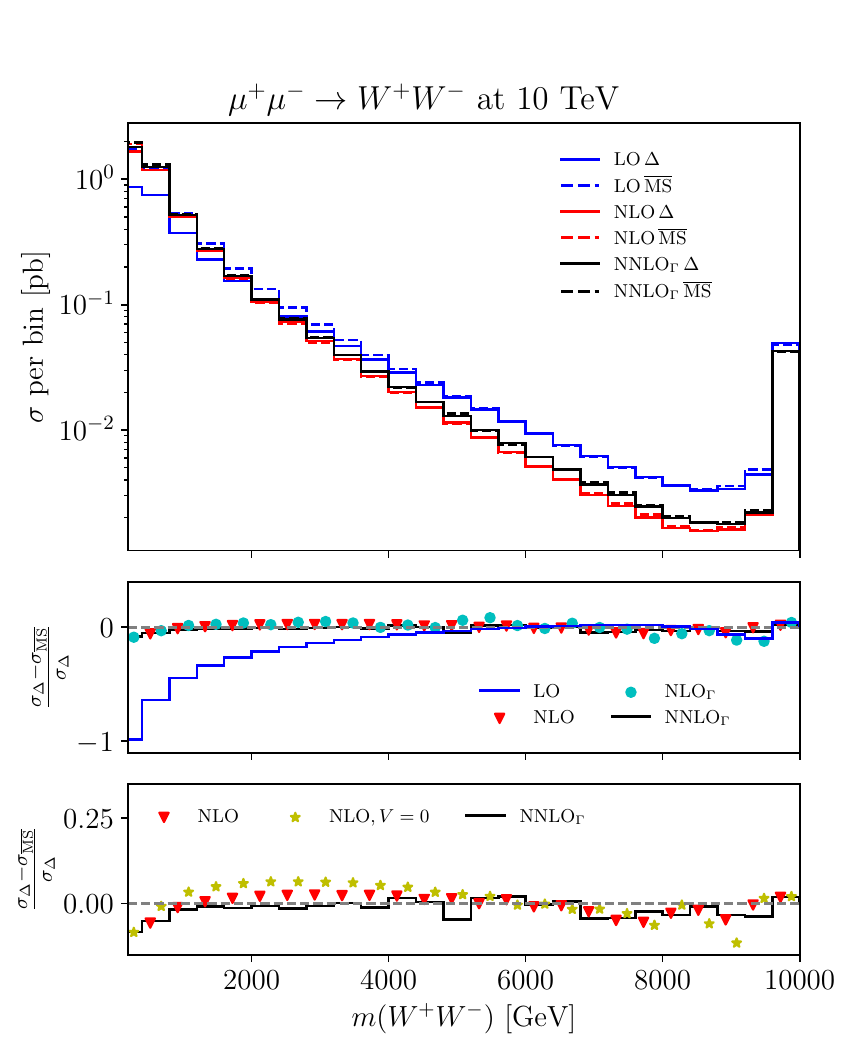}
    \includegraphics[width=0.49\textwidth, angle=0]{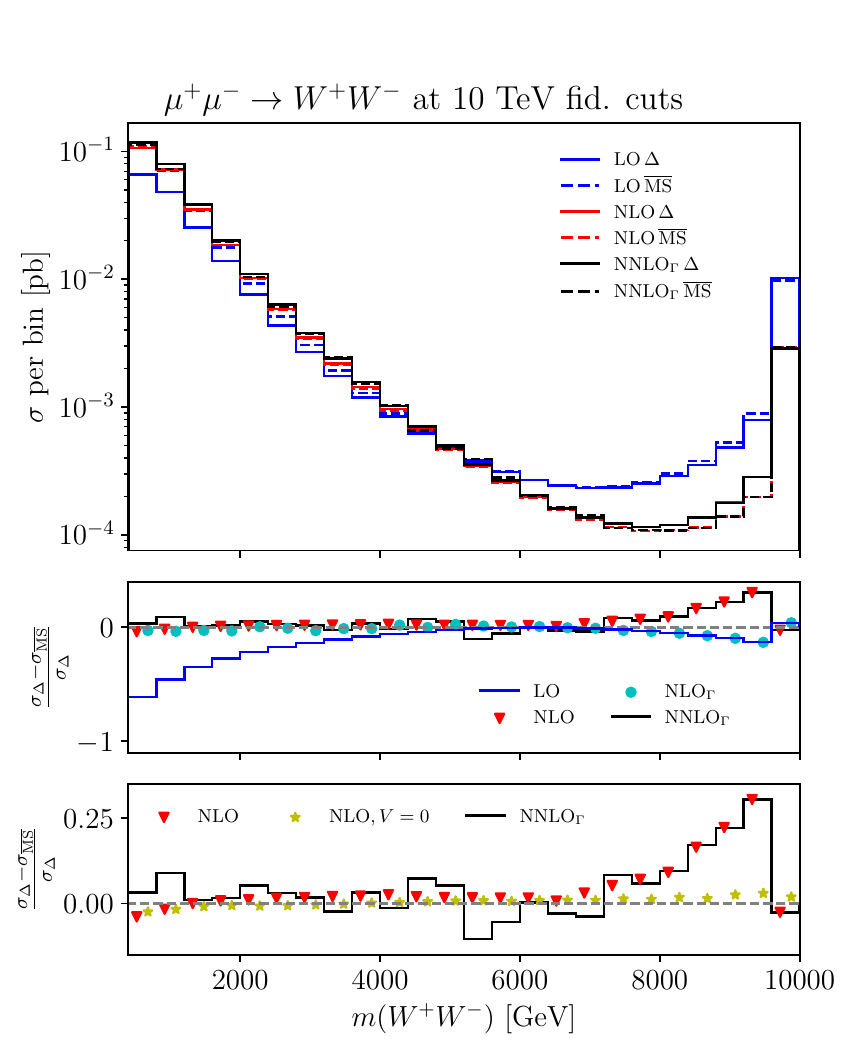}\\	
    \includegraphics[width=0.49\textwidth, angle=0]{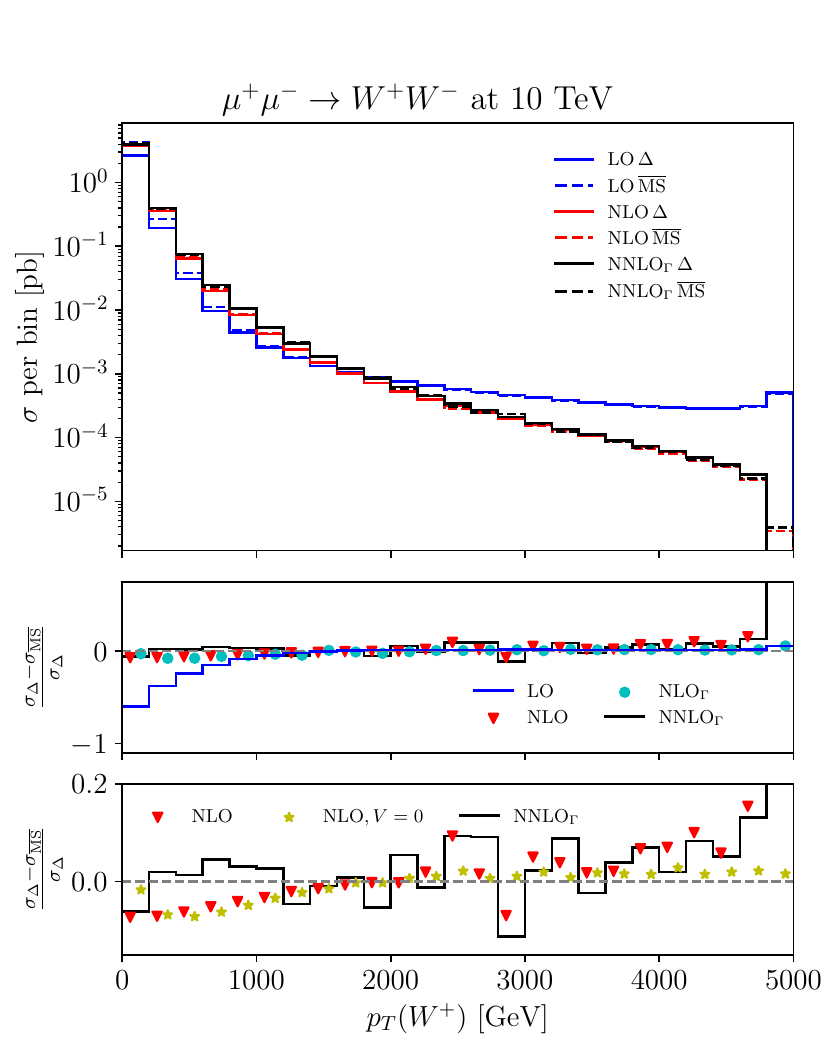}
    \includegraphics[width=0.49\textwidth, angle=0]{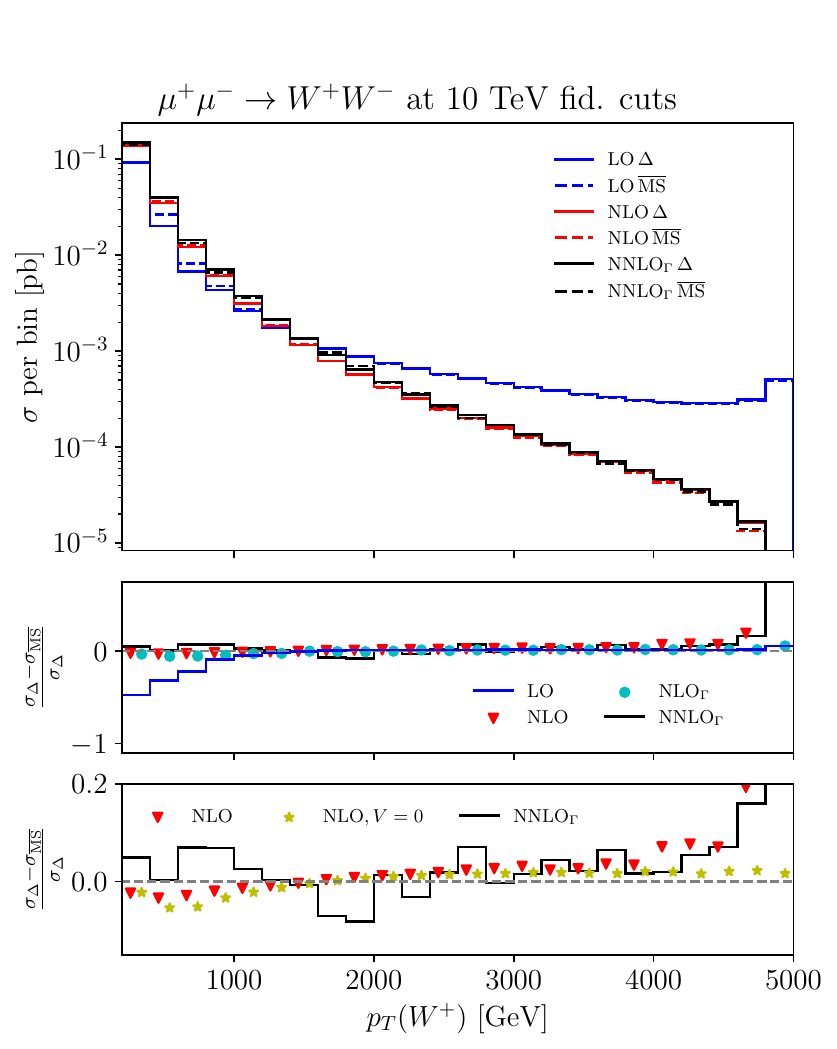}	
	\caption{Same as in fig.~\ref{fig:scheme_tt} for $W^+W^-$ production. } 
	\label{fig:scheme_ww}%
\end{figure*}

Turning now to the presentation of the results, we comment on our predictions 
at a c.m.~energy of 10~TeV, since there the scheme-dependence effects are 
larger. Again, we shall consider $t\bar t$ and $W^+W^-$ production, in 
fig.~\ref{fig:scheme_tt} and fig.~\ref{fig:scheme_ww}, respectively, and 
look at two representative observables, the invariant mass of the pair and 
the transverse momentum of the positively-charged heavy particle. We shall 
consider both the case with and that without the acceptance cuts of 
eq.~(\ref{eq:cuts}). The two figures have the same layout, and show in the top
(bottom) row the invariant-mass (transverse momentum) distribution, without
(left panels) and with (right panels) cuts. Each plot has a main
frame and two insets. The main frame shows absolute rates in the $\Delta$
(solid) and $\overline{\rm MS}$ scheme (dashed), at the LO (blue), NLO (red),
and NNLO$_\Gamma$ (black) accuracy. The two insets display the same quantity, 
namely the relative (w.r.t.~the $\Delta$-scheme result) difference between 
predictions in the two schemes, computed at the different accuracies 
already relevant to the main frame; the bottom inset has a narrower 
$y$ range w.r.t.~to the top one. The colours employed in the insets
are the same as those of the corresponding predictions in the main frame,
but the patterns may be different in order to improve visibility. In
addition to that, in the bottom inset we also show the NLO$_\Gamma$ 
(cyan diamonds) and NLO-no$V$ (yellow stars) relative differences.

We start with $t\bar t$ production, shown in fig.~\ref{fig:scheme_tt}. 
In the invariant-mass distribution, the scheme dependence reduces 
dramatically as the accuracy increases. In particular, at the LO, the 
relative scheme dependence is as large as 50\% at small values of 
$m(t\bar t)$, while  at large values of $m(t\bar t)$ such a
dependence monotonically grows in absolute value up to 10\%, only to
be essentially equal to zero in the rightmost bin of the distribution. 
This behaviour, although with asymptotic values slightly smaller than 10\%, 
appears to be a feature common to all considered accuracies. The negligible 
scheme dependence of the rightmost bin is due to the change of sign of the 
large-$z$ muon PDF in $\overline{\rm MS}$, which leads to an accidental 
cancellation, also driven by the bin size. With the exception of the rightmost
bin, the invariant-mass distribution behaves as one would expect: NLO
corrections reduce the scheme dependence down to 10\% at small values of
$m(t\bar t)$, where the contribution of the virtual matrix elements is
small. On the other hand, since the scheme dependence due to the virtual
matrix elements at the NLO is LO-like, but (in this particular case) opposite
in sign, in the regions where their contribution is sizeable such a reduction
does not take place; specifically, this happens at the high-end of the
spectrum.  This is demonstrated by the results obtained with NLO corrections
that do {\em not} include the virtual matrix elements (see the yellow stars in
the bottom inset), which show the expected reduction of scheme dependence. The
NLO$_\Gamma$ predictions are very close to the NLO ones at small $m(t\bar t)$,
where photon-initiated channels are dominant, and to the LO ones at large
$m(t\bar t)$, since $\delta$NLO$_\Gamma=0$ in that region. Finally, the
NNLO$_\Gamma$ prediction reduces the scheme dependence down to 1\% or below at
small $m(t\bar t)$ values, while it behaves similarly to the NLO one at 
large invariant masses, where photon-induced contributions are unimportant. 
We point out that in the latter region, the very large virtual matrix 
elements are primarily driven by the EWSL, which we have discussed in 
sect.~\ref{sec:results_sdk}. Thus, they are chiefly of weak origin,
whilst the factorisation-scheme dependence pertains only the purely
QED sector of EW corrections. This implies that in order to reduce 
such a dependence at large invariant masses the full NNLO EW results,
or at least its mixed weak-QED part, would be necessary.

The $\pt(t)$ distribution, as was already said, correlates strongly with the
low-$m(t\bar t)$ region at small $p_T(t)$, i.e.~in the region dominated by 
$\gamma\gamma$ fusion, and with the rightmost bin of the $m(t\bar t)$ 
distribution at moderate/large $\pt(t)$. This is why, while at the 
low-end of the $\pt$ spectrum one sees very similar features as the 
corresponding region in the pair invariant mass, at the spectrum high-end 
all scheme dependences are very small (as for the rightmost bin
of $m(t\bar t)$). 

Finally, acceptance cuts have a negligible effect on the scheme-dependence
patterns.

\medskip

Turning to $W^+W^-$ production, shown in fig.~\ref{fig:scheme_ww}, we observe
some striking differences with respect to the $t\bar t$ case just
considered. These are due to the richer phenomenology of this process (in
terms of production modes). The only common feature with $t\bt$ production is
the large scheme dependence at the LO for small $m(W^+W^-)$. Regardless of the
presence of the acceptance cuts, at small values of the invariant mass all
predictions beyond LO display a similar scheme dependence, at the few-percent
level. This can be ascribed to different effects, from the presence of
sizeable, LO-like contributions in the $\dNNLOG$ corrections (e.g.~$ZZ$
fusion), to the non-trivial interplay between virtual corrections and
e.g.~$\mu^+\gamma$ real-emission channels at the NLO. For the case without
cuts, this remains true up to the high-end of the spectrum, where
forward/backward $W$ emissions play an important role. Acceptance cuts remove
these contributions, leading to a behaviour in this region similar to the one
observed for $t\bar t$, with the large NLO scheme dependence exacerbated by
the much larger virtual matrix elements.

As is the case in $t\bt$ production, the $\pt$ distribution display features
which correlate quite well with those of $m(W^+W^-)$. It has to be noted that
at the upper end of this distribution, the virtual contribution is so large
that the cross section becomes negative in the rightmost bin.

In conclusion, from a practical standpoint the $\Delta$ scheme appears
preferable to the $\MSb$ one for high-energy lepton colliders. It provides one
with a closer correspondence with semi-classical photon emission, and is
numerically much more stable w.r.t.~the latter. Nonetheless, the residual
scheme dependence remains a useful diagnostic tool of the robustness of the
calculation, and can be exploited to estimate a part of the theoretical
uncertainty associated with the truncation of the perturbative series.

Phenomenologically, our results confirm that while the overall scheme
dependence is moderate, it is non-negligible in precision applications, and
must be included in the uncertainty budget of future-collider 
analyses\footnote{The reader must pay attention to the fact that
EW-PDF-based simulations do not quote such an uncertainty not because
it is equal to zero, but because it is ill-defined at the LO accuracy
relevant to such simulations unless one follows the prescription
of ref.~\cite{Bertone:2022ktl}; this is not done currently, and therefore this
uncertainty is literally out-of-control there.}. The
ability of our framework to consistently account for this dependence, and to
switch between schemes transparently, is an important asset in view of its
deployment in numerical simulations.


\section{Summary and conclusions}
\label{sec:con}

In this work, we have addressed the challenge of providing accurate and
systematically-improvable predictions for inclusive electroweak processes at
future high-energy lepton colliders, with a particular focus on multi-TeV muon
colliders. We have done this by employing a new approach which we have proposed 
and implemented in a companion paper~\cite{Frixione:2025wsv}, that combines
QED-resummed collinear radiation with fixed-order EW corrections, thereby
overcoming many of the conceptual and practical limitations of the
conventional EW-PDF-based approach.

Our strategy consists in improving NLO EW results, obtained in the context 
of the collinear factorisation theorem and therefore nowadays automated, by
adding double-real contributions that embed, but are not limited to, the
leading NNLO VBF-like topologies, in a way that preserves gauge invariance,
full phase-space coverage, and includes both soft and collinear-enhanced 
diagrams featuring $\gamma/Z$ exchange in the $t$ channels, which in the 
case of the photon are resummed 
via lepton PDFs up to NLL accuracy. The weak component of these diagrams is 
treated exactly at fixed order, thus ensuring the inclusion of 
mass-suppressed and interference effects that are lost in the
traditional VBF approximation. Importantly our framework, by avoiding any 
double counting with NLO EW corrections, opens up a clear path toward future
refinements.

We have demonstrated the power and flexibility of this approach by considering
two representative processes, namely $t\bar{t}$ and $W^+W^-$ inclusive
production at $\sqrt{s} = 3$ and $10$~TeV muon colliders. For both processes,
we have shown that the improvement we have proposed has a significant
quantitative impact across a wide range of kinematic variables, particularly
in regions dominated by forward scattering or large invariant masses. In
parallel, we have studied the same observables with the EWA, compared the
predictions thus obtained with those of the full matrix elements which
underpin our procedure, and assessed the differences w.r.t.~EW-PDF-based
calculations.  We have found large discrepancies at the level of both shapes
and normalisation, as well as a strong dependence on factorisation
scales in the EWA case. This underlines the unreliability of the EW-PDF-based
approach at realistic collider energies.

In contrast, our method retains full control over theoretical
uncertainties, and enables consistent improvements through the inclusion 
of both higher-order fixed terms and resummed contributions. In particular,
we have shown that, in those phase space regions which receive significant
contributions from VBF-like topologies, our predictions exhibit 
uncertainties akin to those of full NNLO calculations, and have an
NLO-type behaviour elsewhere. We have observed that Sudakov 
logarithmic enhancements can naturally be incorporated 
within our framework, and combined with the set of NNLO terms 
which we include, without any ambiguity. Likewise, QCD corrections
can be accounted for in a straightforward manner; we point out that this
is consistent with the fact that lepton PDFs have a strongly-interacting
parton component, i.e.~feature quark and gluon densities.

Our results provide robust evidence that the proposed approach is a superior
and viable alternative to the current approximations used for high-energy
lepton collider physics. By capturing all leading contributions in both the
soft and collinear regions, and by maintaining exact mass dependence throughout,
our method gives reliable predictions not only for inclusive rates but also
for differential distributions relevant to experimental analyses.
We point out that while this approach will be superseded by exact
EW NNLO results (provided that the latter will also include the resummation
of QED collinear dynamics), these are presently out of reach, and will
in any case remain much more challenging computations, to be carried
out in a case-by-case manner for the foreseeable future. Conversely,
efforts towards automating the method we have proposed are already
ongoing (and will be guaranteed to succeed thanks to its process
independence), and will be included in future releases of the public event 
generator \aNLOs; this will be crucial for easier phenomenological 
applications.

This work gives a template for a comprehensive, precise, and accurate framework
at future lepton colliders. While we have emphasised the case of a multi-TeV
muon collider, the method which we have proposed will be equally relevant
to the very high-precision targets of a typical sub-TeV $e^+e^-$ machine.
In future phenomenological studies, it will be 
important to apply this methodology to more exclusive observables and to 
other relevant final states, including those sensitive to new physics.

\section*{Acknowledgements}
We thank Giovanni Stagnitto for discussions and for having provided us with a still-unreleased NLL- and
NLO-accurate version of the muon PDFs of ref.~\cite{Bonvini:2025xxx}. 
DP~and MZ~acknowledge the financial support by the MUR (Italy), with
funds of the European Union (NextGenerationEU), through the PRIN2022
grant 2022EZ3S3F; likewise FM, through the PRIN2022 grant 2022RXEZCJ,
and by the project ``QIHEP--Exploring the foundations of quantum information
in particle physics'', which is financed through the PNRR and is funded by
the European Union -- NextGenerationEU, in the context of the extended
partnership PE00000023 NQSTI -- CUP J33C24001210007.
SF~thanks the TH division of CERN for the hospitality during the course
of this work.

\bibliographystyle{JHEP}
\bibliography{bibs/tot}

\providecommand{\href}[2]{#2}\begingroup\raggedright\begin{thebibliography}{10}

\bibitem{AWAKE:2022aeo}
{\scshape AWAKE} collaboration, E.~Gschwendtner et~al., \emph{{The AWAKE Run 2
  Programme and Beyond}},
  \href{http://dx.doi.org/10.3390/sym14081680}{\emph{Symmetry} {\bf 14} (2022)
  1680}, [\href{http://arxiv.org/abs/2206.06040}{{\tt 2206.06040}}].

\bibitem{InternationalMuonCollider:2025sys}
{\scshape International Muon Collider} collaboration, C.~Accettura et~al.,
  \emph{{The Muon Collider}},  \href{http://arxiv.org/abs/2504.21417}{{\tt
  2504.21417}}.

\bibitem{Buttazzo:2018qqp}
D.~Buttazzo, D.~Redigolo, F.~Sala and A.~Tesi, \emph{{Fusing Vectors into
  Scalars at High Energy Lepton Colliders}},
  \href{http://dx.doi.org/10.1007/JHEP11(2018)144}{\emph{JHEP} {\bf 11} (2018)
  144}, [\href{http://arxiv.org/abs/1807.04743}{{\tt 1807.04743}}].

\bibitem{Costantini:2020stv}
A.~Costantini, F.~De~Lillo, F.~Maltoni, L.~Mantani, O.~Mattelaer, R.~Ruiz
  et~al., \emph{{Vector boson fusion at multi-TeV muon colliders}},
  \href{http://dx.doi.org/10.1007/JHEP09(2020)080}{\emph{JHEP} {\bf 09} (2020)
  080}, [\href{http://arxiv.org/abs/2005.10289}{{\tt 2005.10289}}].

\bibitem{Buttazzo:2020uzc}
D.~Buttazzo, R.~Franceschini and A.~Wulzer, \emph{{Two Paths Towards Precision
  at a Very High Energy Lepton Collider}},
  \href{http://dx.doi.org/10.1007/JHEP05(2021)219}{\emph{JHEP} {\bf 05} (2021)
  219}, [\href{http://arxiv.org/abs/2012.11555}{{\tt 2012.11555}}].

\bibitem{AlAli:2021let}
H.~Al~Ali et~al., \emph{{The Muon Smasher's Guide}},
  \href{http://arxiv.org/abs/2103.14043}{{\tt 2103.14043}}.

\bibitem{Aime:2022flm}
C.~Aime et~al., \emph{{Muon Collider Physics Summary}},
  \href{http://arxiv.org/abs/2203.07256}{{\tt 2203.07256}}.

\bibitem{Accettura:2023ked}
C.~Accettura et~al., \emph{{Towards a muon collider}},
  \href{http://dx.doi.org/10.1140/epjc/s10052-023-11889-x}{\emph{Eur. Phys. J.
  C} {\bf 83} (2023) 864}, [\href{http://arxiv.org/abs/2303.08533}{{\tt
  2303.08533}}].

\bibitem{Chen:2016wkt}
J.~Chen, T.~Han and B.~Tweedie, \emph{{Electroweak Splitting Functions and High
  Energy Showering}},
  \href{http://dx.doi.org/10.1007/JHEP11(2017)093}{\emph{JHEP} {\bf 11} (2017)
  093}, [\href{http://arxiv.org/abs/1611.00788}{{\tt 1611.00788}}].

\bibitem{Han:2020uid}
T.~Han, Y.~Ma and K.~Xie, \emph{{High energy leptonic collisions and
  electroweak parton distribution functions}},
  \href{http://dx.doi.org/10.1103/PhysRevD.103.L031301}{\emph{Phys. Rev. D}
  {\bf 103} (2021) L031301}, [\href{http://arxiv.org/abs/2007.14300}{{\tt
  2007.14300}}].

\bibitem{Han:2021kes}
T.~Han, Y.~Ma and K.~Xie, \emph{{Quark and gluon contents of a lepton at high
  energies}}, \href{http://dx.doi.org/10.1007/JHEP02(2022)154}{\emph{JHEP} {\bf
  02} (2022) 154}, [\href{http://arxiv.org/abs/2103.09844}{{\tt 2103.09844}}].

\bibitem{Garosi:2023bvq}
F.~Garosi, D.~Marzocca and S.~Trifinopoulos, \emph{{LePDF: Standard Model PDFs
  for High-Energy Lepton Colliders}},
  \href{http://arxiv.org/abs/2303.16964}{{\tt 2303.16964}}.

\bibitem{Capdevilla:2024ydp}
R.~Capdevilla, F.~Garosi, D.~Marzocca and B.~Stechauner, \emph{{Testing the
  neutrino content of the muon at muon colliders}},
  \href{http://dx.doi.org/10.1007/JHEP04(2025)168}{\emph{JHEP} {\bf 04} (2025)
  168}, [\href{http://arxiv.org/abs/2410.21383}{{\tt 2410.21383}}].

\bibitem{Marzocca:2024fqb}
D.~Marzocca and A.~Stanzione, \emph{{On the impact of the mixed $Z/ \gamma$ PDF
  at muon colliders}},  \href{http://arxiv.org/abs/2408.13191}{{\tt
  2408.13191}}.

\bibitem{Han:2025wdy}
C.~Han, T.~Li and Y.~Wang, \emph{Advances in electroweak pdfs for future
  collider experiments},
  \href{http://dx.doi.org/10.1103/PhysRevD.111.055001}{\emph{Phys. Rev. D} {\bf
  111} (2025) 055001}, [\href{http://arxiv.org/abs/2503.04567}{{\tt
  2503.04567}}].

\bibitem{Bauer:2018xag}
C.~W. Bauer, D.~Provasoli and B.~R. Webber, \emph{{Standard Model Fragmentation
  Functions at Very High Energies}},
  \href{http://dx.doi.org/10.1007/JHEP11(2018)030}{\emph{JHEP} {\bf 11} (2018)
  030}, [\href{http://arxiv.org/abs/1806.10157}{{\tt 1806.10157}}].

\bibitem{Frixione:2025wsv}
S.~Frixione, F.~Maltoni, D.~Pagani and M.~Zaro, \emph{{Double neutral-current
  corrections to NLO electroweak leptonic cross sections}},
  \href{http://arxiv.org/abs/2506.10732}{{\tt 2506.10732}}.

\bibitem{Frixione:2019lga}
S.~Frixione, \emph{{Initial conditions for electron and photon structure and
  fragmentation functions}},
  \href{http://dx.doi.org/10.1007/JHEP11(2019)158}{\emph{JHEP} {\bf 11} (2019)
  158}, [\href{http://arxiv.org/abs/1909.03886}{{\tt 1909.03886}}].

\bibitem{Bertone:2022ktl}
V.~Bertone, M.~Cacciari, S.~Frixione, G.~Stagnitto, M.~Zaro and X.~Zhao,
  \emph{{Improving methods and predictions at high-energy e$^{+}$e$^{-}$
  colliders within collinear factorisation}},
  \href{http://dx.doi.org/10.1007/JHEP10(2022)089}{\emph{JHEP} {\bf 10} (2022)
  089}, [\href{http://arxiv.org/abs/2207.03265}{{\tt 2207.03265}}].

\bibitem{Frixione:2023gmf}
S.~Frixione and G.~Stagnitto, \emph{{The muon parton distribution functions}},
  \href{http://dx.doi.org/10.1007/JHEP12(2023)170}{\emph{JHEP} {\bf 12} (2023)
  170}, [\href{http://arxiv.org/abs/2309.07516}{{\tt 2309.07516}}].

\bibitem{Bonvini:2025xxx}
M.~Bonvini, S.~Frixione and G.~Stagnitto, \emph{{Improved small-x resummation
  for DGLAP splitting functions: HELL 4.0}},
  \href{http://arxiv.org/abs/25xx.yyyyy}{{\tt 25xx.yyyyy}}.

\bibitem{Denner:2024yut}
A.~Denner and S.~Rode, \emph{{Automated resummation of electroweak Sudakov
  logarithms in diboson production at future colliders}},
  \href{http://dx.doi.org/10.1140/epjc/s10052-024-12879-3}{\emph{Eur. Phys. J.
  C} {\bf 84} (2024) 542}, [\href{http://arxiv.org/abs/2402.10503}{{\tt
  2402.10503}}].

\bibitem{Altarelli:1996gh}
G.~Altarelli, T.~Sjostrand and F.~Zwirner, eds., \emph{{Physics at LEP2: Vol.
  1}}, CERN Yellow Reports: Conference Proceedings, 2, 1996.
\newblock 10.5170/CERN-1996-001-V-1.

\bibitem{Dawson:1984gx}
S.~Dawson, \emph{{The Effective W Approximation}},
  \href{http://dx.doi.org/10.1016/0550-3213(85)90038-0}{\emph{Nucl. Phys. B}
  {\bf 249} (1985) 42--60}.

\bibitem{Kane:1984bb}
G.~L. Kane, W.~W. Repko and W.~B. Rolnick, \emph{{The Effective W+-, Z0
  Approximation for High-Energy Collisions}},
  \href{http://dx.doi.org/10.1016/0370-2693(84)90105-9}{\emph{Phys. Lett. B}
  {\bf 148} (1984) 367--372}.

\bibitem{Kunszt:1987tk}
Z.~Kunszt and D.~E. Soper, \emph{{On the Validity of the Effective $W$
  Approximation}},
  \href{http://dx.doi.org/10.1016/0550-3213(88)90673-6}{\emph{Nucl. Phys. B}
  {\bf 296} (1988) 253--289}.

\bibitem{Dittmaier:2023nac}
S.~Dittmaier, P.~Maierh{\"o}fer, C.~Schwan and R.~Winterhalder,
  \emph{{Like-sign W-boson scattering at the LHC {\textemdash} approximations
  and full next-to-leading-order predictions}},
  \href{http://dx.doi.org/10.1007/JHEP11(2023)022}{\emph{JHEP} {\bf 11} (2023)
  022}, [\href{http://arxiv.org/abs/2308.16716}{{\tt 2308.16716}}].

\bibitem{Alwall:2014hca}
J.~Alwall, R.~Frederix, S.~Frixione, V.~Hirschi, F.~Maltoni, O.~Mattelaer
  et~al., \emph{{The automated computation of tree-level and next-to-leading
  order differential cross sections, and their matching to parton shower
  simulations}}, \href{http://dx.doi.org/10.1007/JHEP07(2014)079}{\emph{JHEP}
  {\bf 07} (2014) 079}, [\href{http://arxiv.org/abs/1405.0301}{{\tt
  1405.0301}}].

\bibitem{Frederix:2018nkq}
R.~Frederix, S.~Frixione, V.~Hirschi, D.~Pagani, H.~S. Shao and M.~Zaro,
  \emph{{The automation of next-to-leading order electroweak calculations}},
  \href{http://dx.doi.org/10.1007/JHEP11(2021)085}{\emph{JHEP} {\bf 07} (2018)
  185}, [\href{http://arxiv.org/abs/1804.10017}{{\tt 1804.10017}}].

\bibitem{Frixione:2021zdp}
S.~Frixione, O.~Mattelaer, M.~Zaro and X.~Zhao, \emph{{Lepton collisions in
  MadGraph5\_aMC@NLO}},  \href{http://arxiv.org/abs/2108.10261}{{\tt
  2108.10261}}.

\bibitem{Ruiz:2021tdt}
R.~Ruiz, A.~Costantini, F.~Maltoni and O.~Mattelaer, \emph{{The Effective
  Vector Boson Approximation in high-energy muon collisions}},
  \href{http://dx.doi.org/10.1007/JHEP06(2022)114}{\emph{JHEP} {\bf 06} (2022)
  114}, [\href{http://arxiv.org/abs/2111.02442}{{\tt 2111.02442}}].

\bibitem{Bigaran:2025rvb}
I.~Bigaran and R.~Ruiz, \emph{{Weak bosons as partons below 10 TeV partonic
  center-of-momentum}},  \href{http://arxiv.org/abs/2502.07878}{{\tt
  2502.07878}}.

\bibitem{Sirlin:1980nh}
A.~Sirlin, \emph{{Radiative Corrections in the SU(2)-L x U(1) Theory: A Simple
  Renormalization Framework}},
  \href{http://dx.doi.org/10.1103/PhysRevD.22.971}{\emph{Phys. Rev. D} {\bf 22}
  (1980) 971--981}.

\bibitem{Bertone:2019hks}
V.~Bertone, M.~Cacciari, S.~Frixione and G.~Stagnitto, \emph{{The partonic
  structure of the electron at the next-to-leading logarithmic accuracy in
  QED}}, \href{http://dx.doi.org/10.1007/JHEP03(2020)135}{\emph{JHEP} {\bf 03}
  (2020) 135}, [\href{http://arxiv.org/abs/1911.12040}{{\tt 1911.12040}}].

\bibitem{Frixione:2012wtz}
S.~Frixione, \emph{{On factorisation schemes for the electron parton
  distribution functions in QED}},
  \href{http://dx.doi.org/10.1007/JHEP07(2021)180}{\emph{JHEP} {\bf 07} (2021)
  180}, [\href{http://arxiv.org/abs/2105.06688}{{\tt 2105.06688}}].

\bibitem{Bredt:2022dmm}
P.~M. Bredt, W.~Kilian, J.~Reuter and P.~Stienemeier, \emph{{NLO electroweak
  corrections to multi-boson processes at a muon collider}},
  \href{http://dx.doi.org/10.1007/JHEP12(2022)138}{\emph{JHEP} {\bf 12} (2022)
  138}, [\href{http://arxiv.org/abs/2208.09438}{{\tt 2208.09438}}].

\bibitem{Ma:2024ayr}
Y.~Ma, D.~Pagani and M.~Zaro, \emph{{EW corrections and Heavy Boson Radiation
  at a high-energy muon collider}},
  \href{http://arxiv.org/abs/2409.09129}{{\tt 2409.09129}}.

\bibitem{Kilian:2007gr}
W.~Kilian, T.~Ohl and J.~Reuter, \emph{{WHIZARD: Simulating Multi-Particle
  Processes at LHC and ILC}},
  \href{http://dx.doi.org/10.1140/epjc/s10052-011-1742-y}{\emph{Eur. Phys. J.
  C} {\bf 71} (2011) 1742}, [\href{http://arxiv.org/abs/0708.4233}{{\tt
  0708.4233}}].

\bibitem{Sudakov:1954sw}
V.~V. Sudakov, \emph{{Vertex parts at very high-energies in quantum
  electrodynamics}}, {\emph{Sov. Phys. JETP} {\bf 3} (1956) 65--71}.

\bibitem{Frixione:2014qaa}
S.~Frixione, V.~Hirschi, D.~Pagani, H.~S. Shao and M.~Zaro, \emph{{Weak
  corrections to Higgs hadroproduction in association with a top-quark pair}},
  \href{http://dx.doi.org/10.1007/JHEP09(2014)065}{\emph{JHEP} {\bf 09} (2014)
  065}, [\href{http://arxiv.org/abs/1407.0823}{{\tt 1407.0823}}].

\bibitem{Czakon:2017wor}
M.~Czakon, D.~Heymes, A.~Mitov, D.~Pagani, I.~Tsinikos and M.~Zaro,
  \emph{{Top-pair production at the LHC through NNLO QCD and NLO EW}},
  \href{http://dx.doi.org/10.1007/JHEP10(2017)186}{\emph{JHEP} {\bf 10} (2017)
  186}, [\href{http://arxiv.org/abs/1705.04105}{{\tt 1705.04105}}].

\bibitem{Ciafaloni:1998xg}
P.~Ciafaloni and D.~Comelli, \emph{{Sudakov enhancement of electroweak
  corrections}},
  \href{http://dx.doi.org/10.1016/S0370-2693(98)01541-X}{\emph{Phys. Lett. B}
  {\bf 446} (1999) 278--284}, [\href{http://arxiv.org/abs/hep-ph/9809321}{{\tt
  hep-ph/9809321}}].

\bibitem{Ciafaloni:1999ub}
P.~Ciafaloni and D.~Comelli, \emph{{Electroweak Sudakov form-factors and
  nonfactorizable soft QED effects at NLC energies}},
  \href{http://dx.doi.org/10.1016/S0370-2693(00)00121-0}{\emph{Phys. Lett. B}
  {\bf 476} (2000) 49--57}, [\href{http://arxiv.org/abs/hep-ph/9910278}{{\tt
  hep-ph/9910278}}].

\bibitem{Ciafaloni:2000df}
M.~Ciafaloni, P.~Ciafaloni and D.~Comelli, \emph{{Bloch-Nordsieck violating
  electroweak corrections to inclusive TeV scale hard processes}},
  \href{http://dx.doi.org/10.1103/PhysRevLett.84.4810}{\emph{Phys. Rev. Lett.}
  {\bf 84} (2000) 4810--4813}, [\href{http://arxiv.org/abs/hep-ph/0001142}{{\tt
  hep-ph/0001142}}].

\bibitem{Bell:2010gi}
G.~Bell, J.~H. Kuhn and J.~Rittinger, \emph{{Electroweak Sudakov Logarithms and
  Real Gauge-Boson Radiation in the TeV Region}},
  \href{http://dx.doi.org/10.1140/epjc/s10052-010-1489-x}{\emph{Eur. Phys. J.
  C} {\bf 70} (2010) 659--671}, [\href{http://arxiv.org/abs/1004.4117}{{\tt
  1004.4117}}].

\bibitem{Manohar:2014vxa}
A.~Manohar, B.~Shotwell, C.~Bauer and S.~Turczyk, \emph{{Non-cancellation of
  electroweak logarithms in high-energy scattering}},
  \href{http://dx.doi.org/10.1016/j.physletb.2014.11.050}{\emph{Phys. Lett. B}
  {\bf 740} (2015) 179--187}, [\href{http://arxiv.org/abs/1409.1918}{{\tt
  1409.1918}}].

\bibitem{Chen:2022msz}
S.~Chen, A.~Glioti, R.~Rattazzi, L.~Ricci and A.~Wulzer, \emph{{Learning from
  radiation at a very high energy lepton collider}},
  \href{http://dx.doi.org/10.1007/JHEP05(2022)180}{\emph{JHEP} {\bf 05} (2022)
  180}, [\href{http://arxiv.org/abs/2202.10509}{{\tt 2202.10509}}].

\bibitem{ElFaham:2024egs}
H.~El~Faham, K.~Mimasu, D.~Pagani, C.~Severi, E.~Vryonidou and M.~Zaro,
  \emph{{Electroweak corrections in the SMEFT: four-fermion operators at high
  energies}},  \href{http://arxiv.org/abs/2412.16076}{{\tt 2412.16076}}.

\bibitem{Pagani:2021vyk}
D.~Pagani and M.~Zaro, \emph{{One-loop electroweak Sudakov logarithms: a
  revisitation and automation}},
  \href{http://dx.doi.org/10.1007/JHEP02(2022)161}{\emph{JHEP} {\bf 02} (2022)
  161}, [\href{http://arxiv.org/abs/2110.03714}{{\tt 2110.03714}}].

\bibitem{Denner:2000jv}
A.~Denner and S.~Pozzorini, \emph{{One loop leading logarithms in electroweak
  radiative corrections. 1. Results}},
  \href{http://dx.doi.org/10.1007/s100520100551}{\emph{Eur. Phys. J. C} {\bf
  18} (2001) 461--480}, [\href{http://arxiv.org/abs/hep-ph/0010201}{{\tt
  hep-ph/0010201}}].

\bibitem{Denner:2001gw}
A.~Denner and S.~Pozzorini, \emph{{One loop leading logarithms in electroweak
  radiative corrections. 2. Factorization of collinear singularities}},
  \href{http://dx.doi.org/10.1007/s100520100721}{\emph{Eur. Phys. J. C} {\bf
  21} (2001) 63--79}, [\href{http://arxiv.org/abs/hep-ph/0104127}{{\tt
  hep-ph/0104127}}].

\bibitem{Pagani:2023wgc}
D.~Pagani, T.~Vitos and M.~Zaro, \emph{{Improving NLO QCD event generators with
  high-energy EW corrections}},
  \href{http://dx.doi.org/10.1140/epjc/s10052-024-12836-0}{\emph{Eur. Phys. J.
  C} {\bf 84} (2024) 514}, [\href{http://arxiv.org/abs/2309.00452}{{\tt
  2309.00452}}].

\bibitem{Bardeen:1978yd}
W.~A. Bardeen, A.~J. Buras, D.~W. Duke and T.~Muta, \emph{{Deep Inelastic
  Scattering Beyond the Leading Order in Asymptotically Free Gauge Theories}},
  \href{http://dx.doi.org/10.1103/PhysRevD.18.3998}{\emph{Phys. Rev. D} {\bf
  18} (1978) 3998}.

\bibitem{Frixione:1995ms}
S.~Frixione, Z.~Kunszt and A.~Signer, \emph{{Three jet cross-sections to
  next-to-leading order}},
  \href{http://dx.doi.org/10.1016/0550-3213(96)00110-1}{\emph{Nucl. Phys.} {\bf
  B467} (1996) 399--442}, [\href{http://arxiv.org/abs/hep-ph/9512328}{{\tt
  hep-ph/9512328}}].

\bibitem{Frixione:1997np}
S.~Frixione, \emph{{A General approach to jet cross-sections in QCD}},
  \href{http://dx.doi.org/10.1016/S0550-3213(97)00574-9}{\emph{Nucl. Phys.}
  {\bf B507} (1997) 295--314}, [\href{http://arxiv.org/abs/hep-ph/9706545}{{\tt
  hep-ph/9706545}}].

\bibitem{Frederix:2009yq}
R.~Frederix, S.~Frixione, F.~Maltoni and T.~Stelzer, \emph{{Automation of
  next-to-leading order computations in QCD: The FKS subtraction}},
  \href{http://dx.doi.org/10.1088/1126-6708/2009/10/003}{\emph{JHEP} {\bf 0910}
  (2009) 003}, [\href{http://arxiv.org/abs/0908.4272}{{\tt 0908.4272}}].

\end{thebibliography}\endgroup

\end{document}